\documentclass[%
 aip,
 amsmath,amssymb,
preprint,%
author-year,%
]{revtex4-2}

\usepackage{graphicx}%
\usepackage{rotating}
\usepackage{dcolumn}%
\usepackage{bm}%
\usepackage[mathlines]{lineno}%

\usepackage[utf8]{inputenc}
\usepackage[T1]{fontenc}
\usepackage{mathptmx}
\usepackage{etoolbox}

\usepackage{epsfig}

\usepackage{nomencl}%

\usepackage{amssymb}
\usepackage{amsthm}

\usepackage{lineno}

\usepackage{etex}
\usepackage{graphicx}
\usepackage{epstopdf, epsfig}
\usepackage[T1]{fontenc}
\usepackage{tikz}
\usepackage{calc}

\usepackage{amssymb}
\usepackage{booktabs}

\usepackage[english]{babel}
\usepackage[T1]{fontenc}
\usepackage[utf8]{inputenc}
\usepackage{tabu}
\usepackage[usestackEOL]{stackengine}
\setstackgap{L}{\normalbaselineskip}
\usepackage{soul} %
\usepackage{notoccite} %
\usepackage{mathtools} %

\usepackage{svg}
\usepackage[hidelinks]{hyperref}

\usepackage{cleveref}
\crefname{figure}{Fig.}{Fig.}
\crefname{equation}{Eq.}{Eq.}

\usepackage{CJKutf8} %

\usepackage{bm}

\usepackage{placeins}

\usepackage{cancel}

\usepackage{clipboard}%

\usepackage{enumitem}

\newcommand{\sci}[1] {\times 10^{#1}} %

\newcommand{\ie}{i.e.\ }%

\newcommand{\eg}{e.g.\ }

\newcommand{\re}{Re}
\newcommand{\str}{St}

\newcommand{\lam}{\lambda}

\newcommand{\uinf}{u'_{\rm \infty}}

\newcommand{\ctm}{\bar{C}_{\rm T,pair}}
\newcommand{\ctms}{\bar{C}_{\rm T,single}}

\newcommand*\diff{\mathop{}\!\mathrm{d}}

\newcommand{\caseNum}{550}

\newcommand{\coa}{\color{black}}

\newcommand{\cob}{\color{black}} %

\newcounter{testa}
\newcommand\countera{\stepcounter{testa}\alph{testa}}

\newcommand{\widthb}{0.32}

\makeatletter
\def\@email#1#2{%
 \endgroup
 \patchcmd{\titleblock@produce}
  {\frontmatter@RRAPformat}
  {\frontmatter@RRAPformat{\produce@RRAP{*#1\href{mailto:#2}{#2}}}\frontmatter@RRAPformat}
  {}{}
}%
\makeatother
\begin{document}

	\hypersetup{hidelinks}

\renewcommand{\thefigure}{\arabic{figure}}
\setcounter{figure}{0}

\renewcommand{\theequation}{\arabic{equation}}
\setcounter{equation}{0}

\renewcommand{\thetable}{\arabic{table}}
\setcounter{table}{0}

\preprint{POF23-AR-02820}

\title[]{How wavelength affects hydrodynamic performance of two accelerating mirror-symmetric undulating hydrofoils}

\author{Zhonglu Lin {\coa (林中路)}}
\affiliation{Key Laboratory of Underwater Acoustic Communication and Marine Information Technology of the Ministry of Education, College of Ocean and Earth Sciences, Xiamen University, Xiamen City, Fujian Province, 361005, China}%
\affiliation{Engineering Department, University of Cambridge, Cambridge City, Cambridgeshire, CB2 1PZ, United Kingdom}%
\affiliation{State Key Laboratory of Marine Environmental Science, College of Ocean and Earth Sciences, Xiamen University, Xiamen City, Fujian Province, 361005, China}

\author{Dongfang Liang {\coa (梁东方)}}
\affiliation{Engineering Department, University of Cambridge, Cambridge City, Cambridgeshire, CB2 1PZ, United Kingdom}%

\author{Amneet Pal Singh Bhalla}
\affiliation{Department of Mechanical Engineering, San Diego State University, San Diego City, California State, 92182-1323, United States}%

\author{Ahmed A. Sheikh Al-Shabab}
\affiliation{School of Aerospace, Transport and Manufacturing, Cranfield University, Cranfield City, Bedfordshire, MK43 0AL, United Kingdom}%

\author{Martin Skote}
\affiliation{School of Aerospace, Transport and Manufacturing, Cranfield University, Cranfield City, Bedfordshire, MK43 0AL, United Kingdom}%

\author{Wei Zheng {\coa (郑炜)}}
\affiliation{Key Laboratory of Underwater Acoustic Communication and Marine Information Technology of the Ministry of Education, College of Ocean and Earth Sciences, Xiamen University, Xiamen City, Fujian Province, 361005, China}%
\affiliation{Pen-Tung Sah Institute of Micro-Nano Science and Technology, Xiamen University, Xiamen City, Fujian Province, 361005, China}
\affiliation{Discipline of Intelligent Instrument and Equipment, Xiamen University, Xiamen City, Fujian Province, 361005, China}
\affiliation{State Key Laboratory of Marine Environmental Science, College of Ocean and Earth Sciences, Xiamen University, Xiamen City, Fujian Province, 361005, China}

\author{Yu Zhang {\coa (张宇)}}%
\affiliation{Key Laboratory of Underwater Acoustic Communication and Marine Information Technology of the Ministry of Education, College of Ocean and Earth Sciences, Xiamen University, Xiamen City, Fujian Province, 361005, China}%
\affiliation{State Key Laboratory of Marine Environmental Science, College of Ocean and Earth Sciences, Xiamen University, Xiamen City, Fujian Province, 361005, China}
\email{yuzhang@xmu.edu.cn}

\date{\today}%

\begin{abstract}
Fish schools are capable of simultaneous linear acceleration. To reveal the underlying hydrodynamic mechanism, we numerically investigate how Reynolds number $ \re = 1000 - 2000 $, Strouhal number $ \str = 0.2 - 0.7 $ and wavelength $ \lam = 0.5 - 2 $ affect the mean net thrust and net propulsive efficiency of two side-by-side hydrofoils undulating in anti-phase.
In total, $ \caseNum $ cases are simulated using immersed boundary method. The thrust increases significantly with wavelength and Strouhal number, yet only slightly with the Reynolds number. We apply a symbolic regression algorithm to formulate this relationship. Furthermore, we find that mirror-symmetric schooling can achieve a \textit{net} thrust more than ten times that of a single swimmer, especially at low Reynolds numbers.
\linelabel{line:abstract_cor}
The highest efficiency is obtained at $ \str = 0.5 $ and $ \lam = 1.2 $, where $ \str $ is consistent with that observed in the linear-accelerating natural swimmers, \eg Crevalle jack. Six distinct flow structures are identified. The highest thrust corresponds to an asymmetric flow pattern, whereas the highest efficiency occurs when the flow is symmetric with converging vortex streets.
\end{abstract}

\begin{CJK*}{UTF8}{gkai} %
\maketitle
\end{CJK*}

\setcounter{section}{0}
\section{Introduction}

\subsection{Background overview}

\linelabel{line:err_1}
Fish swimming have been extensively studied for decades in various disciplines, \eg {\cob morphology \citep{Webb1984}}, animal behaviour \citep{Ashraf2017}, robotics \citep{Li2020} and {\cob especially hydrodynamics \citep{weihs1973hydromechanics,Borazjani2010,Dong2007,Maertens2015,Maertens2017,pan2020computational,Chao2019,Chao2021}}.
The investigation of fish swimming mechanisms can inspire the next-generation biomimetic design of autonomous underwater vehicles (AUVs), as the locomotive performance of commercially available AUVs is yet to match those of natural swimmers \citep{Fish2020}.
The present research focuses on the effects of wavelength on two undulating NACA0012 hydrofoils swimming side-by-side. The results can help understand the underlying mechanism of accelerated fish schools. Simultaneous acceleration of fish school can frequently occur in nature \citep{Partridge1981} to evade predators \citep{Zheng2005,Deng2021} or to conduct collective manoeuvre \citep{Lecheval2018} using vision \citep{Rosenthal2015}, lateral line \citep{Coombs2014} and proprioceptive sensing \citep{Li2021a}. Although the effects of wavelength have been investigated in the context of a single swimmer \citep{Thekkethil2017,Khalid2021,chao2022hydrodynamic}, it remains an open question as to how wavelength kinematics affect accelerating fish schools. The following reviews on fish swimming studies regarding the wavelength effect, acceleration, and side-by-side fish schooling highlight the research gap that can be filled by the present study.

In nature, the swimming body wavelength is not only different across various species with different swimming styles at {\cob steady swimming \citep{Santo2021}}, but also varies with the locomotion phase of a single swimmer \citep{DuClos2019}, \eg starting from rest, linear acceleration and steady swimming.
\cite{Santo2021} recently conducted a comparative study on the kinematics of 44 body-caudal fin (BCF) fish species, focusing on the {steady swimming} phase.
\cite{Santo2021} summarised and compared the wavelengths of different BCF species, considering the four classic swimming styles of \textit{anguilliform}, \textit{subcarangiform}, \textit{carangiform} and \textit{thunniform}. The median wavelength significantly increased from \textit{anguilliform} (0.75 body length) to \textit{thunniform} (1.14 body length), yet the wavelength for the tested species occupies a broad range from 0.5 body length to 1.5 body length.
For conciseness, we abbreviate "body length" as "BL" in the following content.
The wavelength of four swimming styles overlaps from 0.75BL to 1.35BL, indicating this wavelength range may be compatible with swimmers of various body shapes and swimmers of different swimming styles.
The results by \cite{Santo2021} have driven us to focus on a similar range of wavelengths, which {\cob will be presented later in \Cref{sec_problem_setup}}.
\cite{DuClos2019} studied how an \textit{anguilliform} swimmer accelerates from rest, and compared the kinematics during steady swimming and acceleration. They discovered that the wavelength during escape acceleration $ \lam \approx 2 $, much longer than that during steady swimming with $ \lam = 0.8 $.
\cite{Nangia2017a} conducted a meta-analysis regarding the wavelength and Strouhal number for various BCF species, finding a convergence of the ratio of wavelength to the tail amplitude during undulation. \cite{Nangia2017a} utilised the ConstraintIB module of open-source immersed boundary software, IBAMR, which is also used in our present study. 

\linelabel{line:error_2_title}
\subsection{Wavelength effects for a single swimmer during steady-swimming {\cob and linear acceleration}}
\label{sec:wavelength_single}
Many numerical studies focused on the wavelength effect of a single swimmer, both in 2D and 3D.
\cite{chao2022hydrodynamic} recently conducted a thorough investigation of the hydrodynamic performance of a single slender swimmer with various Strouhal numbers $ St = 0.1 - 1 $, Reynolds number $ Re = 50 - 2000 $, and $ \lam = 0.5 - 2 $, and they discovered seven types of wake structures. They were able to condense the simulation results into a few formulas. Their study has inspired the present study's choice of parametric space.
\cite{Khalid2020} studied how a single tethered undulating 2D NACA0012 hydrofoil performs with either \textit{anguilliform} or \textit{carangiform} kinematics in a parametric sweep of $ Re = 100,\ 1000,\ 5000 $, $ St = 0.1 - 0.8 $, and $ \lam = 0.5 - 1.5 $. They found that wavelengths do not necessarily optimise the hydrodynamic performance of natural swimmers.
\cite{Thekkethil2017}, \cite{Thekkethil2018,Thekkethil2020} and \cite{Gupta2021} conducted a series of studies regarding how the wavelength affects the thrust and propulsive efficiency of a single undulating NACA0012 hydrofoil.
\cite{Khalid2021} conducted high fidelity 3D simulations of a steady-swimming American eel at $ St = 0.3 - 0.4 $, investigating the influences of wavelengths at $ \lam = 0.65 - 1.25 $. They found that short wavelengths are more hydrodynamically advantageous for \textit{anguilliform} swimmers during their steady motion.
\cite{Borazjani2008,Borazjani2009} conducted a 3D simulation to study a steady-swimming \textit{carangiform} \citep{Borazjani2008} and an \textit{anguilliform} \citep{Borazjani2009} swimmer tethered in a free stream flow at $ Re = 300,\ 4000 $, whereas the wavelength is configured at $ \lam = 0.642-1.1 $.

{\coa In addition, in the present paper, we apply 2D model rather than 3D considering computational cost to simulate $ \caseNum $ cases. What is more, 2D simulation has proven to reveal fundamental patterns in undulating hydrofoils with various wavelengths in the laminar flow regime \citep{chao2022hydrodynamic}.}

The above-mentioned studies have focused on the steady swimming phase. However, steady-swimming is a rare scenario for fish swimming.  Other conditions include the {\cob starting from rest \citep{Domenici2019}} and the linear acceleration \citep{Akanyeti2017}. The linear acceleration occurs when they travel or hold a position in a variable speed or turbulent flow \cite{Tytell2004b}. While steady-swimming and fast-start \citep{Eaton1977,Tytell2008,Borazjani2012,Borazjani2013} are relatively well studied, linear acceleration is still not well understood in both biological and hydrodynamic aspects for a single swimmer, let alone for fish schools.

Wavelength correlates significantly with acceleration and speed during the linear acceleration phase of fish swimming. The existing biological research almost all focused on a \textit{single} accelerating fish. An overview of the wavelength and swimming styles is depicted in \Cref{fig:4modes}.
\cite{Schwalbe2019} scrutinised the function of red muscle, \ie slow-twitch muscle for sustained activities, during the acceleration of bluegill sunfish \textit{Lepomis macrochirus} and how it affects the fish kinematics. They discovered that the fish's undulation kinematics during acceleration differs from that during steady swimming. Body wavelength decreases significantly during acceleration, yet increases significantly with swimming speed. Their research focused on the bluegill sunfish \textit{Lepomis macrochirus}. At different acceleration levels, the fish body wavelength can range from 0.75BL to 0.9BL. They focused on the fish muscle activation and observation of kinematics.
\cite{Akanyeti2017} conducted both biological and robotic fish experiments to investigate the kinematic characteristics and hydrodynamic performance during linear acceleration. Their investigation was carried out using a \textit{tethered} robotic fish while varying the free stream flow to study the \textit{acceleration} at consecutive instants. Our present problem setup is similar to their experimental configuration. They found that tail-beat frequency, rather than amplitude, is most effective on swimming speed and acceleration. The tail-beat amplitude remains constant during steady swimming or acceleration. Hence, our present study fixes tail-beat amplitude while varying the Strouhal number.
\cite{Tytell2004b} conducted the first quantitative research on the linear acceleration of an \textit{anguilliform} swimmer, focusing on its kinematics and wake hydrodynamics. They found that body wavelength $ \lam $ and tailbeat frequency both significantly increase with steady swimming speed.

\subsection{Side-by-side {\cob steady} swimming with constant wavelength}

Side-by-side fish schooling is relatively well studied, yet most studies have focused on the steady swimming scenario with a fixed wavelength.
\cite{Ashraf2017} conducted a fish schooling experiment, discovering that fish favours the phalanx formation, \ie side-by-side of multiple fish, at relatively high steady-swimming speed. The tested Reynolds number ranges from 1000 to 6000.
\cite{Li2021} investigated the schooling of two robotic fish of \textit{subcarangiform}, swimming steadily side-by-side with rigid linking between them, with the lateral distance fixed at 0.33BL. For schooling swimmers, they discovered maximum speed and efficiency at in-phase and anti-phase conditions, respectively, which are compared with a single swimmer. In the present paper, we also fix the gap distance at 0.33BL.
For side-by-side and anti-phase pitching foils, previous studies reached a consensus that much higher thrust can be produced with efficiency similar to a single swimmer \citep{Dewey2014a,Huera-Huarte2018,Gungor2021,Yucel2022}. Our present paper further investigates the hydrodynamic effects due to various wavelengths.
\cite{Li2020} conducted a thorough study on a tethered two-fish school at a steady swimming phase, combining robotic fish experiments and biological observation. They discovered that the front-back distance and phase difference most significantly affected schooling performance. The lateral distance varies from 0.27 to 0.33 in this study.
\cite{Shrivastava2017} conducted a 2D simulation of three hydrofoils swimming side-by-side. Interaction between the swimmers can be observed at {\coa a} lateral distance less than 1BL with $ St = 0.4 - 0.8 $, $ A_{\rm max} $, $ \lam = 1 $, $ Re = 400 $. 
\cite{Wei2022} simulated two initially side-by-side swimmers passively self-propelling with three degrees of freedom. They configured the initial gap ratio as $ G = 0.4 - 0.9 $, and observed improved schooling performance between the two swimmers.
Therefore, the present lateral distance of 0.33BL should allow sufficient schooling interaction between the two swimmers.
\cite{gungor2022classification} investigated the unsteady hydrodynamics of two pitching foils in side-by-side formation. By drawing wake maps, they discovered three distinct vortex patterns of separated, merged, and transitional-merged wake.

\linelabel{line:err_3_deleted}

\subsection{Present study scope}

{\coa In summary, the above-mentioned literature has inspired us that fish-body wavelength is a key factor affecting the fluid-structure interaction of fish-like swimmers, yet the research regarding wavelength-effect on \textit{schooling} and \textit{accelerating} swimmers are relatively scarce, despite its occurrence in the nature, \eg fish school escaping from predator.}
{\coa Also, we use the term "mirror-symmetric" as an equivalent of "side-by-side and anti-phase", not only for the sake of conciseness but also because we would like to strengthen the physically-interesting phenomenon of flow symmetry and its breaking at certain conditions.}
{\coa More specifically,} although excellent studies have emerged to examine single fish wavelength effects \citep{Santo2021,Khalid2021,Thekkethil2020,DuClos2019,Nangia2017a}, linear acceleration \citep{Tytell2004b,Akanyeti2017,Schwalbe2019} and side-by-side schooling \citep{Wei2022,Li2021,Shrivastava2017,Ashraf2017}, how body wavelength affects the linear acceleration of two side-by-side fishlike swimmers has never been systematically investigated. We aim to present a thorough investigation in the present paper, which can be helpful {\coa in understanding} the fish schooling behaviour and to design the collective locomotion strategy of underwater fish-like AUVs.
The rich physics in side-by-side fish schooling can also be relevant to the flow mediated interaction between two oscillating cylinders \citep{lin2022flow,Lin2019,Lin2018c,Lin2018b,Lin2017a,Lin2016a,Gazzola2012,NAIR2007,Hlamb1932hydrodynamics}.

	\newcommand{\addlabeltrim}[3]{%
	\begin{tikzpicture}
		\node[anchor=south west,inner sep=0] (image) at (0,0) 
		{\includegraphics[width=#1\linewidth, trim={0cm 1.3cm 0cm 0.8cm},clip]{{{#2}}}};%
		\begin{scope}[x={(image.south east)},y={(image.north west)}]
			\node[anchor=south west] at (0.00,0.75) {\footnotesize #3};	%
		\end{scope}
	\end{tikzpicture}%
}
	\newcommand{\addlabelintro}[3]{%
	\begin{tikzpicture}
		\node[anchor=south west,inner sep=0] (image) at (0,0) 
		{\includegraphics[width=#1\linewidth]{{{#2}}}};%
		\begin{scope}[x={(image.south east)},y={(image.north west)}]
			\node[anchor=south west] at (0.3,0.9) {\footnotesize #3};	%
		\end{scope}
	\end{tikzpicture}%
}
\newcounter{testdd}
\setcounter{testdd}{0}
\newcommand\counterdd{\stepcounter{testdd}\alph{testdd}}
\begin{figure*}
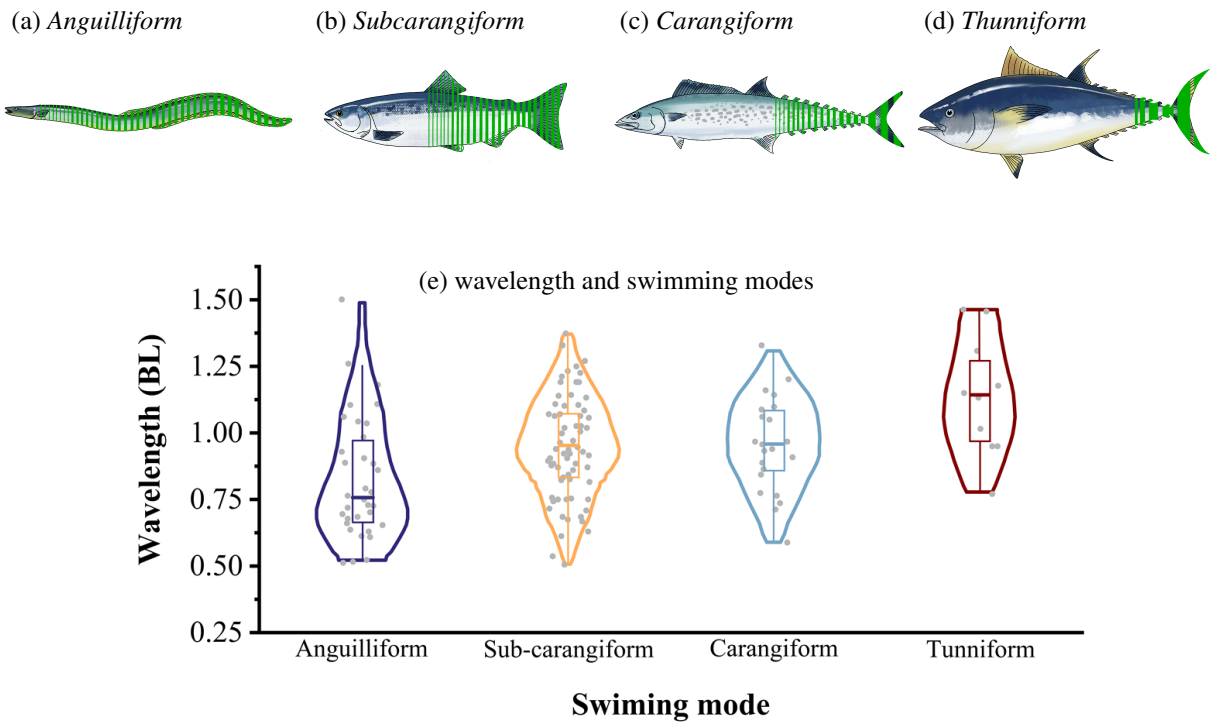

	\centering
	\addlabeltrim{0.24}{1_Eels_green}{(\counterdd) \textit{Anguilliform}}
	\addlabeltrim{0.24}{2_Salomnids_green}{(\counterdd) \textit{Subcarangiform}}
	\addlabeltrim{0.24}{3_Makrell_green}{(\counterdd) \textit{Carangiform}}
	\addlabeltrim{0.24}{4_Tuna_green}{(\counterdd) \textit{Thunniform}}
	
	\addlabelintro{0.8}{intro-swimmingform-wavelength-distribution-steady-swimming}{(\counterdd) wavelength and swimming modes}
	\caption{Four different swimming modes of Body-Caudal-Fin type locomotion
		(a) \textit{Anguilliform} (body undulation, \eg eel)
		(b) \textit{Subcarangiform} (body undulation with caudal fin pitching, \eg salmonid)
		(c) \textit{Carangiform} (minor body undulation with caudal fin pitching, \eg makrell)
		(d) \textit{Thunniform} (mainly caudal fin pitching, \eg tuna). The shaded area demonstrates the body parts with a significant lateral motion to generate thrust (redrawn from figures by \cite{Lindsey1978} and \cite{Sfakiotakis1999}).
		(d) distribution of wavelength with the four swimming modes (adapted from figure by \cite{Santo2021}): These four types have wavelengths ranging from 0.5 to 2 body length for steady swimming conditions across various fish species regardless of the aforementioned body-caudal fin sub-types.}
	\label{fig:4modes}
\end{figure*}

\FloatBarrier
\clearpage

\section{Methodology}
In this section, we present the methodology of the current study.
\Cref{sec_problem_setup} describes the representative problem setup accounting for the schooling swimmers, including geometry, kinematic equation, and non-dimensional analysis.
\Cref{sec_Comput_method} discusses the computational method to implement the problem setup.

\subsection{Problem setup}
\label{sec_problem_setup}

\begin{figure}
	\centering
	\includegraphics[width=1\linewidth]{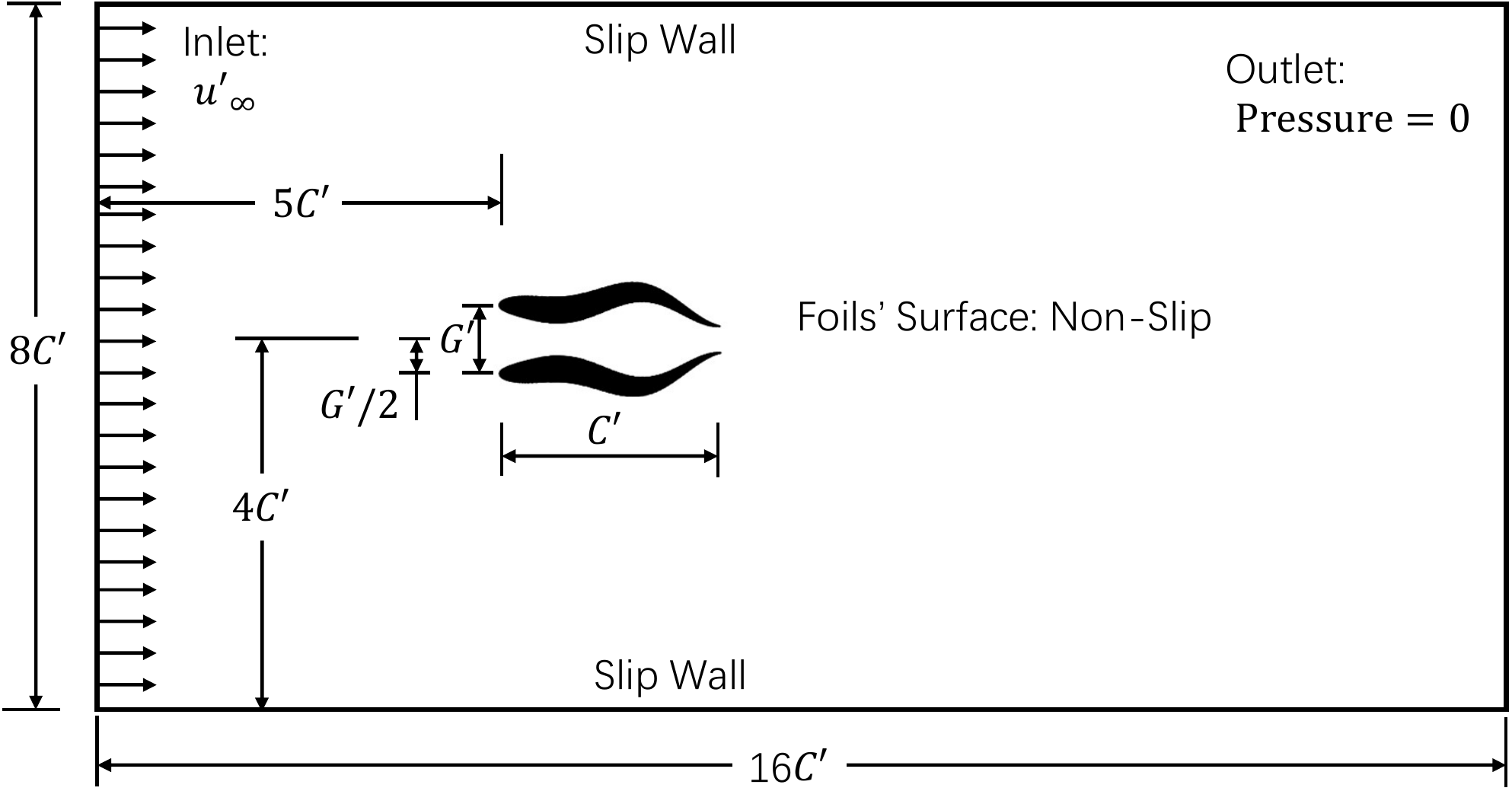}
	\caption{Sketch of the present problem setup: two side-by-side fish swimming in anti-phase with a fixed lateral gap distance $ G = 0.33 $. The two swimmers are mirror symmetric to each other along the horizontal line $ y = 4C $.}
	\label{fig:problemsetup}
\end{figure}

{\coa It is not uncommon to observe simplified models relevant to fish swimming, including a travelling wavy boundary \citep{Ma2019,Wang2021}, filaments \citep{ni2023mode}, and undulating NACA foils \citep{lin2022swimming}.
The present problem setup applies NACA0012 hydrofoil to represent the fish-like swimmer, since the NACA foils have been extensively used as a representation in the previous investigations \citep{Deng2022,Deng2016,Deng2015,Shao2010,Deng2007,yu2021scaling,pan2022effects}.} The complete configuration is summarised in \Cref{fig:problemsetup}. The accelerated fish schooling problem is represented by a two-dimensional form with two wavy hydrofoils undulating side-by-side.
The 2D configuration should adequately describe the present laminar flow regime with $ Re \leq 2000 $ \citep{Gazzola2014,chao2022hydrodynamic}.

The fish body is simplified as a 2D NACA0012 hydrofoil to describe bio-propulsion problems with pitching \citep{Moriche2016} and undulating hydrofoils \citep{Thekkethil2017}.
The geometry of a NACA0012 hydrofoil is similar to that of a \textit{carangiform} or \textit{subcarangiform} swimmer.
The two foils are placed side-by-side while undulating in anti-phase to concentrate on a limited number of variables typical for fish schooling \citep{Ashraf2017}. The kinematics of the swimmers {\coa is} described by the travelling wave equations in the non-dimensional form:

\begin{equation}\label{equ:fish_wave_leader}
	 Y_{\rm 1} = A_{\rm max} X_{\rm 1} \sin \left[ 2\pi \left( \frac{{X}_{\rm 1}}{\lam} - \frac{St }{2 A_{\rm max}} t \right) \right]
\end{equation}
\begin{equation}\label{equ:fish_wave_follower}
	 Y_{\rm 2} = A_{\rm max} X_{\rm 2} \sin \left[ 2\pi \left( \frac{ {X}_{\rm 2} }{\lam} - \frac{St }{2 A_{\rm max}} t \right) + \pi \right]
\end{equation}

This is also a common configuration \citep{Thekkethil2018}, and is chosen here for the convenience of comparison. For completeness, the meaning of the variables is listed as follows: $ Y_{\rm i} = Y'_{\rm i} / C'  $ is the centre-line lateral displacement of each hydrofoil; $ X_{\rm i} = X'_{\rm i} / C' $ is the streamwise position on the centreline; $ i = 1 $ denotes the top swimmer while $ i = 2 $ represents the bottom swimmer. $ t = t' \uinf /C' $ is non-dimensional time; $ \uinf $ is the free-stream velocity; $ A_{\rm max} = A'_{\rm max}/C' $ is non-dimensional tail tip amplitude, where $ C' $ is the fish body length; $ a_{\rm max} $ is the dimensional tail-amplitude; $ \lam = \lam'/C' $ is non-dimensional wavelength, with $ \lam' $ being the dimensional foil undulating wavelength; $ St = 2f'A'_{\rm max} / \uinf $ is Strouhal number, with $ f' $ being the dimensional undulating frequency. Here, dashed alphabets denote dimensional parameters.

In addition, non-dimensional groups to describe a particular case are listed in \Cref{tablecasegroups} together with the investigated parametric space, in which $ \rho' $ is the fluid density; $ f' $ is the undulating frequency; $ \mu' $ is the dynamic viscosity. The non-dimensional lateral and front-back distances are fixed at $ G = G'/C' = 0.33 $ and $ D = D'/C' = 0 $. In summary, only three variables are involved in the present study: Reynolds number $ \re = 1000-2000 $, Strouhal number $ St=0.2-0.7 $, and non-dimensional wavelength $ \lam = 0.5 - 2.0 $.

The outputted metrics of the swimming performance are listed in \Cref{tab:output_para}. Thrust is directly relevant to acceleration, whereas net propulsive efficiency measures how efficiently the input energy is converted to the net thrust for acceleration. The examination of the vorticity field is necessary to examine the flow symmetry and stealth capacity. In \Cref{tab:output_para}, $ F_{\rm T,i} $ is the net thrust on hydrofoils; $ C_{\rm L,i} $ is the instantaneous lateral force coefficient; $ \bm{u} $ is the fluid velocity. $ \omega^* $ is the non-dimensional vorticity, whereas $ P^* $ stands for the non-dimensional pressure.

\begin{ruledtabular}
	\begin{table*}[thb]
		\caption{Non-dimensional input parameters and the involved range of value}
		\centering
		\label{tablecasegroups}
		\begin{tabular}{l c c c}
			Reynolds number & $ \re $ &{ $ { \rho' u'_{\rm \infty} C'}/{\mu'} $} & $ 1000 - 2000 $ \\
			
			Strouhal number  & $ St $ &{ $ {2 f' A'_{\rm max}}/{u'_{\rm \infty}} $} & $ 0.2 - 0.7 $ \\
			
			Wavelength  & $ \lam $ &{  $ {\lambda'}/{C'} $} & $ 0.5 - 2 $ \\

		\end{tabular}
	\end{table*}
\end{ruledtabular}

\begin{ruledtabular}
	\begin{table*}%
		\caption{Non-dimensional output parameters for swimming performance}
		\centering
		\label{tab:output_para}
		\begin{tabular}{l c c c}
			
			Cycle-averaged {net} thrust coefficient & $ \bar{C}_{\rm T,i} $ & $ \frac{1}{T} \int_{\rm t}^{t+T} C_T \diff t =  \frac{1}{T} \int_{\rm t}^{t+T} {2F_{\rm T,i}}/{\rho u^2_{\rm \infty} C} \diff t  $  \\

			Net propulsive efficiency & $ \eta_{\rm i} $ &{ $ {P_{\rm out,i}}/{P_{\rm in,i}} = {\bar{C}_T}/{\bar{C}_P} %
				 $  }  \\

			Fluid vorticity & $ \bm{\omega^*} $ &{ $ \nabla \times \bm{u} $  }  \\
			
			Fluid pressure & $ \bm{P^*} $ &{ $ p / \rho u_{\rm \infty} $  }  \\
			
		\end{tabular}
	\end{table*}
\end{ruledtabular}

\subsection{Simulation method}
\label{sec_Comput_method}

The present paper simulates the problem using a customised version of the ConstraintIB module \citep{Bhalla2013,Griffith2020} implemented in IBAMR \citep{griffith2013ibamr}, which is an open-source immersed boundary method simulation software that depends on several underlying advanced libraries including SAMRAI \citep{Hornung2002,Hornung2006}, PETSc \citep{Balay1997,Balay2010,balay2001petsc}, hypre \citep{falgout2010hypre,Balay1997}, and libmesh \citep{Kirk2006}. It is chosen for its adaptive mesh refinement capacity of the Eulerian background mesh, allowing both computational efficiency and adequate accuracy. The ConstraintIB method has been extensively validated \citep{Bhalla2013,Bhalla2013a,bhalla2014fully,Nangia2017,Nangia2019,Griffith2020,Bhalla2020}. The present customised version has also been validated in \citep{lin2022swimming}. The maximum Reynolds number $ Re \leq 2000 $ in the present study is lower than that in a previous study \citep{lin2022swimming} with $ Re = 5000 $, so here we adopt the same mesh refinement and time step setting that has been verified for mesh independence.
Each numerical simulation {\coa was} run for 20 cycles of undulation.

\subsection{Symbolic regression method}
\label{subsec:symbolic}
The open-source symbolic regression library, PySR \citep{cranmer2023interpretable}, is utilised to automatically extract the interpretable symbolic models for net thrust force, \ie \Cref{equ:SR_predict}, from the data accumulated from the $ \caseNum $ simulated cases. It is based on multi-population evolutionary algorithm with a special evolve-simplify-optimize cycle, being capable to high efficiency parallel computation with integration to deep learning tools. PySR has been proven useful in many studies, including cloud cover formation \citep{grundner2023data}, electron transfer rules \citep{li2023electron} and discovering astrophysical relations \citep{matchev2022analytical}. The present study chose PySR for the customisable configuration that is capable of reducing the regression time, and for the parallelisation that speeds up the data-discovery speed.

\FloatBarrier
\clearpage

\section{Results and discussion}
\label{sec_result}

In this section, we present and discuss our discoveries from $ \caseNum $ cases regarding two side-by-side and anti-phase wavy NACA0012 hydrofoils in the parametric space of Reynolds number $ \re = 1000 - 2000 $, Strouhal number $ \str = 0.2 - 0.7 $, and wavelength $ \lam = 0.5 - 2 $.
\Cref{sec_ctm} discusses how $\str$, $\lam$ and $ \re $ influence the net thrust $ \ctm $ for each swimmer by drawing heat maps while generating a formula for a high-level summary. We also compare the simulated schooling thrust with the analytical formula describing thrust by a single swimmer \citep{chao2022hydrodynamic} finding interesting results.
\Cref{sec_effi} presents the dependence of net propulsive efficiency on $\str$, $\lam$ and $\re$, with extra focus on the cases with the highest efficiency.
In \Cref{sec_flow_map}, we classify observed flow structures into several types, while connecting them to high-thrust or high-efficiency regimes.

\subsection{How $\str$, $\lam$ and $ \re $ affect net thrust $ \ctm $}
\label{sec_ctm}

This subsection discusses how net propulsive efficiency varies with Reynolds number $ Re = 1000 - 2000 $, Strouhal number $ \str = 0.2 - 0.7 $, and wavelength $ \lambda = 0.5 - 2 $. The thrust generally increases with Strouhal number and wavelength, whereas the effect of Reynolds number is marginal, as demonstrated in \Cref{fig:heat_map_ctm}. The contour line of $ \ctm = 0 $ at the white region denotes the steady swimming state, which is often discussed in numerical \citep{Borazjani2008} and experimental \citep{Li2020} studies of fish swimming. With a higher $ \str $ or $ \lam $, the thrust becomes positive $ \ctm > 0 $, \ie the hydrofoil school is accelerating. Conversely, with a lower $ \str $ or $ \lam $, the foils are decelerating with negative thrust. $ Re $ only slightly affects this overall trend. This pattern corresponds well with the single swimmer scenario discussed in \citep{chao2022hydrodynamic}, where the effect of $ Re $ on net thrust becomes insignificant at $ Re > 1000 $ while a positive correlation exists between net thrust and wavelength/Strouhal number.
Here, the representative net thrust is calculated as the average value for the two swimmers, \ie $ \ctm = (\bar{C}_{\rm T,1} + \bar{C}_{\rm T,2})/2 $, where $ \ctm = \bar{C}_{\rm T,1} = \bar{C}_{\rm T,2} $ establishes for the symmetrical cases. $ \bar{C}_{\rm T,i} $ denotes the net thrust on the $ i $th swimmer.

Here, we offer a high-level summary of the mean net thrust $ \ctm $ of the side-by-side and anti-phase scenarios. Following the formal analysis of previous studies regarding a flapping foil \citep{Floryan2017,VanBuren2017,Alam2020} and an undulating foil \citep{chao2022hydrodynamic}, we use the symbolic regression tool PySR \citep{pysr,cranmer2020discovering} to automatically produce an interpretable equation that summarises the $ \ctm $ data in the present study for schooling swimmers, as seen in \Cref{equ:SR_predict}:
\begin{equation}\label{equ:SR_predict}
	\ctm = \re^{0.17} \str^{2.03} \lam^{1.23} - 0.26 \re^{0.19} \str^{1.00} \lam^{0.10} - 6.13 \re^{-0.6}
\end{equation}
The summarising capability of \Cref{equ:SR_predict} can be demonstrated in \Cref{fig:heat_map_ctm}f, with coefficient of determination reaching $ R^2 = 0.953 $. For the convenience of the readers' comparison, here we also copy the $ \ctm $ equation by \citep{chao2022hydrodynamic} for a single swimmer as: %
\begin{equation}\label{equ:chao_single_ctm}
	\ctms = 0.36 \re^{0.208} \str^{3} \lam  - 6.13 \re^{-0.6}
\end{equation}
We can see that compared with the single swimmer thrust $ \ctms $ formula produced by \cite{chao2022hydrodynamic}, the additional mirror-symmetric swimmer casts an interesting effect on the $ \ctm $ of each schooling member.
Wavelength is almost linearly correlated with net thrust as $ \ctm \sim \lam^{1.23} $ when the Strouhal number and wavelength are relatively small, with the scaling exponent being slightly larger than $ \ctm \sim \lam^{1.00} $. So the wavelength is slightly more influential to the thrust during schooling compared with the single swimming condition.
On the other hand, the primary scaling of thrust coefficient with Strouhal number is reduced from $ \ctms \sim \str^3 $ for a single foil \citep{chao2022hydrodynamic} to $ \ctm \sim \str^{2} $ in \Cref{equ:SR_predict} with an additional negative term as $ - 0.26 \re^{0.19} \str^{1.00} \lam^{0.10} $, so the contribution from Strouhal number to net thrust becomes less significant for the present schooling scenario compared with the single swimmer case. In contrast, the scaling exponent of $ 2.03 $ in \Cref{equ:SR_predict} is very close to that of two side-by-side pitching foils, which scale as $ \ctm \sim \str^2 $ \citep{Gungor2021}.
{\coa
In \Cref{equ:SR_predict}, the third term $ -6.13\re^{-0.6} $ indicates that, for stationary hydrofoils, \ie $ St = 0 $, the unseparated boundary layer causes the domination of fluid drag force \citep{chao2022hydrodynamic}. $ \ctm $ increases slowly with $ \re $ due to reduced viscous force. As $ \re $ grows, the third term diminishes towards zero. 

}

Despite minor scaling for $ \str $, side-by-side schooling can lead to significant net thrust amplification $ \ctm/\ctms $ compared with single swimming condition, as seen in \Cref{fig:heat_map_compare_12_ctm}.
Generally speaking, at higher wavelength $ \lam > 0.4 $ and Strouhal number $ \str > 0.7 $, the net thrust/acceleration from schooling can be higher than a single wavy foil.
Furthermore, at $ \lam > 1.1 $ and $ 0.4 < \str < 0.45 $, the net thrust for each schooling swimmer can be more than ten times larger than a single swimmer!
A lower Reynolds number amplifies the schooling advantage for thrust. This advantageous range of $ \str $ and $ \lam $ also corresponds well with the natural \textit{carangiform} fish species with $ \str \approx 0.4 $ \citep{Borazjani2008}.
So a better thrust performance can be another reason to school together in addition to energy conservation \citep{Daghooghi2015,Li2020}.
This observation might be able to explain why fish school together from the perspective of net thrust; such acceleration can be significant for the predator-prey interaction \citep{Triantafyllou2016}, thus being correlated to the survivorship of swimmers.
The maximum amplification factor is obtained at $ \re = 1000 $ reaching $ \ctm/\ctms = 13 $ at $ \str = 0.3 $ and $ \lam = 2 $, as illustrated in \Cref{fig:heat_map_compare_12_ctm}a.
Maximum amplification factor $ \ctm/\ctms $ generally drops with Reynolds number,
as demonstrated in \Cref{fig:heat_map_compare_12_ctm}f.
This result indicates that a higher wavelength $ \lam = 2 $ is advantageous for a single swimmer accelerating from low speed/Reynolds number \citep{DuClos2019} and even more beneficial for each schooling swimmer.
In addition, we should note that schooling can be less advantageous than swimming alone at $ \ctm/\ctms < 1 $, located at $ 0.5 < \lam < 0.7 $ and $ \str > 0.45 $, although in the present parametric space, schooling can yield much better performance in most cases.

\setcounter{testa}{0}
\newcommand{\addlabele}[3]{%
	\begin{tikzpicture}
		\node[anchor=south west,inner sep=0] (image) at (0,0) 
		{\includegraphics[width=#1\textwidth]{#2}};
		\begin{scope}[x={(image.south east)},y={(image.north west)}]
			\node[anchor=south west] at (0.15,0.90) {\footnotesize #3};
		\end{scope}
	\end{tikzpicture}%
}
\begin{figure}
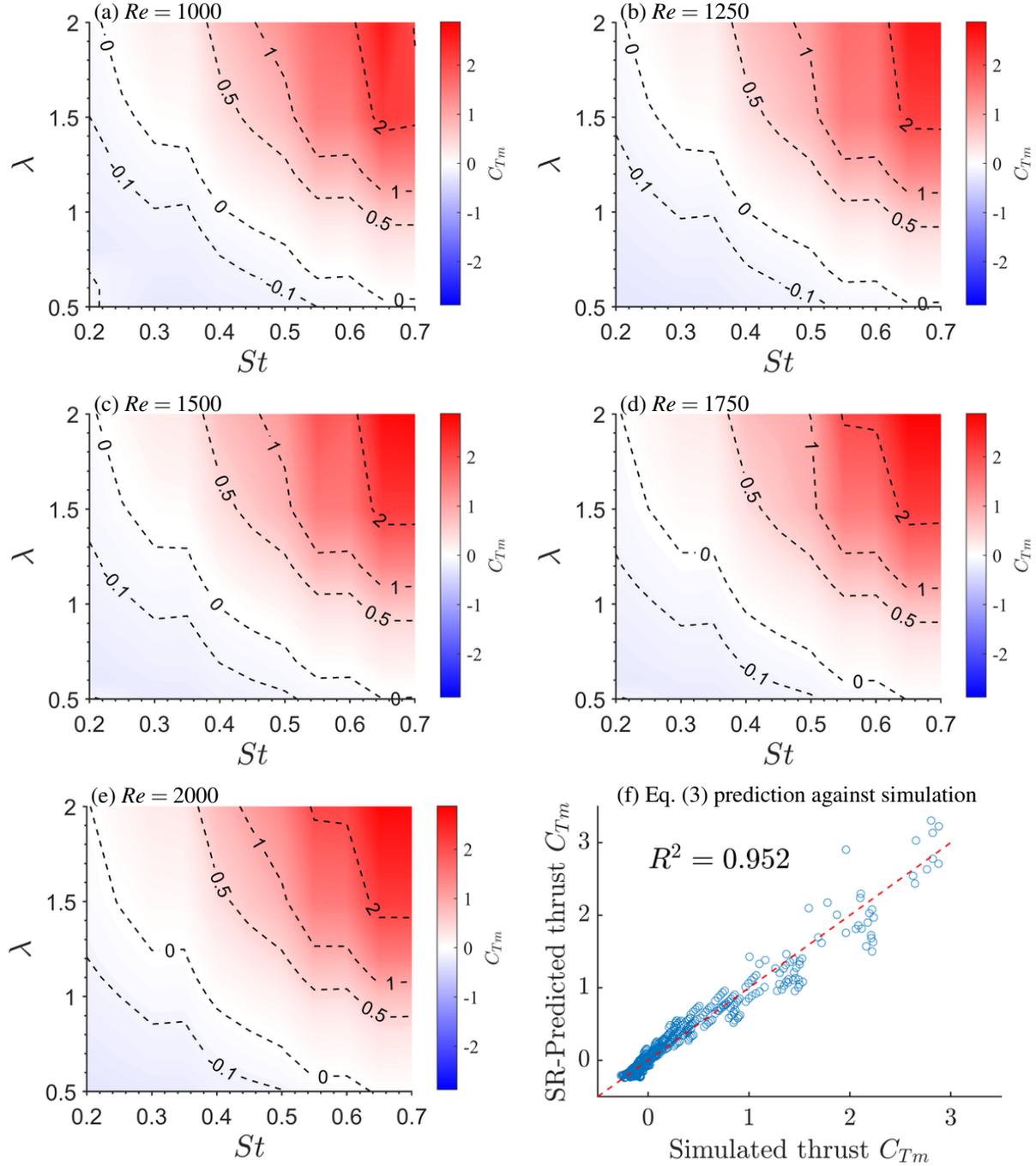

	\centering
	\setcounter{testa}{0}
	\addlabele{0.49}{{{Heat_map_Re_1000__CTm_St_lambda_ppi1_D0_G0d35}}}{(\countera) $ Re = 1000 $}
	\addlabele{0.49}{{{Heat_map_Re_1250__CTm_St_lambda_ppi1_D0_G0d35}}}{(\countera) $ Re = 1250 $}
	\addlabele{0.49}{{{Heat_map_Re_1500__CTm_St_lambda_ppi1_D0_G0d35}}}{(\countera) $ Re = 1500 $}
	\addlabele{0.49}{{{Heat_map_Re_1750__CTm_St_lambda_ppi1_D0_G0d35}}}{(\countera) $ Re = 1750 $}
	\addlabele{0.49}{{{Heat_map_Re_2000__CTm_St_lambda_ppi1_D0_G0d35}}}{(\countera) $ Re = 2000 $}
	\addlabele{0.49}{{{ctm_SR_prediction_accuracy_line}}}{(\countera) \Cref{equ:SR_predict} prediction against simulation}
	\setcounter{testa}{0}
	\caption{Heat map for mean net thrust $ \ctm $ at Strouhal number $ St = 0.2-0.7 $, wavelength $ \lam=0.5-2 $ and Reynolds numbers at 
		(a) $ Re = 1000 $
		(b) $ Re = 1250 $
		(c) $ Re = 1500 $
		(d) $ Re = 1750 $
		(e) $ Re = 2000 $.
		(f) Symbolic regression prediction accuracy comparing simulation results and \Cref{equ:SR_predict}. The thrust on two swimmers is identical due to the symmetrical situation.
	The positive thrust, \ie forward acceleration, is indicated by the positive values with red colour.
	Conversely, the negative thrust, \ie deceleration, is indicated by the negative values with blue colour.
	The contour line of $ \ctm = 0 $ represents the zero net thrust scenarios, \ie steady swimming state.
	Only marginal differences can be observed across various Reynolds numbers.
	}
	\label{fig:heat_map_ctm}
\end{figure}

\setcounter{testa}{0}
\renewcommand{\addlabele}[3]{%
	\begin{tikzpicture}
		\node[anchor=south west,inner sep=0] (image) at (0,0) 
		{\includegraphics[width=#1\textwidth]{#2}};
		\begin{scope}[x={(image.south east)},y={(image.north west)}]
			\node[anchor=south west] at (0.15,0.90) {\footnotesize #3};
		\end{scope}
	\end{tikzpicture}%
}
\renewcommand{\widthb}{0.32}
\begin{figure}
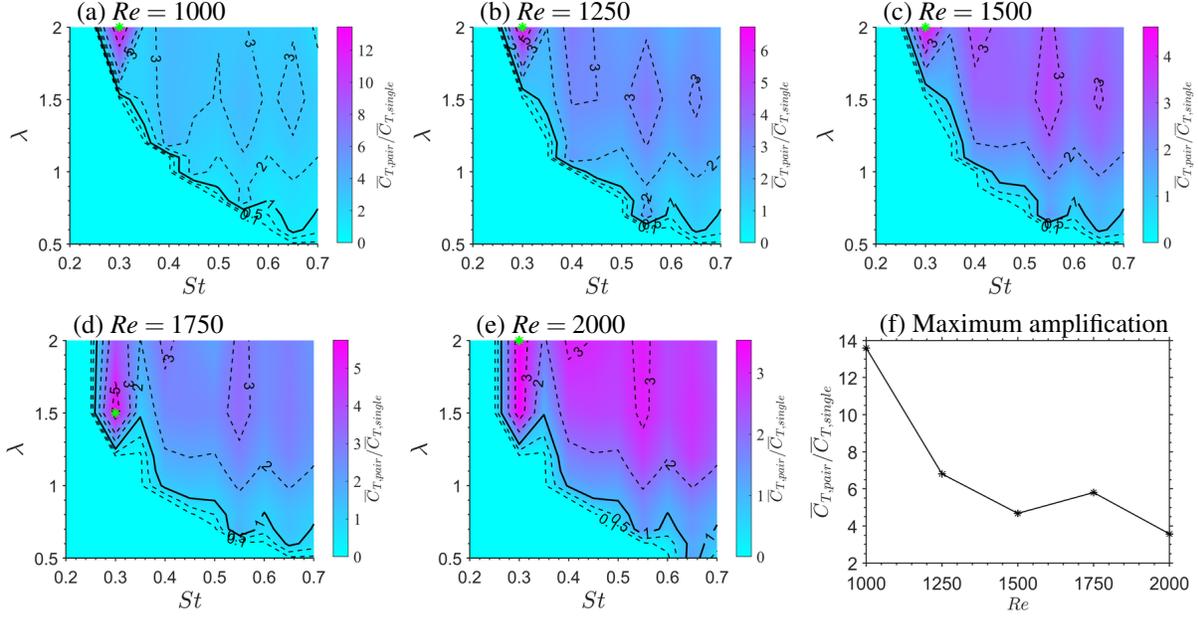

	\centering
	\setcounter{testa}{0}
	\addlabele{\widthb}{{{compare__Heat_map_Re_1000__CTm_St_lambda_ppi1_D0_G0d35}}}{(\countera) $ Re = 1000 $}
	\addlabele{\widthb}{{{compare__Heat_map_Re_1250__CTm_St_lambda_ppi1_D0_G0d35}}}{(\countera) $ Re = 1250 $}
	\addlabele{\widthb}{{{compare__Heat_map_Re_1500__CTm_St_lambda_ppi1_D0_G0d35}}}{(\countera) $ Re = 1500 $}
	\addlabele{\widthb}{{{compare__Heat_map_Re_1750__CTm_St_lambda_ppi1_D0_G0d35}}}{(\countera) $ Re = 1750 $}
	\addlabele{\widthb}{{{compare__Heat_map_Re_2000__CTm_St_lambda_ppi1_D0_G0d35}}}{(\countera) $ Re = 2000 $}
	\addlabele{\widthb}{{{compare__Line_Re_vs_largest_thrust_Amplification}}}{(\countera) Maximum amplification}
	\setcounter{testa}{0}
	\caption{Heat map for \textit{schooling thrust amplification} factor $ \ctm/\ctms $ at Strouhal number $ \str = 0.2-0.7 $, wavelength $ \lam=0.5-2 $ and Reynolds numbers at 
		(a) $ Re = 1000 $
		(b) $ Re = 1250 $
		(c) $ Re = 1500 $
		(d) $ Re = 1750 $
		(e) $ Re = 2000 $
		(f) Maximum thrust amplification due to schooling.
		Here, we only show the results with both $ \ctm > 0 $ and $ \ctms > 0 $; other non-accelerating cases are drawn as zero.
		This map demonstrates how schooling contributes to the thrust/acceleration of the swimmers.
		The thick contour line of $ \ctm/\ctms = 1 $ indicates that thrust from schooling equals that from a single foil.
		$ \ctm/\ctms > 1 $ means that each schooling member produces higher thrust than a single swimmer, and vice versa for $ \ctm/\ctms < 1 $.
		The green marker denotes the location for the highest thrust amplification for each Reynolds number.
		At high $ \str $ and $ \lam $, schooling can produce more thrust. Further, the schooling thrust can be several times higher than a single swimmer, especially at $ \str = 0.3-0.35 $ and $ \lam \leq 1.5 $.
		A lower Reynolds number amplifies the schooling advantage for thrust. This advantageous range of $ \str $ and $ \lam $ also corresponds well with a single \textit{carangiform} fish at $ \str \approx 0.4 $ \citep{Borazjani2008} and other swimming and flying animals at $ 0.2 < \str < 0.4 $ \citep{Taylor2003,Triantafyllou1991}.
		So a better thrust performance can be another reason to school together in addition to energy conservation \citep{Daghooghi2015,Li2020}.
	}
	\label{fig:heat_map_compare_12_ctm}
\end{figure}

\FloatBarrier
\clearpage

\subsection{Dependence of net propulsive efficiency on $\str$, $\lam$ and $\re$ }
\label{sec_effi}

The present paper defines net propulsive efficiency as $ \eta_i = \ctm / \bar{C}_{\rm P} $. We note that this formula measures how efficiently the input power is converted to the net thrust, \ie acceleration \citep{Maertens2015}, as the present study is meant to focus on the problem of acceleration. More discussion and review of the recent development in the efficiency metrics can be found in the appendix from \cite{lin2022swimming}. Since the two swimmers are placed side-by-side in anti-phase, the deforming solid is mirror-symmetric in time and space. Therefore, the resulting flow pattern is symmetrical for most cases in the present study (discussed later in detail). As a result, the thrust for each schooling member can be equivalent to each other $ \eta_1 = \eta_2 $ in most cases. So in the present study, net propulsive efficiency is represented by averaging the values from each of the two schooling swimmers $ \eta = (\eta_1 + \eta_2)/2 $.

Net propulsive efficiency $ \eta $ is generally higher in the range of $ \str > 0.4 $ and $ 0.8 < \lam < 1.5 $, as seen in \Cref{fig:heat_map_eta}.
The highest efficiency is obtained at $ \lam = 1.1, 1.2 $ with $ \str = 0.50, 0.55 $. {\cob The optimal Strouhal number $ St = 0.5 $ matches { the}  observed value for a single linear-accelerating fish, \eg Crevalle jack \citep{Akanyeti2017}, but slightly higher than that for most steady-swimming fish in nature \citep{Borazjani2008}.}
The maximum efficiency and high-efficiency region $ \eta > 24\% $ both increase with Reynolds number. The highest efficiency increases almost linearly with the Reynolds number
, which indicates that thrust generation can be less energy-consuming at a higher swimming/flow speed.
The high-efficiency band is approximately located on the line $ \lam + 3 \str = 2.9  $, which means that to achieve high efficiency, swimmers cannot choose both high wavelength $ \lam > 1.5 $ and high Strouhal number $ \str > 5.5 $ at the same time.

\newcommand{\etapair}{\eta_{\rm pair}}
\newcommand{\etasingle}{\eta_{\rm single}}
\newcommand{\etamp}{\eta_{\rm pair}/\eta_{\rm single}}

Net propulsive efficiency for mirror-symmetric schooling $ \etapair $ can be much higher than that of single swimming $ \etasingle $, reaching $ \etamp = 5 $, as seen in \Cref{fig:heat_map_compare_12_eta}.
The thick line indicates the locations where $ \etamp = 1 $, \ie schooling and single swimming yields identical propelling efficiency. It is seen that schooling can be more efficient at $ 0.25 < \str < 0.55 $ and $ \lam > 1 $. The schooling efficiency can be several times higher than a single swimmer, especially at $ \str = 0.3-0.35 $ and $ \lam \leq 1.5 $. A lower Reynolds number amplifies the schooling advantage for efficiency. In short, mirror-symmetric schooling can be more advantageous at low Reynolds numbers and Strouhal numbers, but higher wave lengths.

In addition, for clarification, given \textit{steady swimming} condition, the Strouhal number of fish swimming ranges from 0.25 to 0.4 \citep{Borazjani2008}. However, in the present study, we focus on the fish swiming with \textit{linear acceleration} condition, which can be slightly different. The optimal Strouhal number for efficient force production can reach $ \str \approx 0.5 $ for certain fish species \citep{Akanyeti2017}. For example, Crevalle jack, Indo-Pacific tarpon, and Mangrove snapper achieve optimal propulsive efficiency at $ St = 0.51, 0.48, 0.48 $, which means that the optimal Strouhal number of 0.5 concluded in the present study actually matches the biological observations for at least some fish species.

\newcommand{\widtha}{0.40}
\begin{figure}
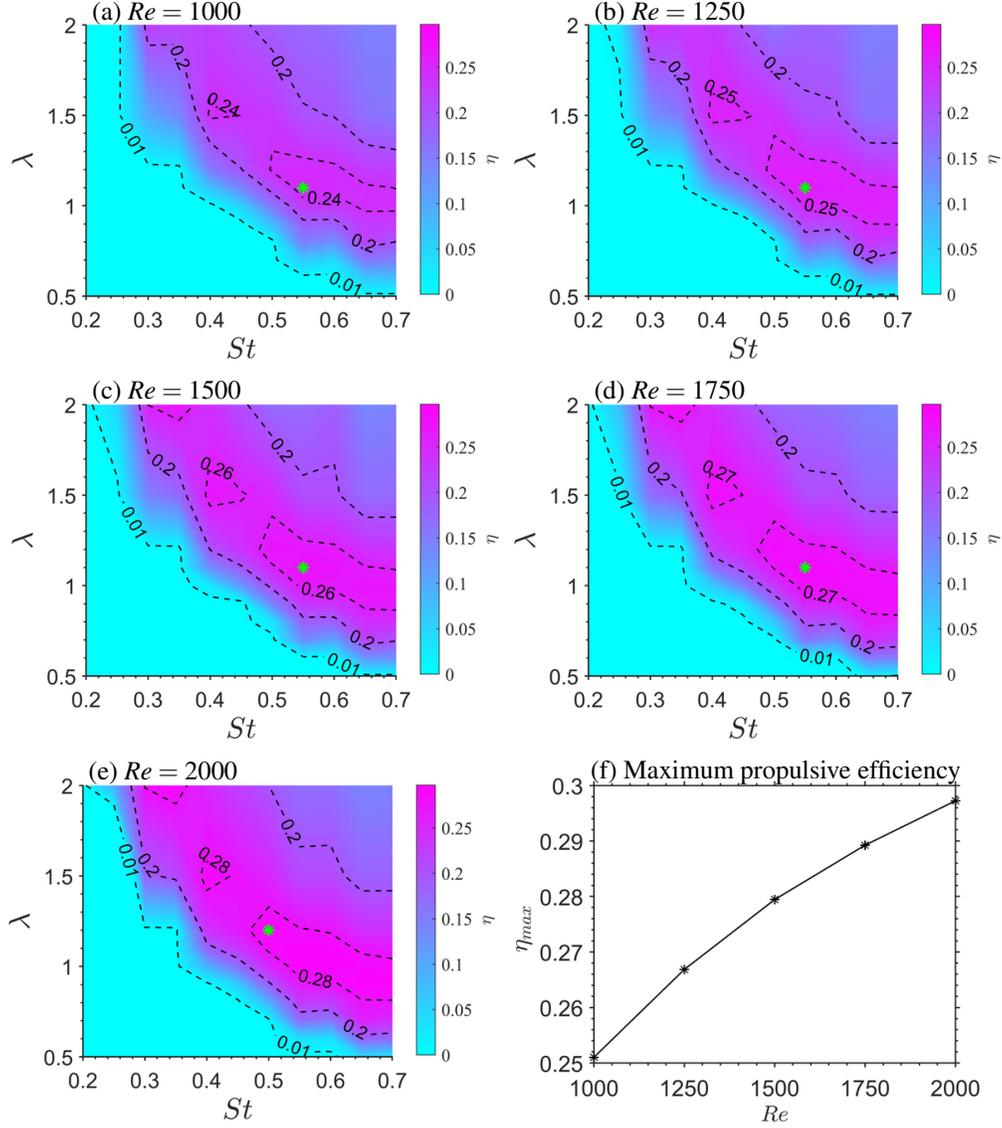

	\centering
	\setcounter{testa}{0}
	\addlabele{\widtha}{{{Heat_map_Re_1000__Effi_St_lambda_ppi1_D0_G0d35}}}{(\countera) $ Re = 1000 $}
	\addlabele{\widtha}{{{Heat_map_Re_1250__Effi_St_lambda_ppi1_D0_G0d35}}}{(\countera) $ Re = 1250 $}
	\addlabele{\widtha}{{{Heat_map_Re_1500__Effi_St_lambda_ppi1_D0_G0d35}}}{(\countera) $ Re = 1500 $}
	\addlabele{\widtha}{{{Heat_map_Re_1750__Effi_St_lambda_ppi1_D0_G0d35}}}{(\countera) $ Re = 1750 $}
	\addlabele{\widtha}{{{Heat_map_Re_2000__Effi_St_lambda_ppi1_D0_G0d35}}}{(\countera) $ Re = 2000 $}
	\addlabele{\widtha}{{{Line_Re_vs_Optimal_Effi}}}{(\countera) Maximum propulsive efficiency}
	\setcounter{testa}{0}
	\caption{Heat map for net propulsive efficiency $ \eta $ at Strouhal number $ St = 0.2-0.7 $, wavelength $ \lam=0.5-2 $ and Reynolds numbers at 
		(a) $ Re = 1000 $
		(b) $ Re = 1250 $
		(c) $ Re = 1500 $
		(d) $ Re = 1750 $
		(e) $ Re = 2000 $.
		(f) Maximum net propulsive efficiency $ \eta_{\rm max} $ at each Reynolds number.
		Due to spatial symmetry at any instant, net propulsive efficiency is identical for each swimmer or two swimmers as a group.
		{\cob The highest efficiency is denoted by the green star marker, located at $ (St, \lam) = (0.55, 1.1) $ for $ Re = 1000-1750 $ and $ (St, \lam) = (0.5, 1.2) $ for $ Re = 2000 $.}
		{\cob The optimal Strouhal number $ St = 0.5 $ matches { the}  observed value for a single linear-accelerating fish, \eg Crevalle jack \citep{Akanyeti2017}, but slightly higher than that for most steady-swimming fish in nature \citep{Borazjani2008}.}		
		The maximum efficiency and high-efficiency region $ \eta > 24\% $ both increase with Reynolds number.
		The cases with negative thrust are drawn as zero.
		The maximum efficiency increases almost linearly with Reynolds number.
	}
	\label{fig:heat_map_eta}
\end{figure}

\setcounter{testa}{0}
\renewcommand{\addlabele}[3]{%
	\begin{tikzpicture}
		\node[anchor=south west,inner sep=0] (image) at (0,0) 
		{\includegraphics[width=#1\textwidth]{#2}};
		\begin{scope}[x={(image.south east)},y={(image.north west)}]
			\node[anchor=south west] at (0.15,0.90) {\footnotesize #3};
		\end{scope}
	\end{tikzpicture}%
}
\renewcommand{\widthb}{0.32}
\begin{figure}
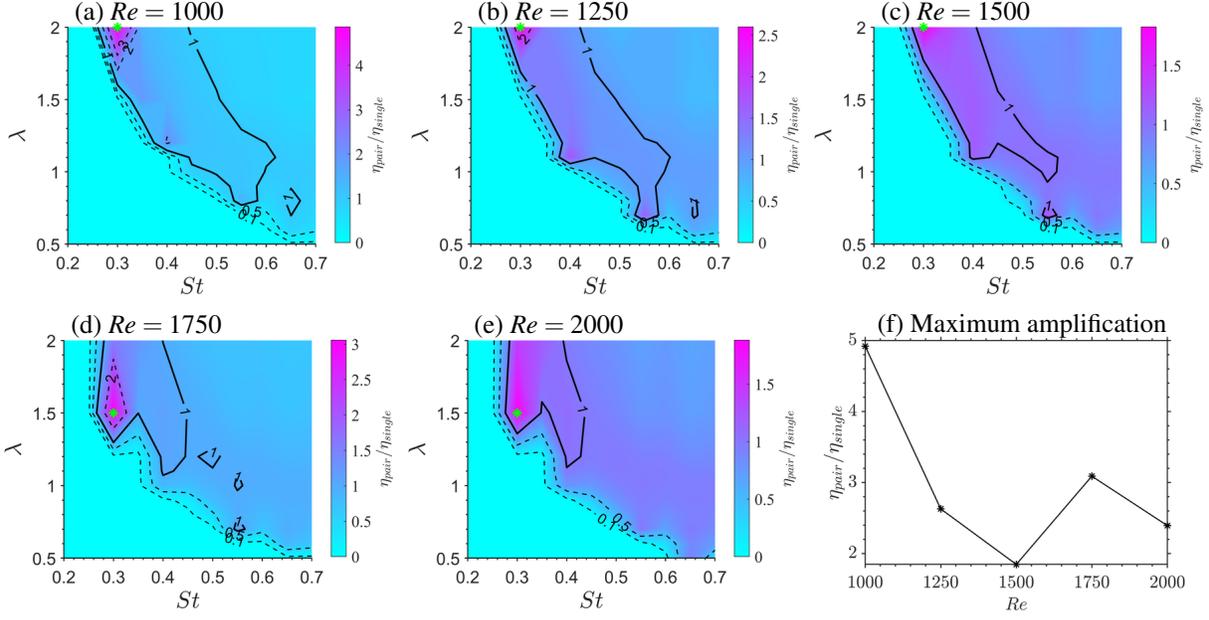

	\centering
	\setcounter{testa}{0}
	\addlabele{\widthb}{{{compare__Heat_map_Re_1000__Eta_St_lambda_ppi1_D0_G0d35}}}{(\countera) $ Re = 1000 $}
	\addlabele{\widthb}{{{compare__Heat_map_Re_1250__Eta_St_lambda_ppi1_D0_G0d35}}}{(\countera) $ Re = 1250 $}
	\addlabele{\widthb}{{{compare__Heat_map_Re_1500__Eta_St_lambda_ppi1_D0_G0d35}}}{(\countera) $ Re = 1500 $}
	\addlabele{\widthb}{{{compare__Heat_map_Re_1750__Eta_St_lambda_ppi1_D0_G0d35}}}{(\countera) $ Re = 1750 $}
	\addlabele{\widthb}{{{compare__Heat_map_Re_2000__Eta_St_lambda_ppi1_D0_G0d35}}}{(\countera) $ Re = 2000 $}
	\addlabele{\widthb}{{{compare__Line_Re_vs_largest_Efficiency_Amplification}}}{(\countera) Maximum amplification}
	\setcounter{testa}{0}
	\caption{Heat map for \textit{schooling efficiency amplification} factor $ \eta_{\rm pair} / \eta_{\rm single} $ at Strouhal number $ \str = 0.2-0.7 $, wavelength $ \lam=0.5-2 $ and Reynolds numbers at 
		(a) $ Re = 1000 $
		(b) $ Re = 1250 $
		(c) $ Re = 1500 $
		(d) $ Re = 1750 $
		(e) $ Re = 2000 $
		(f) Maximum efficiency amplification at each $ \re $ due to schooling.
		Here, we only show the results with both $ \ctm > 0 $ and $ \ctms > 0 $, and the non-accelerating cases are drawn as zero.
		This map demonstrates how schooling contributes to the propulsive efficiency of the swimmers.
		The thick contour line of $ \etamp = 1 $ indicates equivalent propulsive efficiency from schooling and of a single foil.
		$ \etamp > 1 $ means that schooling swimmers produces higher propulsive efficiency than a single swimmer, and vice versa for $ \etamp < 1 $.
		The green marker denotes the location for the highest efficiency amplification for each Reynolds number.
		The schooling efficiency can be several times higher than a single swimmer, especially at $ \str = 0.3-0.35 $ and $ \lam \leq 1.5 $.
		A lower Reynolds number amplifies the schooling advantage for efficiency.
	}
	\label{fig:heat_map_compare_12_eta}
\end{figure}

\FloatBarrier
\clearpage
\subsection{Flow structures maps}
\label{sec_flow_map}

Here, we classify the flow structures of various scenarios based on the overall characteristics, vortex shedding, and flow symmetry.
In the parametric space of the present study, we can classify the flow structures into six types: (a) steady wake, (b) quasi-Karman wake, (c) 2S, (d) 2P-diverge, (e) 2P-converge, and (f) symmetry breaking, as demonstrated in \Cref{fig:example_flow_struct}. The first five types are mirror-symmetric in time and space, whereas the sixth type demonstrates a chaotic flow structure with symmetry breaking. Here, we explain the main characteristics of each type:
\begin{enumerate}[label=(\alph*)]
	\item Steady wake: steady streaming in the far field without vortex formation
	\item Quasi-Karman wake: intermediate state between steady streaming and Karman vortex shedding
	\item 2S: a single vortex shed from each wavy foil per cycle, forming a vortex dipole with the main streaming direction pointing downstream
	\item 2P-diverge: one vortex dipole per cycle per foil, forming two reverse vortex streets towards diverging directions
	\item 2P-converge: similar to 2P-diverge, yet the vortex dipoles are converging instead of diverging
	\item Symmetry breaking: Symmetrical wake breaks, resulting asymmetric flow pattern
\end{enumerate}
This intensity of vorticity and irregularity both increases with the order listed above. For example, the steady wake (a) contains the lowest overall vorticity intensity with stable flow structures, and vice versa for the symmetry breaking (f) case.

Based on this classification, we generate a set of maps to illustrate the distribution of flow structures in the present parametric space, as shown in \Cref{fig:example_flow_struct}; the detailed demonstration for all cases can be found in \Cref{appendix_flow_map_detail}. The markers correspond to the types listed in \Cref{fig:example_flow_struct}. In general, the variation of $ \re = 1000 - 2000 $ is not significantly affecting the flow structure distribution; so we mainly discuss the effects of $ \lam $ and $ \str $. The steady wake (a) is only observed at very low wavelength $ \lam \leq 0.6 $ and Strouhal number $ \str \leq 0.3 $, where the flow is not heavily disturbed. The quasi-Karman wake (b) is observed at slightly higher $ \lam $ and $ \str $, and even higher for the 2S (c) cases. The conversion from quasi-Karman wake (b) to 2S (c) occurs at approximately $ 4.67\str + \lam = 2.367 $. 2P-diverge (d) accounts for the most number of cases in the present study. The transition from 2S (c) to 2P-diverge (d) is observed at approximately $ 4.5\str + \lam = 2.85 $. 2P-converge (e) can be identified at roughly $ 2.67\str + \lam = 3.067 $. The 2P-converge is a boundary condition between the 2P-diverge (d) and the full development of symmetry breaking (f), so only a few cases can be discovered. Furthermore, at $ 2.67\str + \lam > 3.067 $, the flow becomes asymmetric and highly irregular.

The region of high net thrust from \Cref{fig:heat_map_ctm} corresponds to the symmetry breaking (f) region in \Cref{fig:map_flow_struct}. Zero net thrust cases, \ie steady-swimming, partly overlap with the boundary between 2S (c) and 2P-diverge (d). So the positive net thrust, \ie acceleration, mainly corresponds with flow structures of 2P-diverge (d), 2P-converge (e) and symmetry breaking (f). High thrust is found in the symmetry breaking (f) condition, so the irregular flow pattern does not significantly affect the thrust generation.
The region of high net propulsive efficiency from \Cref{fig:heat_map_eta} overlaps with the distribution of the 2P-diverge (d) pattern in \Cref{fig:map_flow_struct}. Cases with the highest efficiency all demonstrate 2P-diverge (d) pattern. Therefore, the structural vortex dipole shedding contributes to higher efficiency. Conversely, the symmetry breaking (f) region cannot yield high efficiency.
While comparing schooling thrust amplification in \Cref{fig:heat_map_compare_12_ctm} and the flow structure map of \Cref{fig:map_flow_struct}, it is interesting to note that the most significant schooling amplification factor $ \ctm/\ctms $ is all located in the 2P-converge (e) region, which corresponds to the s-RKV region of a single swimmer \citep{chao2022hydrodynamic}, featuring a skewed reverse Karman vortex street. This indicates that schooling members can produce more thrust from two converging skewed vortex streets.

\renewcommand{\addlabele}[3]{%
	\begin{tikzpicture}
		\node[anchor=south west,inner sep=0] (image) at (0,0) 
		{\includegraphics[width=#1\textwidth, trim={2cm 0cm 2cm 0cm},clip]{#2}};
		\begin{scope}[x={(image.south east)},y={(image.north west)}]
			\node[anchor=south west] at (0.15,0.90) {\footnotesize #3};
		\end{scope}
	\end{tikzpicture}%
}
\renewcommand{\widtha}{0.32}
\begin{figure}
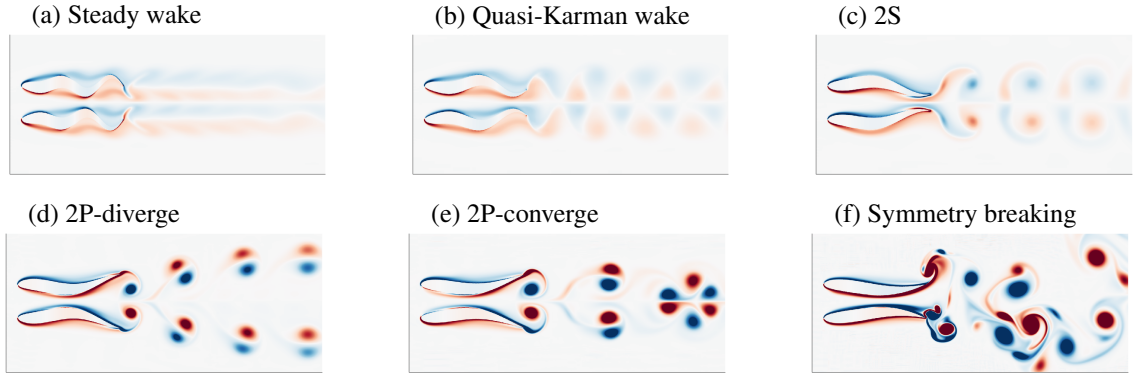

	\centering
	\setcounter{testa}{0}
	\addlabele{\widtha}{{{example_Flow_structure_Re_1000_St_0.2_Lambda_0.5}}}{(\countera) Steady wake}
	\addlabele{\widtha}{{{example_Flow_structure_Re_1000_St_0.25_Lambda_0.7}}}{(\countera) Quasi-Karman wake}
	\addlabele{\widtha}{{{example_Flow_structure_Re_1000_St_0.35_Lambda_1.1}}}{(\countera) 2S}
	\addlabele{\widtha}{{{example_Flow_structure_Re_1000_St_0.45_Lambda_1.2}}}{(\countera) 2P-diverge}
	\addlabele{\widtha}{{{example_Flow_structure_Re_1000_St_0.5_Lambda_1.5}}}{(\countera) 2P-converge}
	\addlabele{\widtha}{{{example_Flow_structure_Re_1000_St_0.7_Lambda_2}}}{(\countera) Symmetry breaking}
	\setcounter{testa}{0}
	\caption{Representative examples for six flow structures identified as:
		(a) steady wake: steady streaming in the far field wake
		(b) quasi-Karman wake: periodically disturbed wake but no distinct vortex
		(c) 2S: one vortex from each foil
		(d) 2P-diverge: a pair of vortices \textit{diverging} in the wake
		(e) 2P-converge: a pair of vortices \textit{converging} in the wake
		(f) Symmetry breaking: an unstable flow that is asymmetric
		Here, the red colour denotes positive vorticity (counter-clockwise) with the blue colour representing the negative vorticity (clockwise).
	}
	\label{fig:example_flow_struct}
\end{figure}
\setcounter{testa}{0}

\renewcommand{\addlabele}[3]{%
	\begin{tikzpicture}
		\node[anchor=south west,inner sep=0] (image) at (0,0) 
		{\includegraphics[width=#1\textwidth, trim={0.75cm 0cm 1cm 0cm},clip]{#2}};
		\begin{scope}[x={(image.south east)},y={(image.north west)}]
			\node[anchor=south west] at (0.15,0.90) {\footnotesize #3};
		\end{scope}
	\end{tikzpicture}%
}
\newcommand{\addlabelcutlegend}[2]{%
	\begin{tikzpicture}
		\node[anchor=south west,inner sep=0] (image) at (0,0) 
		{\includegraphics[width=#1\linewidth, trim={7cm 2.5cm 0.5cm 2.5cm},clip]{#2}};	%
		\begin{scope}[x={(image.south east)},y={(image.north west)}]
		\end{scope}
	\end{tikzpicture}%
}
\renewcommand{\widtha}{0.32}
\begin{figure}
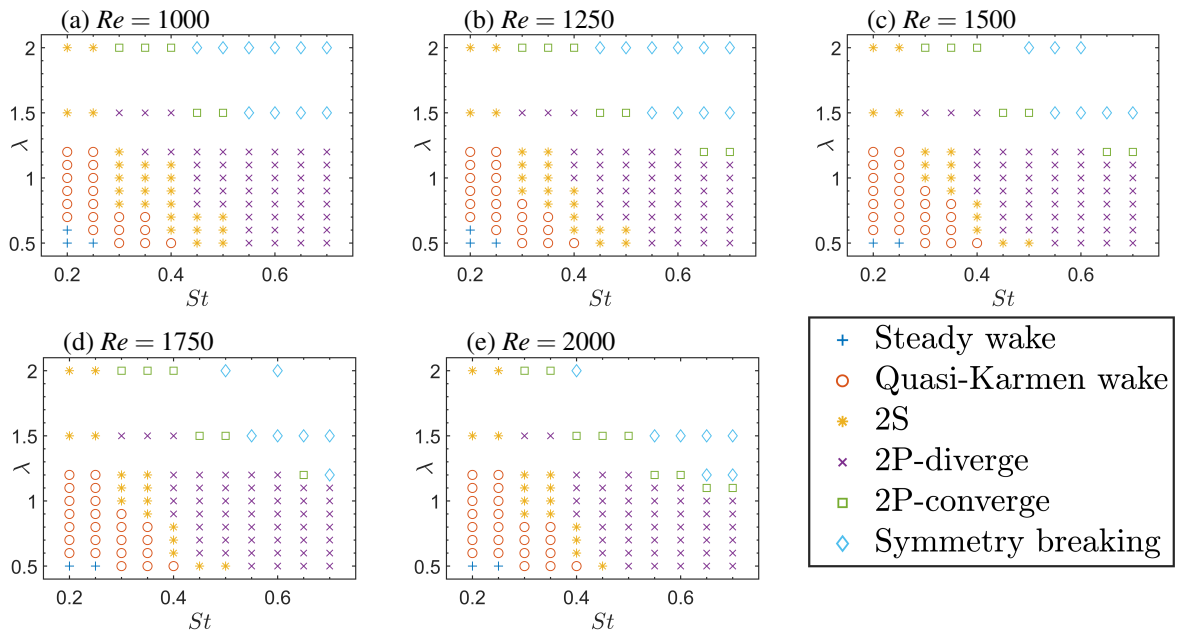

	\centering
	\setcounter{testa}{0}
	\addlabele{\widtha}{{{mapping_Flow_structure_Re_1000}}}{(\countera) $ Re = 1000 $}
	\addlabele{\widtha}{{{mapping_Flow_structure_Re_1250}}}{(\countera) $ Re = 1250 $}
	\addlabele{\widtha}{{{mapping_Flow_structure_Re_1500}}}{(\countera) $ Re = 1500 $}
	\addlabele{\widtha}{{{mapping_Flow_structure_Re_1750}}}{(\countera) $ Re = 1750 $}
	\addlabele{\widtha}{{{mapping_Flow_structure_Re_2000}}}{(\countera) $ Re = 2000 $}
	\addlabelcutlegend{0.31}{{{legend_for_mapping_Flow_structure}}}
	\setcounter{testa}{0}
	\caption{Flow structure classification at Strouhal number $ St = 0.2-0.7 $, wavelength $ \lam=0.5-2 $ and Reynolds numbers at 
		(a) $ Re = 1000 $
		(b) $ Re = 1250 $
		(c) $ Re = 1500 $
		(d) $ Re = 1750 $
		(e) $ Re = 2000 $.
		The alphabetic marker types correspond to the classification shown in \Cref{fig:example_flow_struct}.
		}
	\label{fig:map_flow_struct}
\end{figure}
\setcounter{testa}{0}

\FloatBarrier
\clearpage

{ \cob

\subsection{Investigation of Fluid Mechanism}
\label{sec:mechanism}
The Reynolds number, wavelength, and Strouhal number show substantial variations that ultimately result in a distinct difference in the flow structure and corresponding underlying mechanisms. Concurrent with the variation in flow structure, significant changes in performance metrics, \textit{i.e.}, thrust and efficiency, are observed. In this section, to delve deeper into the mechanisms, three typical cases are selected for a detailed analysis and comparative study. The cases are analysed using fluid vorticity, net force, pressure distribution, and fluid velocity vectors.

The selected cases for our investigation are presented in \Cref{tab:study3cases}. Each case represents a consistent increment in input values: the Reynolds number increases by 500, the Strouhal number by 0.1, and the wavelength by 0.3. These cases are chosen to represent different flow structures, namely: 2S in the first case, 2P-diverge in the second case, and symmetry breaking in the third case. To facilitate comparison, we maintain identical colour scales for vorticity, pressure, and velocity vectors across these cases.

\begin{ruledtabular}
	\begin{table*}[thb]
		\caption{\cob Representative cases for in-depth study and comparison}
		\centering
		\label{tab:study3cases}
		\begin{tabular}{c c c c c c c c c}
			
			No.\ & $ \re $ & $ \str $ & $ \lam $ & Flow Structure & $ \ctm $ & $ \eta_{\rm pair} $ & Vorticity \& Force & Pressure \& Velocity \\
			
			1st & $ 1000 $ & $ 0.4 $ & $ 0.9 $ & 2S (\Cref{fig:example_flow_struct}c) & $ -0.059 $ & N/A  & \Cref{fig:vor_s0d4} & \Cref{fig:p_u_s0d4} \\
			
			2nd & $ 1500 $ & $ 0.5 $ & $ 1.2 $ & 2P-diverge (\Cref{fig:example_flow_struct}d) & $ 0.403 $ & $ 27.5\% $ & \Cref{fig:vor_s0d5} & \Cref{fig:p_u_s0d5}\\
			
			3rd & $ 2000 $ & $ 0.6 $ & $ 1.5 $ & Sym-breaking (\Cref{fig:example_flow_struct}f) & $ 1.527 $ & $ 21.4\% $ & \Cref{fig:vor_s0d6} & \Cref{fig:p_u_s0d6}
		\end{tabular}
	\end{table*}
\end{ruledtabular}

In the first case at $ \re = 1000 $, $ \str = 0.4 $, and $ \lam = 0.9 $, each hydrofoil sheds two vortices per undulation cycle. The generated thrust force profiles are nearly identical between the hydrofoils, with peak thrust corresponding to the shedding of stronger vortices (\Cref{fig:vor_s0d4}i-\ref{fig:vor_s0d4}j). The primary pressure fluctuation occurs between the swimmers (\Cref{fig:p_u_s0d4}), with high thrust related to high downstream fluid velocity and rapid dissipation of negative pressure downstream. A two-row vortex array accounts for velocity vector fluctuations in the wake.

The similarity in the thrust force profiles between the two swimmers (\Cref{fig:vor_s0d4}i-\ref{fig:vor_s0d4}j) is noteworthy, with two peaks at instants c and g associated with maximum instantaneous thrust. Interestingly, the shedding of the stronger vortex corresponds to the higher thrust peak at instant c, while the formation of a smaller, short-lived vortex results in a smaller thrust peak at instant g. On the contrary, the thrust trough at instant h corresponds to the complete shedding of the minor vortex, which lacks the necessary strength to generate sufficient thrust.
Significant pressure fluctuations primarily occur in the gap between the two swimmers, as illustrated in \Cref{fig:p_u_s0d4}. The maximum positive pressure, which aligns with low thrust, is attained at the instant a in the gap, resulting in a high lift force that separates the swimmers. In contrast, at instants e, the pressure in the gap hits a negative maximum, where the thrust is close to the local minimum, and the lift turns positive, drawing the swimmers together.

Furthermore, high thrust at instants b is achieved when strong positive pressure is exerted at the posterior part of the swimmers, propelling them forward. At instant c, the swimmers are "sucked" forward due to strong negative pressure at their front. In the downstream area, the vortex shedding creates negative pressure (\Cref{fig:p_u_s0d4}g) which dissipates rapidly within one cycle. By analysing the velocity field, we discovered that high thrust aligns with high fluid velocity towards the downstream direction, "propelling" the swimmers forward. The two-row vortex array in the wake introduces velocity vector fluctuations that follow a similar pattern.

In the second case, with $ \re = 1500 $, $ \str = 0.5 $, and $ \lam = 1.2 $, we identify a distinctive "2P-diverge" flow pattern in the wake of the two swimmers, wherein the outer vortex exhibits greater persistence and influences the vortex direction alignment. Instances of simultaneous high pressures at the ends of the swimmers produce high net thrust.

Observation of the vorticity distribution (\Cref{fig:vor_s0d5}) reveals that two distinct rows of vortex dipoles are propelled from the swimmers' posterior part, forming the unique "2P-diverge" flow structure pattern, contrasting markedly with the "2S" pattern of the first case. The vortex shedding mechanism is similar to the first case, with each undulation cycle of the hydrofoils generating two vortices. However, the outer vortex persists rather than dissipating, resulting in vortex dipoles propelled away from the mirror-symmetric axis between the swimmers.
The outer vortex's rapid dissipation compared to the inner one can be ascertained from the vorticity and pressure distribution (\Cref{fig:vor_s0d5,fig:p_u_s0d5}). As the dissipation progresses, the vortex dipoles' moving direction gradually aligns with the free-stream velocity.
Mirroring the first case, the time history of the thrust force features two peaks and one trough. The major and minor peaks are located at instants d and h, respectively. It is intriguing that these instants of high net thrust coincide with high positive and negative pressure at the swimmers' posterior and anterior parts, respectively (\Cref{fig:p_u_s0d5}). This observation suggests that the swimmer is simultaneously "pushed" and "pulled" by the positive and negative pressure at different locations, contributing to high instantaneous thrust. The larger peak at instant d occurs when positive pressure is on the outside, and negative pressure is in the gap. The thrust at instant h is weaker due to the formation of the vortex pair near the tails, which slightly reduces the pressure in the gap.

In the third case, at $ \re = 2000 $, $ \str = 0.6 $, and $ \lam = 1.5 $, despite irregular wake flow and symmetry breaking, the fluid-structure interaction mechanism aligns largely with the second case exhibiting the 2P-diverge pattern. The near-tail flow field maintains substantial regularity and symmetry.

In the third case, as depicted in \Cref{fig:vor_s0d6,fig:p_u_s0d6}, the flow structure exhibits irregularity with symmetry breaking, although the vortices near the tails remain largely regular and symmetric. Like to the second case, each hydrofoil generates a pair of vortices in a complete cycle. The vortex dipoles shed from each foil rapidly converge whilst interacting with previously generated vortices, leading to irregularity in the flow field. This pattern differs from the second case, where the vortex sheddings symmetrically diverge rather than quickly impinge on each other.
Despite the irregular wake flow, the net thrust and lift force time history between the two swimmers remain strikingly identical, as seen in \Cref{fig:vor_s0d6}i and \ref{fig:vor_s0d6}j. Furthermore, the correlation between pressure distribution and force variation strongly resembles to the second case with the 2P-diverge pattern. High thrust is observed when the swimmers are "pushed" and "pulled" by the positive and negative pressure at the posterior and anterior parts of the hydrofoil body. However, the vortex pair in the gap slightly mitigates the thrust at instant g, as seen in \Cref{fig:p_u_s0d6}g.
Regarding the wake region, here, the pressure disturbance is significantly more pronounced compared with the previous cases, due to the strong negative pressure created by the recirculation and interaction of the vortices.

\newcommand{\addlabelvor}[3]{%
	\begin{tikzpicture}
		\node[anchor=south west,inner sep=0] (image) at (0,0) 
		{\includegraphics[width=#1\linewidth, trim={6.75cm 4.5cm 1.5cm 15cm},clip]{{{#2}}}};%
		\begin{scope}[x={(image.south east)},y={(image.north west)}]
			\node[anchor=south west] at (0.05,0.95) {\footnotesize #3};	%
		\end{scope}
	\end{tikzpicture}%
}
\newcommand{\addlabelnotrim}[3]{%
	\begin{tikzpicture}
		\node[anchor=south west,inner sep=0] (image) at (0,0) 
		{\includegraphics[width=#1\linewidth, trim={0cm 0cm 0cm 0cm},clip]{{{#2}}}};%
		\begin{scope}[x={(image.south east)},y={(image.north west)}]
			\node[anchor=south west] at (-0.01,-0.03) {\footnotesize #3};	%
		\end{scope}
	\end{tikzpicture}%
}
\newcommand{\addlabela}[3]{%
	\begin{tikzpicture}
		\node[anchor=south west,inner sep=0] (image) at (0,0) 
		{\includegraphics[width=#1\linewidth, trim={0cm 0cm 0cm 0cm},clip]{#2}};
		\begin{scope}[x={(image.south east)},y={(image.north west)}]
			\node[anchor=east] at (0.045,0.75) {#3};	%
		\end{scope}
	\end{tikzpicture}%
}
\renewcommand{\widthb}{0.40}
\begin{figure}
	\centering
	\addlabelvor{\widthb}{g0.33d0.00p1.00e1000.0s0.400r421_L_00.900_vorticity___1_}{(\countera) $ t/T = 10.000 $}
	\addlabelvor{\widthb}{g0.33d0.00p1.00e1000.0s0.400r421_L_00.900_vorticity___11_}{(\countera) $ t/T = 10.125 $}
	\addlabelvor{\widthb}{g0.33d0.00p1.00e1000.0s0.400r421_L_00.900_vorticity___21_}{(\countera) $ t/T = 10.250 $}
	\addlabelvor{\widthb}{g0.33d0.00p1.00e1000.0s0.400r421_L_00.900_vorticity___31_}{(\countera) $ t/T = 10.375 $}
	\addlabelvor{\widthb}{g0.33d0.00p1.00e1000.0s0.400r421_L_00.900_vorticity___41_}{(\countera) $ t/T = 10.500 $}
	\addlabelvor{\widthb}{g0.33d0.00p1.00e1000.0s0.400r421_L_00.900_vorticity___51_}{(\countera) $ t/T = 10.625 $}
	\addlabelvor{\widthb}{g0.33d0.00p1.00e1000.0s0.400r421_L_00.900_vorticity___61_}{(\countera) $ t/T = 10.750 $}
	\addlabelvor{\widthb}{g0.33d0.00p1.00e1000.0s0.400r421_L_00.900_vorticity___71_}{(\countera) $ t/T = 10.875 $}
	
	\addlabela{0.55}{{{legend_vorticity}}}{$\bm{\omega}^*$:}
	
	\addlabelnotrim{\widthb}{Time_History_of_Leader_G=0.33_D=0.00_Lam=0.90_anti_phi}{(i) Bottom swimmer}
	\addlabelnotrim{\widthb}{Time_History_of_Follower_G=0.33_D=0.00_Lam=0.90_anti_phi}{(j) Top swimmer}
	\caption{\cob Vorticity contours and hydrofoil deformation with $ \re = 1000 $, $ \str = 0.4 $, $ \lam = 0.9 $, at instants of a typical period
		(a-h) $ t/T = 10.000-10.875 $. Time histories of thrust and lift coefficient for the (i) Bottom and (j) Top swimmers.}
	\label{fig:vor_s0d4}
\end{figure}
\setcounter{testa}{0}

\newcommand{\addlabelb}[3]{%
	\begin{tikzpicture}
		\node[anchor=south west,inner sep=0] (image) at (0,0) 
		{\includegraphics[width=#1\linewidth, trim={0cm 0cm 0cm 0cm},clip]{#2}}; %
		\begin{scope}[x={(image.south east)},y={(image.north west)}]
			\node[anchor=east] at (0.045,0.75) {#3};	%
		\end{scope}
	\end{tikzpicture}%
}
\begin{figure}
	\centering
	\addlabelvor{\widthb}{g0.33d0.00p1.00e1000.0s0.400r421_L_00.900_pressure_U_vector___1_}{(\countera) $ t/T = 10.000 $}
	\addlabelvor{\widthb}{g0.33d0.00p1.00e1000.0s0.400r421_L_00.900_pressure_U_vector___11_}{(\countera) $ t/T = 10.125 $}
	\addlabelvor{\widthb}{g0.33d0.00p1.00e1000.0s0.400r421_L_00.900_pressure_U_vector___21_}{(\countera) $ t/T = 10.250 $}
	\addlabelvor{\widthb}{g0.33d0.00p1.00e1000.0s0.400r421_L_00.900_pressure_U_vector___31_}{(\countera) $ t/T = 10.375 $}
	\addlabelvor{\widthb}{g0.33d0.00p1.00e1000.0s0.400r421_L_00.900_pressure_U_vector___41_}{(\countera) $ t/T = 10.500 $}
	\addlabelvor{\widthb}{g0.33d0.00p1.00e1000.0s0.400r421_L_00.900_pressure_U_vector___51_}{(\countera) $ t/T = 10.625 $}
	\addlabelvor{\widthb}{g0.33d0.00p1.00e1000.0s0.400r421_L_00.900_pressure_U_vector___61_}{(\countera) $ t/T = 10.750 $}
	\addlabelvor{\widthb}{g0.33d0.00p1.00e1000.0s0.400r421_L_00.900_pressure_U_vector___71_}{(\countera) $ t/T = 10.875 $}
	
	\addlabelb{0.55}{{{legend_pressure}}}{$\bm{P}^*$:}
	\caption{\cob Pressure contours, velocity vectors and hydrofoil deformation with $ \re = 1000 $, $ \str = 0.4 $, $ \lam = 0.9 $, at instants of a typical period
		(a-h) $ t/T = 10.00-10.875 $. Time histories of thrust and lift coefficient for the (i) Bottom and (j) Top swimmers.}
	\label{fig:p_u_s0d4}
\end{figure}
\setcounter{testa}{0}

\renewcommand{\widthb}{0.40}
\begin{figure}
	\centering
	\addlabelvor{\widthb}{g0.33d0.00p1.00e1500.0s0.500r433_L_01.200_vorticity___1_}{(\countera) $ t/T = 10.000 $}
	\addlabelvor{\widthb}{g0.33d0.00p1.00e1500.0s0.500r433_L_01.200_vorticity___11_}{(\countera) $ t/T = 10.125 $}
	\addlabelvor{\widthb}{g0.33d0.00p1.00e1500.0s0.500r433_L_01.200_vorticity___21_}{(\countera) $ t/T = 10.250 $}
	\addlabelvor{\widthb}{g0.33d0.00p1.00e1500.0s0.500r433_L_01.200_vorticity___31_}{(\countera) $ t/T = 10.375 $}
	\addlabelvor{\widthb}{g0.33d0.00p1.00e1500.0s0.500r433_L_01.200_vorticity___41_}{(\countera) $ t/T = 10.500 $}
	\addlabelvor{\widthb}{g0.33d0.00p1.00e1500.0s0.500r433_L_01.200_vorticity___51_}{(\countera) $ t/T = 10.625 $}
	\addlabelvor{\widthb}{g0.33d0.00p1.00e1500.0s0.500r433_L_01.200_vorticity___61_}{(\countera) $ t/T = 10.750 $}
	\addlabelvor{\widthb}{g0.33d0.00p1.00e1500.0s0.500r433_L_01.200_vorticity___71_}{(\countera) $ t/T = 10.875 $}
	
	\addlabela{0.55}{{{legend_vorticity}}}{$\bm{\omega}^*$:}
	
	\addlabelnotrim{\widthb}{Time_History_of_Leader_G=0.33_D=0.00_Lam=1.20_anti_phi}{(i) Bottom swimmer}
	\addlabelnotrim{\widthb}{Time_History_of_Follower_G=0.33_D=0.00_Lam=1.20_anti_phi}{(j) Top swimmer}
	\caption{\cob Vorticity contours and hydrofoil deformation with $ \re = 1500 $, $ \str = 0.5 $, $ \lam = 1.2 $, at instants of a typical period
		(a-h) $ t/T = 10.00-10.875 $. Time histories of thrust and lift coefficient for the (i) Bottom and (j) Top swimmers.}
	\label{fig:vor_s0d5}
\end{figure}
\setcounter{testa}{0}

\begin{figure}
	\centering
	\addlabelvor{\widthb}{g0.33d0.00p1.00e1500.0s0.500r433_L_01.200_pressure_U_vector___1_}{(\countera) $ t/T = 10.000 $}
	\addlabelvor{\widthb}{g0.33d0.00p1.00e1500.0s0.500r433_L_01.200_pressure_U_vector___11_}{(\countera) $ t/T = 10.125 $}
	\addlabelvor{\widthb}{g0.33d0.00p1.00e1500.0s0.500r433_L_01.200_pressure_U_vector___21_}{(\countera) $ t/T = 10.250 $}
	\addlabelvor{\widthb}{g0.33d0.00p1.00e1500.0s0.500r433_L_01.200_pressure_U_vector___31_}{(\countera) $ t/T = 10.375 $}
	\addlabelvor{\widthb}{g0.33d0.00p1.00e1500.0s0.500r433_L_01.200_pressure_U_vector___41_}{(\countera) $ t/T = 10.500 $}
	\addlabelvor{\widthb}{g0.33d0.00p1.00e1500.0s0.500r433_L_01.200_pressure_U_vector___51_}{(\countera) $ t/T = 10.625 $}
	\addlabelvor{\widthb}{g0.33d0.00p1.00e1500.0s0.500r433_L_01.200_pressure_U_vector___61_}{(\countera) $ t/T = 10.750 $}
	\addlabelvor{\widthb}{g0.33d0.00p1.00e1500.0s0.500r433_L_01.200_pressure_U_vector___71_}{(\countera) $ t/T = 10.875 $}
	
	\addlabelb{0.55}{{{legend_pressure}}}{$\bm{P}^*$:}
	\caption{\cob Pressure contours, velocity vectors and hydrofoil deformation with $ \re = 1500 $, $ \str = 0.5 $, $ \lam = 1.2 $, at instants of a typical period
		(a-h) $ t/T = 10.00-10.875 $. Time histories of thrust and lift coefficient for the (i) Bottom and (j) Top swimmers.}
	\label{fig:p_u_s0d5}
\end{figure}
\setcounter{testa}{0}

\renewcommand{\widthb}{0.40}
\begin{figure}
	\centering
	\addlabelvor{\widthb}{g0.33d0.00p1.00e2000.0s0.600r445_L_01.500_vorticity___1_}{(\countera) $ t/T = 10.000 $}
	\addlabelvor{\widthb}{g0.33d0.00p1.00e2000.0s0.600r445_L_01.500_vorticity___11_}{(\countera) $ t/T = 10.125 $}
	\addlabelvor{\widthb}{g0.33d0.00p1.00e2000.0s0.600r445_L_01.500_vorticity___21_}{(\countera) $ t/T = 10.250 $}
	\addlabelvor{\widthb}{g0.33d0.00p1.00e2000.0s0.600r445_L_01.500_vorticity___31_}{(\countera) $ t/T = 10.375 $}
	\addlabelvor{\widthb}{g0.33d0.00p1.00e2000.0s0.600r445_L_01.500_vorticity___41_}{(\countera) $ t/T = 10.500 $}
	\addlabelvor{\widthb}{g0.33d0.00p1.00e2000.0s0.600r445_L_01.500_vorticity___51_}{(\countera) $ t/T = 10.625 $}
	\addlabelvor{\widthb}{g0.33d0.00p1.00e2000.0s0.600r445_L_01.500_vorticity___61_}{(\countera) $ t/T = 10.750 $}
	\addlabelvor{\widthb}{g0.33d0.00p1.00e2000.0s0.600r445_L_01.500_vorticity___71_}{(\countera) $ t/T = 10.875 $}
	
	\addlabela{0.55}{{{legend_vorticity}}}{$\bm{\omega}^*$:}
	
	\addlabelnotrim{\widthb}{Time_History_of_Leader_G=0.33_D=0.00_Lam=1.50_anti_phi}{(i) Bottom swimmer}
	\addlabelnotrim{\widthb}{Time_History_of_Follower_G=0.33_D=0.00_Lam=1.50_anti_phi}{(j) Top swimmer}
	\caption{\cob Vorticity contours and hydrofoil deformation with $ \re = 2000 $, $ \str = 0.6 $, $ \lam = 1.5 $, at instants of a typical period
		(a-h) $ t/T = 10.00-10.875 $. Time histories of thrust and lift coefficient for the (i) Bottom and (j) Top swimmers.}
	\label{fig:vor_s0d6}
\end{figure}
\setcounter{testa}{0}

\begin{figure}
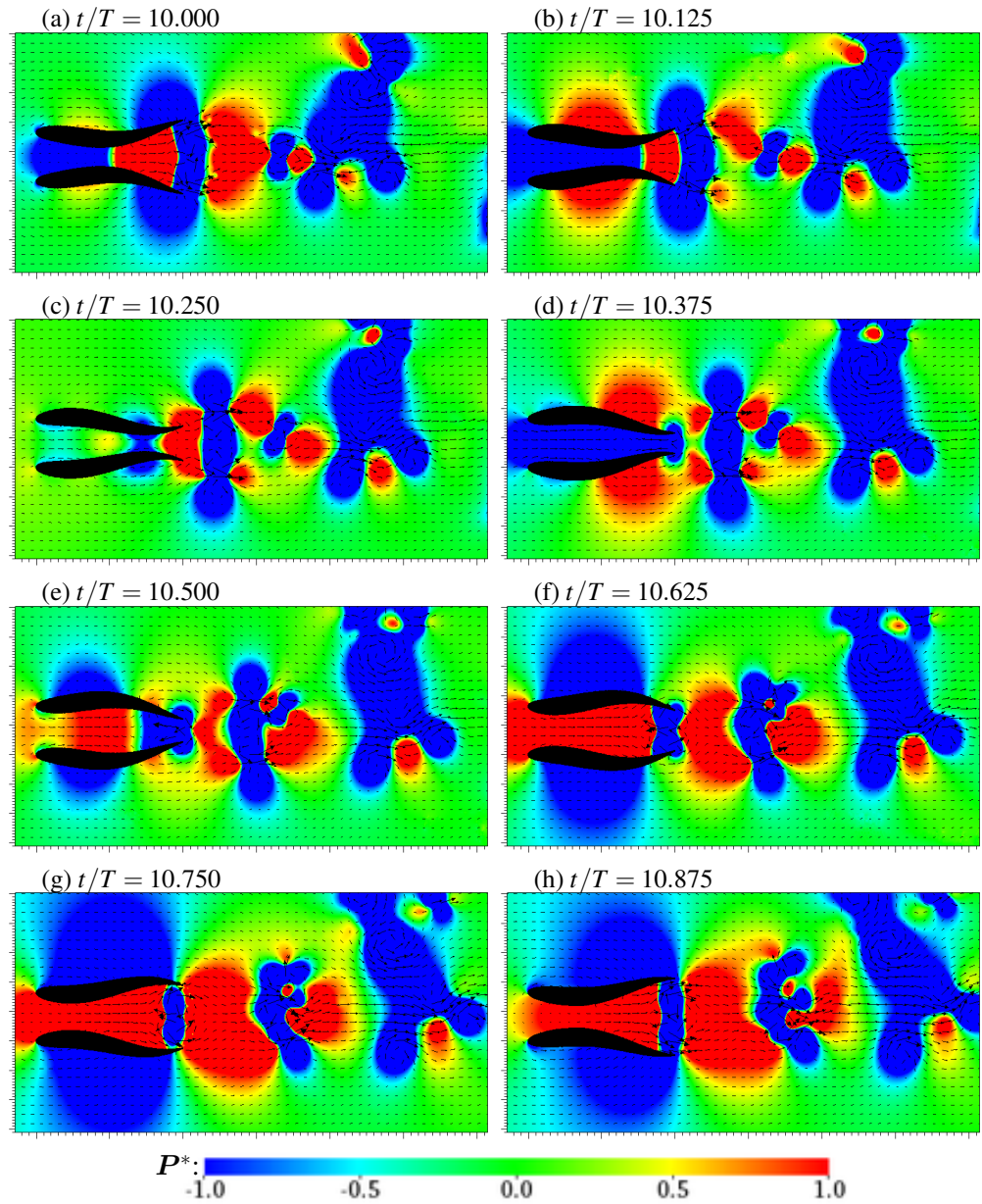

	\centering
	\addlabelvor{\widthb}{g0.33d0.00p1.00e2000.0s0.600r445_L_01.500_pressure_U_vector___1_}{(\countera) $ t/T = 10.000 $}
	\addlabelvor{\widthb}{g0.33d0.00p1.00e2000.0s0.600r445_L_01.500_pressure_U_vector___11_}{(\countera) $ t/T = 10.125 $}
	\addlabelvor{\widthb}{g0.33d0.00p1.00e2000.0s0.600r445_L_01.500_pressure_U_vector___21_}{(\countera) $ t/T = 10.250 $}
	\addlabelvor{\widthb}{g0.33d0.00p1.00e2000.0s0.600r445_L_01.500_pressure_U_vector___31_}{(\countera) $ t/T = 10.375 $}
	\addlabelvor{\widthb}{g0.33d0.00p1.00e2000.0s0.600r445_L_01.500_pressure_U_vector___41_}{(\countera) $ t/T = 10.500 $}
	\addlabelvor{\widthb}{g0.33d0.00p1.00e2000.0s0.600r445_L_01.500_pressure_U_vector___51_}{(\countera) $ t/T = 10.625 $}
	\addlabelvor{\widthb}{g0.33d0.00p1.00e2000.0s0.600r445_L_01.500_pressure_U_vector___61_}{(\countera) $ t/T = 10.750 $}
	\addlabelvor{\widthb}{g0.33d0.00p1.00e2000.0s0.600r445_L_01.500_pressure_U_vector___71_}{(\countera) $ t/T = 10.875 $}
	
	\addlabelb{0.55}{{{legend_pressure}}}{$\bm{P}^*$:}
	\caption{\cob Pressure contours, velocity vectors and hydrofoil deformation with $ \re = 2000 $, $ \str = 0.6 $, $ \lam = 1.5 $, at instants of a typical period
	(a-h) $ t/T = 10.00-10.875 $. Time histories of thrust and lift coefficient for the (i) Bottom and (j) Top swimmers.}
	\label{fig:p_u_s0d6}
\end{figure}
\setcounter{testa}{0}

}

\FloatBarrier
\clearpage
\section{Conclusions}
The effects of fish body wavelength on its linear acceleration during side-by-side schooling conditions have not been studied in detail previously.
In the present paper, we conducted a systematic numerical study, involving $ \caseNum $ cases of two linearly-accelerating side-by-side wavy NACA0012 hydrofoils swimming in anti-phase. We examined the net thrust distribution, net propulsive efficiency, and flow structures by drawing heat maps and proposing empirical formulas.
The simulation is conducted on a customised version of the ConstraintIB \citep{Bhalla2013,Griffith2020} module from the IBAMR \citep{griffith2013ibamr} open-source library.
The parametric space is tested for Strouhal number $\str = 0.2 - 0.7$, wavelength $\lam = 0.5 - 2$ and Reynolds number $ \re = 1000 - 2000 $. The lateral gap distance and maximum tail amplitude are fixed at $ G = 0.33 $ and $ A_{\rm max} = 0.1 $, respectively. These ranges are chosen based on BCF swimmers in nature \citep{Gazzola2014,Santo2021}.

Here, we summarise the discoveries as follows:
\begin{enumerate}
	\item We propose an equation as a high-level summary of the mean net thrust on each undulating swimmer: $ \ctm = \re^{0.17} \str^{2.03} \lam^{1.23} - 0.26 \re^{0.19} \str^{1.00} \lam^{0.10} - 6.13 \re^{-0.6} $.
	\item Mean net thrust increases with wavelength $ \lam $ and Strouhal number $ \str $, yet only slightly with Reynolds number $ \re $ {\coa in the present laminar regime}. When $ \lam $ and $ \str $ are relatively small, the thrust increases almost linearly with wavelength $ \ctm \sim \lam^{1.23} $ while scaling with Strouhal number as $ \ctm \sim \str^{2} $, where the scaling-exponent being {\coa two-thirds} of that for a single swimmer $ \ctm \sim \str^{3} $ \citep{chao2022hydrodynamic}.
	\item Side-by-side anti-phase schooling can enhance the thrust by more than ten times, as compared with a single swimmer at $ \lam \geq 1.5 $ and $ \str = 0.3 $.
	\item High net propulsive efficiency $ \eta $ is achieved at $ \str > 0.4 $ and $ 0.8 < \lam < 1.5 $, with the highest obtained at about $ \str = 0.5 $ and $ \lam = 1.1 $, which are consistent with the biological observations.
	\item We classify the flow structures into six distinct types based on their flow characteristics: (a) steady wake, (b) quasi-Karman wake, (c) 2S, (d) 2P-diverge, (e) 2P-converge, (f) symmetry breaking.
	\item Wavelength $ \lam $ and Strouhal number $ \str $ are more influential than Reynolds number $ \re $ in determining the flow structures in the tested parametric space. The highest net thrust is accompanied with symmetry breaking, whereas the high-efficiency regime corresponds to a 2P-diverge type wake.
	\item {\cob Instantaneous high thrust occurs when differential pressure at the rear and front of the hydrofoil body simultaneously "pushes" and "pulls" the swimmers, respectively, while the presence of a vortex pair in the intermediary space diminishes thrust when the tails move outwards.}
\end{enumerate}

Based on these results, we can make a few interesting comments that {\coa help understand} the hydrodynamically-relevant behaviour of biological swimmers in nature or for designing a schooling strategy for biomimetic robotic underwater vehicles.
Side-by-side schooling can produce much higher thrust and, therefore, higher acceleration than a single BCF swimmer. This schooling advantage is most prominent at low Reynolds numbers, implying that the fish's swimming agility improves in school. Hence, schooling may be preferred due to better survivorship associated with faster acceleration.
Also, in the context of schooling conditions considered here, the wavelength and Strouhal number for best efficiency are $ \lam \approx 1.2 $ and $ \str \approx 0.5 $, respectively, which match the observed values for a single swimmer \citep{Borazjani2008}.
In future, Floquet analysis can be further applied to shed light on the symmetry breaking during acceleration.

{\coa \section{Limitations and future works}
\label{sec:limitation}
The present study involves only 2D simulations, whereas high-fidelity 3D models can be useful to real 3D flow patterns for a few cases, \eg maximum thrust or efficiency.
The present study applies a tethered configuration to approximate various instants of linear acceleration, which is not the most accurate and intuitive method. We will conduct more self-propelling simulations in the future.
Also, the conclusions regarding Reynolds number are limited to the present laminar parametric space involving $ \re = 1000-2000 $. The effects of $ \re $ in transition and turbulent regimes remain to be explored in the future.
What is more, further analysis of the underlying mechanism should require examination of added mass and added damping, etc.

}

\begin{acknowledgments}
This work was funded by
China Postdoctoral Science Foundation (Grant No. 2021M691865)
and by
Science and Technology Major Project of Fujian Province in China (Grant No. 2021NZ033016).
We appreciate the US National Science Foundation award OAC 1931368 (A.P.S.B) for supporting the IBAMR library.
This work was also financially supported by the National Natural Science Foundation of China (Grant Nos. 12074323; 42106181), the Natural Science Foundation of Fujian Province of China (No. 2022J02003), the China National Postdoctoral Program for Innovative Talents (Grant No. BX2021168) and the Outstanding Postdoctoral Scholarship, State Key Laboratory of Marine Environmental Science at Xiamen University.

\end{acknowledgments}

\section*{Data Availability Statement}
The data that support the findings of this study are available from the corresponding author upon reasonable request.

\bibliography{library}

\begin{thebibliography}{99}%
\makeatletter
\providecommand \@ifxundefined [1]{%
 \@ifx{#1\undefined}
}%
\providecommand \@ifnum [1]{%
 \ifnum #1\expandafter \@firstoftwo
 \else \expandafter \@secondoftwo
 \fi
}%
\providecommand \@ifx [1]{%
 \ifx #1\expandafter \@firstoftwo
 \else \expandafter \@secondoftwo
 \fi
}%
\providecommand \natexlab [1]{#1}%
\providecommand \enquote  [1]{``#1''}%
\providecommand \bibnamefont  [1]{#1}%
\providecommand \bibfnamefont [1]{#1}%
\providecommand \citenamefont [1]{#1}%
\providecommand \href@noop [0]{\@secondoftwo}%
\providecommand \href [0]{\begingroup \@sanitize@url \@href}%
\providecommand \@href[1]{\@@startlink{#1}\@@href}%
\providecommand \@@href[1]{\endgroup#1\@@endlink}%
\providecommand \@sanitize@url [0]{\catcode `\\12\catcode `\$12\catcode
  `\&12\catcode `\#12\catcode `\^12\catcode `\_12\catcode `\%12\relax}%
\providecommand \@@startlink[1]{}%
\providecommand \@@endlink[0]{}%
\providecommand \url  [0]{\begingroup\@sanitize@url \@url }%
\providecommand \@url [1]{\endgroup\@href {#1}{\urlprefix }}%
\providecommand \urlprefix  [0]{URL }%
\providecommand \Eprint [0]{\href }%
\providecommand \doibase [0]{https://doi.org/}%
\providecommand \selectlanguage [0]{\@gobble}%
\providecommand \bibinfo  [0]{\@secondoftwo}%
\providecommand \bibfield  [0]{\@secondoftwo}%
\providecommand \translation [1]{[#1]}%
\providecommand \BibitemOpen [0]{}%
\providecommand \bibitemStop [0]{}%
\providecommand \bibitemNoStop [0]{.\EOS\space}%
\providecommand \EOS [0]{\spacefactor3000\relax}%
\providecommand \BibitemShut  [1]{\csname bibitem#1\endcsname}%
\let\auto@bib@innerbib\@empty
\bibitem [{\citenamefont {Akanyeti}\ \emph {et~al.}(2017)\citenamefont
  {Akanyeti}, \citenamefont {Putney}, \citenamefont {Yanagitsuru},
  \citenamefont {Lauder}, \citenamefont {Stewart},\ and\ \citenamefont
  {Liao}}]{Akanyeti2017}%
  \BibitemOpen
  \bibfield  {author} {\bibinfo {author} {\bibnamefont {Akanyeti},
  \bibfnamefont {O.}}, \bibinfo {author} {\bibnamefont {Putney}, \bibfnamefont
  {J.}}, \bibinfo {author} {\bibnamefont {Yanagitsuru}, \bibfnamefont {Y.~R.}},
  \bibinfo {author} {\bibnamefont {Lauder}, \bibfnamefont {G.~V.}}, \bibinfo
  {author} {\bibnamefont {Stewart}, \bibfnamefont {W.~J.}}, and\ \bibinfo
  {author} {\bibnamefont {Liao}, \bibfnamefont {J.~C.}},\ }\bibfield  {title}
  {\enquote {\bibinfo {title} {{Accelerating fishes increase propulsive
  efficiency by modulating vortex ring geometry}},}\ }\href
  {https://doi.org/10.1073/pnas.1705968115} {\bibfield  {journal} {\bibinfo
  {journal} {Proceedings of the National Academy of Sciences of the United
  States of America}\ }\textbf {\bibinfo {volume} {114}} (\bibinfo {year}
  {2017}),\ 10.1073/pnas.1705968115}\BibitemShut {NoStop}%
\bibitem [{\citenamefont {Alam}\ and\ \citenamefont
  {Muhammad}(2020)}]{Alam2020}%
  \BibitemOpen
  \bibfield  {author} {\bibinfo {author} {\bibnamefont {Alam}, \bibfnamefont
  {M.~M.}}and\ \bibinfo {author} {\bibnamefont {Muhammad}, \bibfnamefont
  {Z.}},\ }\bibfield  {title} {\enquote {\bibinfo {title} {{Dynamics of flow
  around a pitching hydrofoil}},}\ }\href
  {https://doi.org/10.1016/j.jfluidstructs.2020.103151} {\bibfield  {journal}
  {\bibinfo  {journal} {Journal of Fluids and Structures}\ }\textbf {\bibinfo
  {volume} {99}} (\bibinfo {year} {2020}),\
  10.1016/j.jfluidstructs.2020.103151}\BibitemShut {NoStop}%
\bibitem [{\citenamefont {Ashraf}\ \emph {et~al.}(2017)\citenamefont {Ashraf},
  \citenamefont {Bradshaw}, \citenamefont {Ha}, \citenamefont {Halloy},
  \citenamefont {Godoy-Diana},\ and\ \citenamefont {Thiria}}]{Ashraf2017}%
  \BibitemOpen
  \bibfield  {author} {\bibinfo {author} {\bibnamefont {Ashraf}, \bibfnamefont
  {I.}}, \bibinfo {author} {\bibnamefont {Bradshaw}, \bibfnamefont {H.}},
  \bibinfo {author} {\bibnamefont {Ha}, \bibfnamefont {T.~T.}}, \bibinfo
  {author} {\bibnamefont {Halloy}, \bibfnamefont {J.}}, \bibinfo {author}
  {\bibnamefont {Godoy-Diana}, \bibfnamefont {R.}}, and\ \bibinfo {author}
  {\bibnamefont {Thiria}, \bibfnamefont {B.}},\ }\bibfield  {title} {\enquote
  {\bibinfo {title} {{Simple phalanx pattern leads to energy saving in cohesive
  fish schooling}},}\ }\href {https://doi.org/10.1073/pnas.1706503114}
  {\bibfield  {journal} {\bibinfo  {journal} {Proceedings of the National
  Academy of Sciences of the United States of America}\ }\textbf {\bibinfo
  {volume} {114}} (\bibinfo {year} {2017}),\
  10.1073/pnas.1706503114}\BibitemShut {NoStop}%
\bibitem [{\citenamefont {Balay}\ \emph {et~al.}(2001)\citenamefont {Balay},
  \citenamefont {Abhyankar}, \citenamefont {Adams}, \citenamefont {Brown},
  \citenamefont {Brune}, \citenamefont {Buschelman}, \citenamefont {Dalcin},
  \citenamefont {Eijkhout}, \citenamefont {Gropp}, \citenamefont {Kaushik},\
  and\ \citenamefont {Others}}]{balay2001petsc}%
  \BibitemOpen
  \bibfield  {author} {\bibinfo {author} {\bibnamefont {Balay}, \bibfnamefont
  {S.}}, \bibinfo {author} {\bibnamefont {Abhyankar}, \bibfnamefont {S.}},
  \bibinfo {author} {\bibnamefont {Adams}, \bibfnamefont {M.~F.}}, \bibinfo
  {author} {\bibnamefont {Brown}, \bibfnamefont {J.}}, \bibinfo {author}
  {\bibnamefont {Brune}, \bibfnamefont {P.}}, \bibinfo {author} {\bibnamefont
  {Buschelman}, \bibfnamefont {K.}}, \bibinfo {author} {\bibnamefont {Dalcin},
  \bibfnamefont {L.}}, \bibinfo {author} {\bibnamefont {Eijkhout},
  \bibfnamefont {V.}}, \bibinfo {author} {\bibnamefont {Gropp}, \bibfnamefont
  {W.~D.}}, \bibinfo {author} {\bibnamefont {Kaushik}, \bibfnamefont {D.}},
  and\ \bibinfo {author} {\bibnamefont {Others},},\ }\href@noop {} {\enquote
  {\bibinfo {title} {{PETSc web page}},}\ } (\bibinfo {year}
  {2001})\BibitemShut {NoStop}%
\bibitem [{\citenamefont {Balay}\ \emph {et~al.}(2010)\citenamefont {Balay},
  \citenamefont {Buschelman}, \citenamefont {Eijkhout}, \citenamefont {Gropp},
  \citenamefont {Kaushik}, \citenamefont {Knepley}, \citenamefont {Mcinnes},
  \citenamefont {Smith},\ and\ \citenamefont {Zhang}}]{Balay2010}%
  \BibitemOpen
  \bibfield  {author} {\bibinfo {author} {\bibnamefont {Balay}, \bibfnamefont
  {S.}}, \bibinfo {author} {\bibnamefont {Buschelman}, \bibfnamefont {K.}},
  \bibinfo {author} {\bibnamefont {Eijkhout}, \bibfnamefont {V.}}, \bibinfo
  {author} {\bibnamefont {Gropp}, \bibfnamefont {W.}}, \bibinfo {author}
  {\bibnamefont {Kaushik}, \bibfnamefont {D.}}, \bibinfo {author} {\bibnamefont
  {Knepley}, \bibfnamefont {M.}}, \bibinfo {author} {\bibnamefont {Mcinnes},
  \bibfnamefont {L.~C.}}, \bibinfo {author} {\bibnamefont {Smith},
  \bibfnamefont {B.}}, and\ \bibinfo {author} {\bibnamefont {Zhang},
  \bibfnamefont {H.}},\ }\bibfield  {title} {\enquote {\bibinfo {title} {{PETSc
  Users Manual}},}\ }\href@noop {} {\bibfield  {journal} {\bibinfo  {journal}
  {ReVision}\ }\textbf {\bibinfo {volume} {2}} (\bibinfo {year}
  {2010})}\BibitemShut {NoStop}%
\bibitem [{\citenamefont {Balay}\ \emph {et~al.}(1997)\citenamefont {Balay},
  \citenamefont {Gropp}, \citenamefont {McInnes},\ and\ \citenamefont
  {Smith}}]{Balay1997}%
  \BibitemOpen
  \bibfield  {author} {\bibinfo {author} {\bibnamefont {Balay}, \bibfnamefont
  {S.}}, \bibinfo {author} {\bibnamefont {Gropp}, \bibfnamefont {W.~D.}},
  \bibinfo {author} {\bibnamefont {McInnes}, \bibfnamefont {L.~C.}}, and\
  \bibinfo {author} {\bibnamefont {Smith}, \bibfnamefont {B.~F.}},\ }\bibfield
  {title} {\enquote {\bibinfo {title} {{Efficient Management of Parallelism in
  Object-Oriented Numerical Software Libraries}},}\ }in\ \href
  {https://doi.org/10.1007/978-1-4612-1986-6_8} {\emph {\bibinfo {booktitle}
  {Modern Software Tools for Scientific Computing}}}\ (\bibinfo {year}
  {1997})\BibitemShut {NoStop}%
\bibitem [{\citenamefont {Bhalla}\ \emph {et~al.}(2013)\citenamefont {Bhalla},
  \citenamefont {Bale}, \citenamefont {Griffith},\ and\ \citenamefont
  {Patankar}}]{Bhalla2013}%
  \BibitemOpen
  \bibfield  {author} {\bibinfo {author} {\bibnamefont {Bhalla}, \bibfnamefont
  {A.~P.~S.}}, \bibinfo {author} {\bibnamefont {Bale}, \bibfnamefont {R.}},
  \bibinfo {author} {\bibnamefont {Griffith}, \bibfnamefont {B.~E.}}, and\
  \bibinfo {author} {\bibnamefont {Patankar}, \bibfnamefont {N.~A.}},\
  }\bibfield  {title} {\enquote {\bibinfo {title} {{A unified mathematical
  framework and an adaptive numerical method for fluid-structure interaction
  with rigid, deforming, and elastic bodies}},}\ }\href
  {https://doi.org/10.1016/j.jcp.2013.04.033} {\bibfield  {journal} {\bibinfo
  {journal} {Journal of Computational Physics}\ }\textbf {\bibinfo {volume}
  {250}} (\bibinfo {year} {2013}),\ 10.1016/j.jcp.2013.04.033}\BibitemShut
  {NoStop}%
\bibitem [{\citenamefont {Bhalla}\ \emph {et~al.}(2014)\citenamefont {Bhalla},
  \citenamefont {Bale}, \citenamefont {Griffith},\ and\ \citenamefont
  {Patankar}}]{bhalla2014fully}%
  \BibitemOpen
  \bibfield  {author} {\bibinfo {author} {\bibnamefont {Bhalla}, \bibfnamefont
  {A.~P.~S.}}, \bibinfo {author} {\bibnamefont {Bale}, \bibfnamefont {R.}},
  \bibinfo {author} {\bibnamefont {Griffith}, \bibfnamefont {B.~E.}}, and\
  \bibinfo {author} {\bibnamefont {Patankar}, \bibfnamefont {N.~A.}},\
  }\bibfield  {title} {\enquote {\bibinfo {title} {{Fully resolved immersed
  electrohydrodynamics for particle motion, electrolocation, and
  self-propulsion}},}\ }\href@noop {} {\bibfield  {journal} {\bibinfo
  {journal} {Journal of Computational Physics}\ }\textbf {\bibinfo {volume}
  {256}},\ \bibinfo {pages} {88--108} (\bibinfo {year} {2014})}\BibitemShut
  {NoStop}%
\bibitem [{\citenamefont {Bhalla}, \citenamefont {Griffith},\ and\
  \citenamefont {Patankar}(2013)}]{Bhalla2013a}%
  \BibitemOpen
  \bibfield  {author} {\bibinfo {author} {\bibnamefont {Bhalla}, \bibfnamefont
  {A.~P.~S.}}, \bibinfo {author} {\bibnamefont {Griffith}, \bibfnamefont
  {B.~E.}}, and\ \bibinfo {author} {\bibnamefont {Patankar}, \bibfnamefont
  {N.~A.}},\ }\bibfield  {title} {\enquote {\bibinfo {title} {{A Forced Damped
  Oscillation Framework for Undulatory Swimming Provides New Insights into How
  Propulsion Arises in Active and Passive Swimming}},}\ }\href
  {https://doi.org/10.1371/journal.pcbi.1003097} {\bibfield  {journal}
  {\bibinfo  {journal} {PLoS Computational Biology}\ }\textbf {\bibinfo
  {volume} {9}} (\bibinfo {year} {2013}),\
  10.1371/journal.pcbi.1003097}\BibitemShut {NoStop}%
\bibitem [{\citenamefont {Bhalla}\ \emph {et~al.}(2020)\citenamefont {Bhalla},
  \citenamefont {Nangia}, \citenamefont {Dafnakis}, \citenamefont {Bracco},\
  and\ \citenamefont {Mattiazzo}}]{Bhalla2020}%
  \BibitemOpen
  \bibfield  {author} {\bibinfo {author} {\bibnamefont {Bhalla}, \bibfnamefont
  {A.~P.~S.}}, \bibinfo {author} {\bibnamefont {Nangia}, \bibfnamefont {N.}},
  \bibinfo {author} {\bibnamefont {Dafnakis}, \bibfnamefont {P.}}, \bibinfo
  {author} {\bibnamefont {Bracco}, \bibfnamefont {G.}}, and\ \bibinfo {author}
  {\bibnamefont {Mattiazzo}, \bibfnamefont {G.}},\ }\bibfield  {title}
  {\enquote {\bibinfo {title} {{Simulating water-entry/exit problems using
  Eulerian–Lagrangian and fully-Eulerian fictitious domain methods within the
  open-source IBAMR library}},}\ }\href
  {https://doi.org/10.1016/j.apor.2019.101932} {\bibfield  {journal} {\bibinfo
  {journal} {Applied Ocean Research}\ }\textbf {\bibinfo {volume} {94}}
  (\bibinfo {year} {2020}),\ 10.1016/j.apor.2019.101932}\BibitemShut {NoStop}%
\bibitem [{\citenamefont {Borazjani}(2013)}]{Borazjani2013}%
  \BibitemOpen
  \bibfield  {author} {\bibinfo {author} {\bibnamefont {Borazjani},
  \bibfnamefont {I.}},\ }\bibfield  {title} {\enquote {\bibinfo {title} {{The
  functional role of caudal and anal/dorsal fins during the C-start of a
  bluegill sunfish}},}\ }\href {https://doi.org/10.1242/jeb.079434} {\bibfield
  {journal} {\bibinfo  {journal} {Journal of Experimental Biology}\ }\textbf
  {\bibinfo {volume} {216}} (\bibinfo {year} {2013}),\
  10.1242/jeb.079434}\BibitemShut {NoStop}%
\bibitem [{\citenamefont {Borazjani}\ and\ \citenamefont
  {Sotiropoulos}(2008)}]{Borazjani2008}%
  \BibitemOpen
  \bibfield  {author} {\bibinfo {author} {\bibnamefont {Borazjani},
  \bibfnamefont {I.}}and\ \bibinfo {author} {\bibnamefont {Sotiropoulos},
  \bibfnamefont {F.}},\ }\bibfield  {title} {\enquote {\bibinfo {title}
  {{Numerical investigation of the hydrodynamics of carangiform swimming in the
  transitional and inertial flow regimes}},}\ }\href
  {https://doi.org/10.1242/jeb.015644} {\bibfield  {journal} {\bibinfo
  {journal} {Journal of Experimental Biology}\ }\textbf {\bibinfo {volume}
  {211}} (\bibinfo {year} {2008}),\ 10.1242/jeb.015644}\BibitemShut {NoStop}%
\bibitem [{\citenamefont {Borazjani}\ and\ \citenamefont
  {Sotiropoulos}(2009)}]{Borazjani2009}%
  \BibitemOpen
  \bibfield  {author} {\bibinfo {author} {\bibnamefont {Borazjani},
  \bibfnamefont {I.}}and\ \bibinfo {author} {\bibnamefont {Sotiropoulos},
  \bibfnamefont {F.}},\ }\bibfield  {title} {\enquote {\bibinfo {title}
  {{Numerical investigation of the hydrodynamics of anguilliform swimming in
  the transitional and inertial flow regimes}},}\ }\href
  {https://doi.org/10.1242/jeb.025007} {\bibfield  {journal} {\bibinfo
  {journal} {Journal of Experimental Biology}\ }\textbf {\bibinfo {volume}
  {212}} (\bibinfo {year} {2009}),\ 10.1242/jeb.025007}\BibitemShut {NoStop}%
\bibitem [{\citenamefont {Borazjani}\ and\ \citenamefont
  {Sotiropoulos}(2010)}]{Borazjani2010}%
  \BibitemOpen
  \bibfield  {author} {\bibinfo {author} {\bibnamefont {Borazjani},
  \bibfnamefont {I.}}and\ \bibinfo {author} {\bibnamefont {Sotiropoulos},
  \bibfnamefont {F.}},\ }\bibfield  {title} {\enquote {\bibinfo {title} {{On
  the role of form and kinematics on the hydrodynamics of self-propelled
  body/caudal fin swimming}},}\ }\href {https://doi.org/10.1242/jeb.030932}
  {\bibfield  {journal} {\bibinfo  {journal} {Journal of Experimental Biology}\
  }\textbf {\bibinfo {volume} {213}} (\bibinfo {year} {2010}),\
  10.1242/jeb.030932}\BibitemShut {NoStop}%
\bibitem [{\citenamefont {Borazjani}\ \emph {et~al.}(2012)\citenamefont
  {Borazjani}, \citenamefont {Sotiropoulos}, \citenamefont {Tytell},\ and\
  \citenamefont {Lauder}}]{Borazjani2012}%
  \BibitemOpen
  \bibfield  {author} {\bibinfo {author} {\bibnamefont {Borazjani},
  \bibfnamefont {I.}}, \bibinfo {author} {\bibnamefont {Sotiropoulos},
  \bibfnamefont {F.}}, \bibinfo {author} {\bibnamefont {Tytell}, \bibfnamefont
  {E.~D.}}, and\ \bibinfo {author} {\bibnamefont {Lauder}, \bibfnamefont
  {G.~V.}},\ }\bibfield  {title} {\enquote {\bibinfo {title} {{Hydrodynamics of
  the bluegill sunfish C-start escape response: Three-dimensional simulations
  and comparison with experimental data}},}\ }\href
  {https://doi.org/10.1242/jeb.063016} {\bibfield  {journal} {\bibinfo
  {journal} {Journal of Experimental Biology}\ }\textbf {\bibinfo {volume}
  {215}} (\bibinfo {year} {2012}),\ 10.1242/jeb.063016}\BibitemShut {NoStop}%
\bibitem [{\citenamefont {Chao}, \citenamefont {Alam},\ and\ \citenamefont
  {Cheng}(2022)}]{chao2022hydrodynamic}%
  \BibitemOpen
  \bibfield  {author} {\bibinfo {author} {\bibnamefont {Chao}, \bibfnamefont
  {L.-M.}}, \bibinfo {author} {\bibnamefont {Alam}, \bibfnamefont {M.~M.}},
  and\ \bibinfo {author} {\bibnamefont {Cheng}, \bibfnamefont {L.}},\
  }\bibfield  {title} {\enquote {\bibinfo {title} {{Hydrodynamic performance of
  slender swimmer: effect of travelling wavelength}},}\ }\href@noop {}
  {\bibfield  {journal} {\bibinfo  {journal} {Journal of Fluid Mechanics}\
  }\textbf {\bibinfo {volume} {947}},\ \bibinfo {pages} {A8} (\bibinfo {year}
  {2022})}\BibitemShut {NoStop}%
\bibitem [{\citenamefont {Chao}, \citenamefont {Alam},\ and\ \citenamefont
  {Ji}(2021)}]{Chao2021}%
  \BibitemOpen
  \bibfield  {author} {\bibinfo {author} {\bibnamefont {Chao}, \bibfnamefont
  {L.~M.}}, \bibinfo {author} {\bibnamefont {Alam}, \bibfnamefont {M.~M.}},
  and\ \bibinfo {author} {\bibnamefont {Ji}, \bibfnamefont {C.}},\ }\bibfield
  {title} {\enquote {\bibinfo {title} {{Drag-thrust transition and wake
  structures of a pitching foil undergoing asymmetric oscillation}},}\ }\href
  {https://doi.org/10.1016/j.jfluidstructs.2021.103289} {\bibfield  {journal}
  {\bibinfo  {journal} {Journal of Fluids and Structures}\ }\textbf {\bibinfo
  {volume} {103}} (\bibinfo {year} {2021}),\
  10.1016/j.jfluidstructs.2021.103289}\BibitemShut {NoStop}%
\bibitem [{\citenamefont {Chao}\ \emph {et~al.}(2019)\citenamefont {Chao},
  \citenamefont {Pan}, \citenamefont {Zhang},\ and\ \citenamefont
  {Yan}}]{Chao2019}%
  \BibitemOpen
  \bibfield  {author} {\bibinfo {author} {\bibnamefont {Chao}, \bibfnamefont
  {L.~M.}}, \bibinfo {author} {\bibnamefont {Pan}, \bibfnamefont {G.}},
  \bibinfo {author} {\bibnamefont {Zhang}, \bibfnamefont {D.}}, and\ \bibinfo
  {author} {\bibnamefont {Yan}, \bibfnamefont {G.~X.}},\ }\bibfield  {title}
  {\enquote {\bibinfo {title} {{On the two staggered swimming fish}},}\ }\href
  {https://doi.org/10.1016/j.chaos.2019.04.028} {\bibfield  {journal} {\bibinfo
   {journal} {Chaos, Solitons and Fractals}\ }\textbf {\bibinfo {volume} {123}}
  (\bibinfo {year} {2019}),\ 10.1016/j.chaos.2019.04.028}\BibitemShut {NoStop}%
\bibitem [{\citenamefont {Coombs}\ and\ \citenamefont
  {Montgomery}(2014)}]{Coombs2014}%
  \BibitemOpen
  \bibfield  {author} {\bibinfo {author} {\bibnamefont {Coombs}, \bibfnamefont
  {S.}}and\ \bibinfo {author} {\bibnamefont {Montgomery}, \bibfnamefont {J.}},\
  }\bibfield  {title} {\enquote {\bibinfo {title} {{The role of flow and the
  lateral line in the multisensory guidance of orienting behaviors}},}\ }in\
  \href {https://doi.org/10.1007/978-3-642-41446-6_3} {\emph {\bibinfo
  {booktitle} {Flow Sensing in Air and Water: Behavioral, Neural and
  Engineering Principles of Operation}}}\ (\bibinfo {year} {2014})\BibitemShut
  {NoStop}%
\bibitem [{\citenamefont {Cranmer}(2020)}]{pysr}%
  \BibitemOpen
  \bibfield  {author} {\bibinfo {author} {\bibnamefont {Cranmer}, \bibfnamefont
  {M.}},\ }\href {https://doi.org/10.5281/zenodo.4041459} {\enquote {\bibinfo
  {title} {{PySR: Fast \& Parallelized Symbolic Regression in Python/Julia}},}\
  } (\bibinfo {year} {2020})\BibitemShut {NoStop}%
\bibitem [{\citenamefont {Cranmer}(2023)}]{cranmer2023interpretable}%
  \BibitemOpen
  \bibfield  {author} {\bibinfo {author} {\bibnamefont {Cranmer}, \bibfnamefont
  {M.}},\ }\href@noop {} {\enquote {\bibinfo {title} {{Interpretable Machine
  Learning for Science with PySR and SymbolicRegression.jl}},}\ } (\bibinfo
  {year} {2023}),\ \Eprint {https://arxiv.org/abs/2305.01582} {arXiv:2305.01582
  [astro-ph.IM]} \BibitemShut {NoStop}%
\bibitem [{\citenamefont {Cranmer}\ \emph {et~al.}(2020)\citenamefont
  {Cranmer}, \citenamefont {Sanchez-Gonzalez}, \citenamefont {Battaglia},
  \citenamefont {Xu}, \citenamefont {Cranmer}, \citenamefont {Spergel},\ and\
  \citenamefont {Ho}}]{cranmer2020discovering}%
  \BibitemOpen
  \bibfield  {author} {\bibinfo {author} {\bibnamefont {Cranmer}, \bibfnamefont
  {M.}}, \bibinfo {author} {\bibnamefont {Sanchez-Gonzalez}, \bibfnamefont
  {A.}}, \bibinfo {author} {\bibnamefont {Battaglia}, \bibfnamefont {P.}},
  \bibinfo {author} {\bibnamefont {Xu}, \bibfnamefont {R.}}, \bibinfo {author}
  {\bibnamefont {Cranmer}, \bibfnamefont {K.}}, \bibinfo {author} {\bibnamefont
  {Spergel}, \bibfnamefont {D.}}, and\ \bibinfo {author} {\bibnamefont {Ho},
  \bibfnamefont {S.}},\ }\bibfield  {title} {\enquote {\bibinfo {title}
  {{Discovering Symbolic Models from Deep Learning with Inductive Biases}},}\
  }\href@noop {} {\bibfield  {journal} {\bibinfo  {journal} {NeurIPS 2020}\ }
  (\bibinfo {year} {2020})},\ \Eprint {https://arxiv.org/abs/2006.11287}
  {arXiv:2006.11287 [cs.LG]} \BibitemShut {NoStop}%
\bibitem [{\citenamefont {Daghooghi}\ and\ \citenamefont
  {Borazjani}(2015)}]{Daghooghi2015}%
  \BibitemOpen
  \bibfield  {author} {\bibinfo {author} {\bibnamefont {Daghooghi},
  \bibfnamefont {M.}}and\ \bibinfo {author} {\bibnamefont {Borazjani},
  \bibfnamefont {I.}},\ }\bibfield  {title} {\enquote {\bibinfo {title} {{The
  hydrodynamic advantages of synchronized swimming in a rectangular
  pattern}},}\ }\href {https://doi.org/10.1088/1748-3190/10/5/056018}
  {\bibfield  {journal} {\bibinfo  {journal} {Bioinspiration and Biomimetics}\
  }\textbf {\bibinfo {volume} {10}} (\bibinfo {year} {2015}),\
  10.1088/1748-3190/10/5/056018}\BibitemShut {NoStop}%
\bibitem [{\citenamefont {Deng}\ and\ \citenamefont {Liu}(2021)}]{Deng2021}%
  \BibitemOpen
  \bibfield  {author} {\bibinfo {author} {\bibnamefont {Deng}, \bibfnamefont
  {J.}}and\ \bibinfo {author} {\bibnamefont {Liu}, \bibfnamefont {D.}},\
  }\bibfield  {title} {\enquote {\bibinfo {title} {{Spontaneous response of a
  self-organized fish school to a predator}},}\ }\href
  {https://doi.org/10.1088/1748-3190/abfd7f} {\bibfield  {journal} {\bibinfo
  {journal} {Bioinspiration and Biomimetics}\ }\textbf {\bibinfo {volume} {16}}
  (\bibinfo {year} {2021}),\ 10.1088/1748-3190/abfd7f}\BibitemShut {NoStop}%
\bibitem [{\citenamefont {Deng}, \citenamefont {Shao},\ and\ \citenamefont
  {Yu}(2007)}]{Deng2007}%
  \BibitemOpen
  \bibfield  {author} {\bibinfo {author} {\bibnamefont {Deng}, \bibfnamefont
  {J.}}, \bibinfo {author} {\bibnamefont {Shao}, \bibfnamefont {X.~M.}}, and\
  \bibinfo {author} {\bibnamefont {Yu}, \bibfnamefont {Z.~S.}},\ }\bibfield
  {title} {\enquote {\bibinfo {title} {{Hydrodynamic studies on two traveling
  wavy foils in tandem arrangement}},}\ }\href
  {https://doi.org/10.1063/1.2814259} {\bibfield  {journal} {\bibinfo
  {journal} {Physics of Fluids}\ }\textbf {\bibinfo {volume} {19}} (\bibinfo
  {year} {2007}),\ 10.1063/1.2814259}\BibitemShut {NoStop}%
\bibitem [{\citenamefont {Deng}\ \emph {et~al.}(2016)\citenamefont {Deng},
  \citenamefont {Sun}, \citenamefont {{Lubao Teng}}, \citenamefont {Pan},\ and\
  \citenamefont {Shao}}]{Deng2016}%
  \BibitemOpen
  \bibfield  {author} {\bibinfo {author} {\bibnamefont {Deng}, \bibfnamefont
  {J.}}, \bibinfo {author} {\bibnamefont {Sun}, \bibfnamefont {L.}}, \bibinfo
  {author} {\bibnamefont {{Lubao Teng}},}, \bibinfo {author} {\bibnamefont
  {Pan}, \bibfnamefont {D.}}, and\ \bibinfo {author} {\bibnamefont {Shao},
  \bibfnamefont {X.}},\ }\bibfield  {title} {\enquote {\bibinfo {title} {{The
  correlation between wake transition and propulsive efficiency of a flapping
  foil: A numerical study}},}\ }\href {https://doi.org/10.1063/1.4961566}
  {\bibfield  {journal} {\bibinfo  {journal} {Physics of Fluids}\ }\textbf
  {\bibinfo {volume} {28}} (\bibinfo {year} {2016}),\
  10.1063/1.4961566}\BibitemShut {NoStop}%
\bibitem [{\citenamefont {Deng}\ \emph {et~al.}(2015)\citenamefont {Deng},
  \citenamefont {Teng}, \citenamefont {Pan},\ and\ \citenamefont
  {Shao}}]{Deng2015}%
  \BibitemOpen
  \bibfield  {author} {\bibinfo {author} {\bibnamefont {Deng}, \bibfnamefont
  {J.}}, \bibinfo {author} {\bibnamefont {Teng}, \bibfnamefont {L.}}, \bibinfo
  {author} {\bibnamefont {Pan}, \bibfnamefont {D.}}, and\ \bibinfo {author}
  {\bibnamefont {Shao}, \bibfnamefont {X.}},\ }\bibfield  {title} {\enquote
  {\bibinfo {title} {{Inertial effects of the semi-passive flapping foil on its
  energy extraction efficiency}},}\ }\href {https://doi.org/10.1063/1.4921384}
  {\bibfield  {journal} {\bibinfo  {journal} {Physics of Fluids}\ }\textbf
  {\bibinfo {volume} {27}} (\bibinfo {year} {2015}),\
  10.1063/1.4921384}\BibitemShut {NoStop}%
\bibitem [{\citenamefont {Deng}\ \emph {et~al.}(2022)\citenamefont {Deng},
  \citenamefont {Wang}, \citenamefont {Kandel},\ and\ \citenamefont
  {Teng}}]{Deng2022}%
  \BibitemOpen
  \bibfield  {author} {\bibinfo {author} {\bibnamefont {Deng}, \bibfnamefont
  {J.}}, \bibinfo {author} {\bibnamefont {Wang}, \bibfnamefont {S.}}, \bibinfo
  {author} {\bibnamefont {Kandel}, \bibfnamefont {P.}}, and\ \bibinfo {author}
  {\bibnamefont {Teng}, \bibfnamefont {L.}},\ }\bibfield  {title} {\enquote
  {\bibinfo {title} {{Effects of free surface on a flapping-foil based ocean
  current energy extractor}},}\ }\href
  {https://doi.org/10.1016/j.renene.2021.09.098} {\bibfield  {journal}
  {\bibinfo  {journal} {Renewable Energy}\ }\textbf {\bibinfo {volume} {181}}
  (\bibinfo {year} {2022}),\ 10.1016/j.renene.2021.09.098}\BibitemShut
  {NoStop}%
\bibitem [{\citenamefont {Dewey}\ \emph {et~al.}(2014)\citenamefont {Dewey},
  \citenamefont {Quinn}, \citenamefont {Boschitsch},\ and\ \citenamefont
  {Smits}}]{Dewey2014a}%
  \BibitemOpen
  \bibfield  {author} {\bibinfo {author} {\bibnamefont {Dewey}, \bibfnamefont
  {P.~A.}}, \bibinfo {author} {\bibnamefont {Quinn}, \bibfnamefont {D.~B.}},
  \bibinfo {author} {\bibnamefont {Boschitsch}, \bibfnamefont {B.~M.}}, and\
  \bibinfo {author} {\bibnamefont {Smits}, \bibfnamefont {A.~J.}},\ }\bibfield
  {title} {\enquote {\bibinfo {title} {{Propulsive performance of unsteady
  tandem hydrofoils in a side-by-side configuration}},}\ }\href
  {https://doi.org/10.1063/1.4871024} {\bibfield  {journal} {\bibinfo
  {journal} {Physics of Fluids}\ }\textbf {\bibinfo {volume} {26}} (\bibinfo
  {year} {2014}),\ 10.1063/1.4871024}\BibitemShut {NoStop}%
\bibitem [{\citenamefont {Domenici}\ and\ \citenamefont
  {Hale}(2019)}]{Domenici2019}%
  \BibitemOpen
  \bibfield  {author} {\bibinfo {author} {\bibnamefont {Domenici},
  \bibfnamefont {P.}}and\ \bibinfo {author} {\bibnamefont {Hale}, \bibfnamefont
  {M.~E.}},\ }\href {https://doi.org/10.1242/jeb.166009} {\enquote {\bibinfo
  {title} {{Escape responses of fish: A review of the diversity in motor
  control, kinematics and behaviour}},}\ } (\bibinfo {year} {2019})\BibitemShut
  {NoStop}%
\bibitem [{\citenamefont {Dong}\ and\ \citenamefont {Lu}(2007)}]{Dong2007}%
  \BibitemOpen
  \bibfield  {author} {\bibinfo {author} {\bibnamefont {Dong}, \bibfnamefont
  {G.~J.}}and\ \bibinfo {author} {\bibnamefont {Lu}, \bibfnamefont {X.~Y.}},\
  }\bibfield  {title} {\enquote {\bibinfo {title} {{Characteristics of flow
  over traveling wavy foils in a side-by-side arrangement}},}\ }\href
  {https://doi.org/10.1063/1.2736083} {\bibfield  {journal} {\bibinfo
  {journal} {Physics of Fluids}\ }\textbf {\bibinfo {volume} {19}} (\bibinfo
  {year} {2007}),\ 10.1063/1.2736083}\BibitemShut {NoStop}%
\bibitem [{\citenamefont {{Du Clos}}\ \emph {et~al.}(2019)\citenamefont {{Du
  Clos}}, \citenamefont {Dabiri}, \citenamefont {Costello}, \citenamefont
  {Colin}, \citenamefont {Morgan}, \citenamefont {Fogerson},\ and\
  \citenamefont {Gemmell}}]{DuClos2019}%
  \BibitemOpen
  \bibfield  {author} {\bibinfo {author} {\bibnamefont {{Du Clos}},
  \bibfnamefont {K.~T.}}, \bibinfo {author} {\bibnamefont {Dabiri},
  \bibfnamefont {J.~O.}}, \bibinfo {author} {\bibnamefont {Costello},
  \bibfnamefont {J.~H.}}, \bibinfo {author} {\bibnamefont {Colin},
  \bibfnamefont {S.~P.}}, \bibinfo {author} {\bibnamefont {Morgan},
  \bibfnamefont {J.~R.}}, \bibinfo {author} {\bibnamefont {Fogerson},
  \bibfnamefont {S.~M.}}, and\ \bibinfo {author} {\bibnamefont {Gemmell},
  \bibfnamefont {B.~J.}},\ }\bibfield  {title} {\enquote {\bibinfo {title}
  {{Thrust generation during steady swimming and acceleration from rest in
  anguilliform swimmers}},}\ }\href {https://doi.org/10.1242/jeb.212464}
  {\bibfield  {journal} {\bibinfo  {journal} {Journal of Experimental Biology}\
  }\textbf {\bibinfo {volume} {222}} (\bibinfo {year} {2019}),\
  10.1242/jeb.212464}\BibitemShut {NoStop}%
\bibitem [{\citenamefont {Eaton}, \citenamefont {Bombardieri},\ and\
  \citenamefont {Meyer}(1977)}]{Eaton1977}%
  \BibitemOpen
  \bibfield  {author} {\bibinfo {author} {\bibnamefont {Eaton}, \bibfnamefont
  {R.~C.}}, \bibinfo {author} {\bibnamefont {Bombardieri}, \bibfnamefont
  {R.~A.}}, and\ \bibinfo {author} {\bibnamefont {Meyer}, \bibfnamefont
  {D.~L.}},\ }\bibfield  {title} {\enquote {\bibinfo {title} {{The
  Mauthner-initiated startle response in teleost fish.}}}\ }\href
  {https://doi.org/10.1242/jeb.66.1.65} {\bibfield  {journal} {\bibinfo
  {journal} {Journal of Experimental Biology}\ }\textbf {\bibinfo {volume}
  {66}} (\bibinfo {year} {1977}),\ 10.1242/jeb.66.1.65}\BibitemShut {NoStop}%
\bibitem [{\citenamefont {Falgout}\ \emph {et~al.}(2010)\citenamefont
  {Falgout}, \citenamefont {Cleary}, \citenamefont {Jones}, \citenamefont
  {Chow}, \citenamefont {Henson}, \citenamefont {Baldwin}, \citenamefont
  {Brown}, \citenamefont {Vassilevski},\ and\ \citenamefont
  {Yang}}]{falgout2010hypre}%
  \BibitemOpen
  \bibfield  {author} {\bibinfo {author} {\bibnamefont {Falgout}, \bibfnamefont
  {R.}}, \bibinfo {author} {\bibnamefont {Cleary}, \bibfnamefont {A.}},
  \bibinfo {author} {\bibnamefont {Jones}, \bibfnamefont {J.}}, \bibinfo
  {author} {\bibnamefont {Chow}, \bibfnamefont {E.}}, \bibinfo {author}
  {\bibnamefont {Henson}, \bibfnamefont {V.}}, \bibinfo {author} {\bibnamefont
  {Baldwin}, \bibfnamefont {C.}}, \bibinfo {author} {\bibnamefont {Brown},
  \bibfnamefont {P.}}, \bibinfo {author} {\bibnamefont {Vassilevski},
  \bibfnamefont {P.}}, and\ \bibinfo {author} {\bibnamefont {Yang},
  \bibfnamefont {U.~M.}},\ }\bibfield  {title} {\enquote {\bibinfo {title}
  {{HYPRE: High Performance Preconditioners}},}\ }\href@noop {} {\bibfield
  {journal} {\bibinfo  {journal} {Users Manual. Version}\ }\textbf {\bibinfo
  {volume} {1}} (\bibinfo {year} {2010})}\BibitemShut {NoStop}%
\bibitem [{\citenamefont {Fish}(2020)}]{Fish2020}%
  \BibitemOpen
  \bibfield  {author} {\bibinfo {author} {\bibnamefont {Fish}, \bibfnamefont
  {F.~E.}},\ }\bibfield  {title} {\enquote {\bibinfo {title} {{Advantages of
  aquatic animals as models for bio-inspired drones over present AUV
  technology}},}\ }\href {https://doi.org/10.1088/1748-3190/ab5a34} {\bibfield
  {journal} {\bibinfo  {journal} {Bioinspiration and Biomimetics}\ }\textbf
  {\bibinfo {volume} {15}} (\bibinfo {year} {2020}),\
  10.1088/1748-3190/ab5a34}\BibitemShut {NoStop}%
\bibitem [{\citenamefont {Floryan}\ \emph {et~al.}(2017)\citenamefont
  {Floryan}, \citenamefont {{Van Buren}}, \citenamefont {Rowley},\ and\
  \citenamefont {Smits}}]{Floryan2017}%
  \BibitemOpen
  \bibfield  {author} {\bibinfo {author} {\bibnamefont {Floryan}, \bibfnamefont
  {D.}}, \bibinfo {author} {\bibnamefont {{Van Buren}}, \bibfnamefont {T.}},
  \bibinfo {author} {\bibnamefont {Rowley}, \bibfnamefont {C.~W.}}, and\
  \bibinfo {author} {\bibnamefont {Smits}, \bibfnamefont {A.~J.}},\ }\bibfield
  {title} {\enquote {\bibinfo {title} {{Scaling the propulsive performance of
  heaving and pitching foils}},}\ }\href {https://doi.org/10.1017/jfm.2017.302}
  {\bibfield  {journal} {\bibinfo  {journal} {Journal of Fluid Mechanics}\
  }\textbf {\bibinfo {volume} {822}} (\bibinfo {year} {2017}),\
  10.1017/jfm.2017.302}\BibitemShut {NoStop}%
\bibitem [{\citenamefont {Gazzola}, \citenamefont {Argentina},\ and\
  \citenamefont {Mahadevan}(2014)}]{Gazzola2014}%
  \BibitemOpen
  \bibfield  {author} {\bibinfo {author} {\bibnamefont {Gazzola}, \bibfnamefont
  {M.}}, \bibinfo {author} {\bibnamefont {Argentina}, \bibfnamefont {M.}}, and\
  \bibinfo {author} {\bibnamefont {Mahadevan}, \bibfnamefont {L.}},\ }\bibfield
   {title} {\enquote {\bibinfo {title} {{Scaling macroscopic aquatic
  locomotion}},}\ }\href {https://doi.org/10.1038/nphys3078} {\bibfield
  {journal} {\bibinfo  {journal} {Nature Physics}\ }\textbf {\bibinfo {volume}
  {10}} (\bibinfo {year} {2014}),\ 10.1038/nphys3078}\BibitemShut {NoStop}%
\bibitem [{\citenamefont {Gazzola}\ \emph {et~al.}(2012)\citenamefont
  {Gazzola}, \citenamefont {Mimeau}, \citenamefont {Tchieu},\ and\
  \citenamefont {Koumoutsakos}}]{Gazzola2012}%
  \BibitemOpen
  \bibfield  {author} {\bibinfo {author} {\bibnamefont {Gazzola}, \bibfnamefont
  {M.}}, \bibinfo {author} {\bibnamefont {Mimeau}, \bibfnamefont {C.}},
  \bibinfo {author} {\bibnamefont {Tchieu}, \bibfnamefont {A.~A.}}, and\
  \bibinfo {author} {\bibnamefont {Koumoutsakos}, \bibfnamefont {P.}},\
  }\bibfield  {title} {\enquote {\bibinfo {title} {{Flow Mediated Interactions
  Between Two Cylinders at Finite $Re$ Numbers}},}\ }\href
  {https://doi.org/10.1063/1.4704195} {\bibfield  {journal} {\bibinfo
  {journal} {Physics of Fluids}\ }\textbf {\bibinfo {volume} {24}},\ \bibinfo
  {pages} {043103} (\bibinfo {year} {2012})}\BibitemShut {NoStop}%
\bibitem [{\citenamefont {Griffith}(2013)}]{griffith2013ibamr}%
  \BibitemOpen
  \bibfield  {author} {\bibinfo {author} {\bibnamefont {Griffith},
  \bibfnamefont {B.~E.}},\ }\bibfield  {title} {\enquote {\bibinfo {title}
  {{IBAMR: An adaptive and distributed-memory parallel implementation of the
  immersed boundary method}},}\ }\href@noop {} {\bibfield  {journal} {\bibinfo
  {journal} {URL https://ibamr. github. io/about}\ } (\bibinfo {year}
  {2013})}\BibitemShut {NoStop}%
\bibitem [{\citenamefont {Griffith}\ and\ \citenamefont
  {Patankar}(2020)}]{Griffith2020}%
  \BibitemOpen
  \bibfield  {author} {\bibinfo {author} {\bibnamefont {Griffith},
  \bibfnamefont {B.~E.}}and\ \bibinfo {author} {\bibnamefont {Patankar},
  \bibfnamefont {N.~A.}},\ }\href
  {https://doi.org/10.1146/annurev-fluid-010719-060228} {\enquote {\bibinfo
  {title} {{Immersed Methods for Fluid-Structure Interaction}},}\ } (\bibinfo
  {year} {2020})\BibitemShut {NoStop}%
\bibitem [{\citenamefont {Grundner}\ \emph {et~al.}(2023)\citenamefont
  {Grundner}, \citenamefont {Beucler}, \citenamefont {Gentine},\ and\
  \citenamefont {Eyring}}]{grundner2023data}%
  \BibitemOpen
  \bibfield  {author} {\bibinfo {author} {\bibnamefont {Grundner},
  \bibfnamefont {A.}}, \bibinfo {author} {\bibnamefont {Beucler}, \bibfnamefont
  {T.}}, \bibinfo {author} {\bibnamefont {Gentine}, \bibfnamefont {P.}}, and\
  \bibinfo {author} {\bibnamefont {Eyring}, \bibfnamefont {V.}},\ }\bibfield
  {title} {\enquote {\bibinfo {title} {{Data-Driven Equation Discovery of a
  Cloud Cover Parameterization}},}\ }\href@noop {} {\bibfield  {journal}
  {\bibinfo  {journal} {arXiv preprint arXiv:2304.08063}\ } (\bibinfo {year}
  {2023})}\BibitemShut {NoStop}%
\bibitem [{\citenamefont {Gungor}\ and\ \citenamefont
  {Hemmati}(2021)}]{Gungor2021}%
  \BibitemOpen
  \bibfield  {author} {\bibinfo {author} {\bibnamefont {Gungor}, \bibfnamefont
  {A.}}and\ \bibinfo {author} {\bibnamefont {Hemmati}, \bibfnamefont {A.}},\
  }\bibfield  {title} {\enquote {\bibinfo {title} {{The scaling and performance
  of side-by-side pitching hydrofoils}},}\ }\href
  {https://doi.org/10.1016/j.jfluidstructs.2021.103320} {\bibfield  {journal}
  {\bibinfo  {journal} {Journal of Fluids and Structures}\ }\textbf {\bibinfo
  {volume} {104}} (\bibinfo {year} {2021}),\
  10.1016/j.jfluidstructs.2021.103320}\BibitemShut {NoStop}%
\bibitem [{\citenamefont {Gungor}, \citenamefont {Khalid},\ and\ \citenamefont
  {Hemmati}(2022)}]{gungor2022classification}%
  \BibitemOpen
  \bibfield  {author} {\bibinfo {author} {\bibnamefont {Gungor}, \bibfnamefont
  {A.}}, \bibinfo {author} {\bibnamefont {Khalid}, \bibfnamefont {M.~S.~U.}},
  and\ \bibinfo {author} {\bibnamefont {Hemmati}, \bibfnamefont {A.}},\
  }\bibfield  {title} {\enquote {\bibinfo {title} {{Classification of vortex
  patterns of oscillating foils in side-by-side configurations}},}\ }\href@noop
  {} {\bibfield  {journal} {\bibinfo  {journal} {Journal of Fluid Mechanics}\
  }\textbf {\bibinfo {volume} {951}},\ \bibinfo {pages} {A37} (\bibinfo {year}
  {2022})}\BibitemShut {NoStop}%
\bibitem [{\citenamefont {Gupta}\ \emph {et~al.}(2021)\citenamefont {Gupta},
  \citenamefont {Thekkethil}, \citenamefont {Agrawal}, \citenamefont
  {Hourigan}, \citenamefont {Thompson},\ and\ \citenamefont
  {Sharma}}]{Gupta2021}%
  \BibitemOpen
  \bibfield  {author} {\bibinfo {author} {\bibnamefont {Gupta}, \bibfnamefont
  {S.}}, \bibinfo {author} {\bibnamefont {Thekkethil}, \bibfnamefont {N.}},
  \bibinfo {author} {\bibnamefont {Agrawal}, \bibfnamefont {A.}}, \bibinfo
  {author} {\bibnamefont {Hourigan}, \bibfnamefont {K.}}, \bibinfo {author}
  {\bibnamefont {Thompson}, \bibfnamefont {M.~C.}}, and\ \bibinfo {author}
  {\bibnamefont {Sharma}, \bibfnamefont {A.}},\ }\bibfield  {title} {\enquote
  {\bibinfo {title} {{Body-caudal fin fish-inspired self-propulsion study on
  burst-and-coast and continuous swimming of a hydrofoil model}},}\ }\href
  {https://doi.org/10.1063/5.0061417} {\bibfield  {journal} {\bibinfo
  {journal} {Physics of Fluids}\ }\textbf {\bibinfo {volume} {33}} (\bibinfo
  {year} {2021}),\ 10.1063/5.0061417}\BibitemShut {NoStop}%
\bibitem [{\citenamefont {Hornung}\ and\ \citenamefont
  {Kohn}(2002)}]{Hornung2002}%
  \BibitemOpen
  \bibfield  {author} {\bibinfo {author} {\bibnamefont {Hornung}, \bibfnamefont
  {R.~D.}}and\ \bibinfo {author} {\bibnamefont {Kohn}, \bibfnamefont {S.~R.}},\
  }\bibfield  {title} {\enquote {\bibinfo {title} {{Managing application
  complexity in the SAMRAI object-oriented framework}},}\ }\href
  {https://doi.org/10.1002/cpe.652} {\bibfield  {journal} {\bibinfo  {journal}
  {Concurrency and Computation: Practice and Experience}\ }\textbf {\bibinfo
  {volume} {14}} (\bibinfo {year} {2002}),\ 10.1002/cpe.652}\BibitemShut
  {NoStop}%
\bibitem [{\citenamefont {Hornung}, \citenamefont {Wissink},\ and\
  \citenamefont {Kohn}(2006)}]{Hornung2006}%
  \BibitemOpen
  \bibfield  {author} {\bibinfo {author} {\bibnamefont {Hornung}, \bibfnamefont
  {R.~D.}}, \bibinfo {author} {\bibnamefont {Wissink}, \bibfnamefont {A.~M.}},
  and\ \bibinfo {author} {\bibnamefont {Kohn}, \bibfnamefont {S.~R.}},\
  }\bibfield  {title} {\enquote {\bibinfo {title} {{Managing complex data and
  geometry in parallel structured AMR applications}},}\ }\href
  {https://doi.org/10.1007/s00366-006-0038-6} {\bibfield  {journal} {\bibinfo
  {journal} {Engineering with Computers}\ }\textbf {\bibinfo {volume} {22}}
  (\bibinfo {year} {2006}),\ 10.1007/s00366-006-0038-6}\BibitemShut {NoStop}%
\bibitem [{\citenamefont {Huera-Huarte}(2018)}]{Huera-Huarte2018}%
  \BibitemOpen
  \bibfield  {author} {\bibinfo {author} {\bibnamefont {Huera-Huarte},
  \bibfnamefont {F.~J.}},\ }\bibfield  {title} {\enquote {\bibinfo {title}
  {{Propulsive performance of a pair of pitching foils in staggered
  configurations}},}\ }\href
  {https://doi.org/10.1016/j.jfluidstructs.2018.04.024} {\bibfield  {journal}
  {\bibinfo  {journal} {Journal of Fluids and Structures}\ }\textbf {\bibinfo
  {volume} {81}} (\bibinfo {year} {2018}),\
  10.1016/j.jfluidstructs.2018.04.024}\BibitemShut {NoStop}%
\bibitem [{\citenamefont {Khalid}\ \emph {et~al.}(2021)\citenamefont {Khalid},
  \citenamefont {Wang}, \citenamefont {Akhtar}, \citenamefont {Dong},
  \citenamefont {Liu},\ and\ \citenamefont {Hemmati}}]{Khalid2021}%
  \BibitemOpen
  \bibfield  {author} {\bibinfo {author} {\bibnamefont {Khalid}, \bibfnamefont
  {M.~S.~U.}}, \bibinfo {author} {\bibnamefont {Wang}, \bibfnamefont {J.}},
  \bibinfo {author} {\bibnamefont {Akhtar}, \bibfnamefont {I.}}, \bibinfo
  {author} {\bibnamefont {Dong}, \bibfnamefont {H.}}, \bibinfo {author}
  {\bibnamefont {Liu}, \bibfnamefont {M.}}, and\ \bibinfo {author}
  {\bibnamefont {Hemmati}, \bibfnamefont {A.}},\ }\bibfield  {title} {\enquote
  {\bibinfo {title} {{Why do anguilliform swimmers perform undulation with
  wavelengths shorter than their bodylengths?}}}\ }\href
  {https://doi.org/10.1063/5.0040473} {\bibfield  {journal} {\bibinfo
  {journal} {Physics of Fluids}\ }\textbf {\bibinfo {volume} {33}} (\bibinfo
  {year} {2021}),\ 10.1063/5.0040473}\BibitemShut {NoStop}%
\bibitem [{\citenamefont {Khalid}\ \emph {et~al.}(2020)\citenamefont {Khalid},
  \citenamefont {Wang}, \citenamefont {Dong},\ and\ \citenamefont
  {Liu}}]{Khalid2020}%
  \BibitemOpen
  \bibfield  {author} {\bibinfo {author} {\bibnamefont {Khalid}, \bibfnamefont
  {M.~S.~U.}}, \bibinfo {author} {\bibnamefont {Wang}, \bibfnamefont {J.}},
  \bibinfo {author} {\bibnamefont {Dong}, \bibfnamefont {H.}}, and\ \bibinfo
  {author} {\bibnamefont {Liu}, \bibfnamefont {M.}},\ }\bibfield  {title}
  {\enquote {\bibinfo {title} {{Flow transitions and mapping for undulating
  swimmers}},}\ }\href {https://doi.org/10.1103/PhysRevFluids.5.063104}
  {\bibfield  {journal} {\bibinfo  {journal} {Physical Review Fluids}\ }\textbf
  {\bibinfo {volume} {5}} (\bibinfo {year} {2020}),\
  10.1103/PhysRevFluids.5.063104}\BibitemShut {NoStop}%
\bibitem [{\citenamefont {Kirk}\ \emph {et~al.}(2006)\citenamefont {Kirk},
  \citenamefont {Peterson}, \citenamefont {Stogner},\ and\ \citenamefont
  {Carey}}]{Kirk2006}%
  \BibitemOpen
  \bibfield  {author} {\bibinfo {author} {\bibnamefont {Kirk}, \bibfnamefont
  {B.~S.}}, \bibinfo {author} {\bibnamefont {Peterson}, \bibfnamefont {J.~W.}},
  \bibinfo {author} {\bibnamefont {Stogner}, \bibfnamefont {R.~H.}}, and\
  \bibinfo {author} {\bibnamefont {Carey}, \bibfnamefont {G.~F.}},\ }\bibfield
  {title} {\enquote {\bibinfo {title} {{libMesh : a C++ library for parallel
  adaptive mesh refinement/coarsening simulations}},}\ }\href
  {https://doi.org/10.1007/s00366-006-0049-3} {\bibfield  {journal} {\bibinfo
  {journal} {Engineering with Computers}\ }\textbf {\bibinfo {volume} {22}}
  (\bibinfo {year} {2006}),\ 10.1007/s00366-006-0049-3}\BibitemShut {NoStop}%
\bibitem [{\citenamefont {Lamb}(1932)}]{Hlamb1932hydrodynamics}%
  \BibitemOpen
  \bibfield  {author} {\bibinfo {author} {\bibnamefont {Lamb}, \bibfnamefont
  {H.}},\ }\href@noop {} {\emph {\bibinfo {title} {{Hydrodynamics}}}},\
  \bibinfo {edition} {6th}\ ed.\ (\bibinfo  {publisher} {Cambridge University
  Press},\ \bibinfo {address} {Cambridge},\ \bibinfo {year} {1932})\ p.\
  \bibinfo {pages} {182}\BibitemShut {NoStop}%
\bibitem [{\citenamefont {Lecheval}\ \emph {et~al.}(2018)\citenamefont
  {Lecheval}, \citenamefont {Jiang}, \citenamefont {Tichit}, \citenamefont
  {Sire}, \citenamefont {Hemelrijk},\ and\ \citenamefont
  {Theraulaz}}]{Lecheval2018}%
  \BibitemOpen
  \bibfield  {author} {\bibinfo {author} {\bibnamefont {Lecheval},
  \bibfnamefont {V.}}, \bibinfo {author} {\bibnamefont {Jiang}, \bibfnamefont
  {L.}}, \bibinfo {author} {\bibnamefont {Tichit}, \bibfnamefont {P.}},
  \bibinfo {author} {\bibnamefont {Sire}, \bibfnamefont {C.}}, \bibinfo
  {author} {\bibnamefont {Hemelrijk}, \bibfnamefont {C.~K.}}, and\ \bibinfo
  {author} {\bibnamefont {Theraulaz}, \bibfnamefont {G.}},\ }\bibfield  {title}
  {\enquote {\bibinfo {title} {{Social conformity and propagation of
  information in collective u-turns of fish schools}},}\ }\href
  {https://doi.org/10.1098/rspb.2018.0251} {\bibfield  {journal} {\bibinfo
  {journal} {Proceedings of the Royal Society B: Biological Sciences}\ }\textbf
  {\bibinfo {volume} {285}} (\bibinfo {year} {2018}),\
  10.1098/rspb.2018.0251}\BibitemShut {NoStop}%
\bibitem [{\citenamefont {Li}\ \emph {et~al.}(2021{\natexlab{a}})\citenamefont
  {Li}, \citenamefont {Liu}, \citenamefont {Deng}, \citenamefont {Lutz},\ and\
  \citenamefont {Xie}}]{Li2021a}%
  \BibitemOpen
  \bibfield  {author} {\bibinfo {author} {\bibnamefont {Li}, \bibfnamefont
  {L.}}, \bibinfo {author} {\bibnamefont {Liu}, \bibfnamefont {D.}}, \bibinfo
  {author} {\bibnamefont {Deng}, \bibfnamefont {J.}}, \bibinfo {author}
  {\bibnamefont {Lutz}, \bibfnamefont {M.~J.}}, and\ \bibinfo {author}
  {\bibnamefont {Xie}, \bibfnamefont {G.}},\ }\bibfield  {title} {\enquote
  {\bibinfo {title} {{Fish can save energy via proprioceptive sensing}},}\
  }\href {https://doi.org/10.1088/1748-3190/ac165e} {\bibfield  {journal}
  {\bibinfo  {journal} {Bioinspiration and Biomimetics}\ }\textbf {\bibinfo
  {volume} {16}} (\bibinfo {year} {2021}{\natexlab{a}}),\
  10.1088/1748-3190/ac165e}\BibitemShut {NoStop}%
\bibitem [{\citenamefont {Li}\ \emph {et~al.}(2020)\citenamefont {Li},
  \citenamefont {Nagy}, \citenamefont {Graving}, \citenamefont {Bak-Coleman},
  \citenamefont {Xie},\ and\ \citenamefont {Couzin}}]{Li2020}%
  \BibitemOpen
  \bibfield  {author} {\bibinfo {author} {\bibnamefont {Li}, \bibfnamefont
  {L.}}, \bibinfo {author} {\bibnamefont {Nagy}, \bibfnamefont {M.}}, \bibinfo
  {author} {\bibnamefont {Graving}, \bibfnamefont {J.~M.}}, \bibinfo {author}
  {\bibnamefont {Bak-Coleman}, \bibfnamefont {J.}}, \bibinfo {author}
  {\bibnamefont {Xie}, \bibfnamefont {G.}}, and\ \bibinfo {author}
  {\bibnamefont {Couzin}, \bibfnamefont {I.~D.}},\ }\bibfield  {title}
  {\enquote {\bibinfo {title} {{Vortex phase matching as a strategy for
  schooling in robots and in fish}},}\ }\href
  {https://doi.org/10.1038/s41467-020-19086-0} {\bibfield  {journal} {\bibinfo
  {journal} {Nature Communications}\ } (\bibinfo {year} {2020}),\
  10.1038/s41467-020-19086-0}\BibitemShut {NoStop}%
\bibitem [{\citenamefont {Li}\ \emph {et~al.}(2021{\natexlab{b}})\citenamefont
  {Li}, \citenamefont {Ravi}, \citenamefont {Xie},\ and\ \citenamefont
  {Couzin}}]{Li2021}%
  \BibitemOpen
  \bibfield  {author} {\bibinfo {author} {\bibnamefont {Li}, \bibfnamefont
  {L.}}, \bibinfo {author} {\bibnamefont {Ravi}, \bibfnamefont {S.}}, \bibinfo
  {author} {\bibnamefont {Xie}, \bibfnamefont {G.}}, and\ \bibinfo {author}
  {\bibnamefont {Couzin}, \bibfnamefont {I.~D.}},\ }\bibfield  {title}
  {\enquote {\bibinfo {title} {{Using a robotic platform to study the influence
  of relative tailbeat phase on the energetic costs of side-by-side swimming in
  fish}},}\ }\href {https://doi.org/10.1098/rspa.2020.0810} {\bibfield
  {journal} {\bibinfo  {journal} {Proceedings of the Royal Society A:
  Mathematical, Physical and Engineering Sciences}\ }\textbf {\bibinfo {volume}
  {477}} (\bibinfo {year} {2021}{\natexlab{b}}),\
  10.1098/rspa.2020.0810}\BibitemShut {NoStop}%
\bibitem [{\citenamefont {Li}\ \emph {et~al.}(2023)\citenamefont {Li},
  \citenamefont {Wang}, \citenamefont {Li}, \citenamefont {Ye}, \citenamefont
  {Zhang}, \citenamefont {Yin}, \citenamefont {Jia}, \citenamefont {Hou},
  \citenamefont {Wang}, \citenamefont {Ding},\ and\ \citenamefont
  {Others}}]{li2023electron}%
  \BibitemOpen
  \bibfield  {author} {\bibinfo {author} {\bibnamefont {Li}, \bibfnamefont
  {Y.}}, \bibinfo {author} {\bibnamefont {Wang}, \bibfnamefont {H.}}, \bibinfo
  {author} {\bibnamefont {Li}, \bibfnamefont {Y.}}, \bibinfo {author}
  {\bibnamefont {Ye}, \bibfnamefont {H.}}, \bibinfo {author} {\bibnamefont
  {Zhang}, \bibfnamefont {Y.}}, \bibinfo {author} {\bibnamefont {Yin},
  \bibfnamefont {R.}}, \bibinfo {author} {\bibnamefont {Jia}, \bibfnamefont
  {H.}}, \bibinfo {author} {\bibnamefont {Hou}, \bibfnamefont {B.}}, \bibinfo
  {author} {\bibnamefont {Wang}, \bibfnamefont {C.}}, \bibinfo {author}
  {\bibnamefont {Ding}, \bibfnamefont {H.}}, and\ \bibinfo {author}
  {\bibnamefont {Others},},\ }\bibfield  {title} {\enquote {\bibinfo {title}
  {{Electron transfer rules of minerals under pressure informed by machine
  learning}},}\ }\href@noop {} {\bibfield  {journal} {\bibinfo  {journal}
  {Nature Communications}\ }\textbf {\bibinfo {volume} {14}},\ \bibinfo {pages}
  {1815} (\bibinfo {year} {2023})}\BibitemShut {NoStop}%
\bibitem [{\citenamefont {Lin}\ \emph {et~al.}(2023)\citenamefont {Lin},
  \citenamefont {Bhalla}, \citenamefont {Griffith}, \citenamefont {Sheng},
  \citenamefont {Li}, \citenamefont {Liang},\ and\ \citenamefont
  {Zhang}}]{lin2022swimming}%
  \BibitemOpen
  \bibfield  {author} {\bibinfo {author} {\bibnamefont {Lin}, \bibfnamefont
  {Z.}}, \bibinfo {author} {\bibnamefont {Bhalla}, \bibfnamefont {A.~P.~S.}},
  \bibinfo {author} {\bibnamefont {Griffith}, \bibfnamefont {B.~E.}}, \bibinfo
  {author} {\bibnamefont {Sheng}, \bibfnamefont {Z.}}, \bibinfo {author}
  {\bibnamefont {Li}, \bibfnamefont {H.}}, \bibinfo {author} {\bibnamefont
  {Liang}, \bibfnamefont {D.}}, and\ \bibinfo {author} {\bibnamefont {Zhang},
  \bibfnamefont {Y.}},\ }\bibfield  {title} {\enquote {\bibinfo {title} {{How
  swimming style and schooling affect the hydrodynamics of two accelerating
  wavy hydrofoils}},}\ }\href
  {https://doi.org/https://doi.org/10.1016/j.oceaneng.2022.113314} {\bibfield
  {journal} {\bibinfo  {journal} {Ocean Engineering}\ }\textbf {\bibinfo
  {volume} {268}},\ \bibinfo {pages} {113314} (\bibinfo {year}
  {2023})}\BibitemShut {NoStop}%
\bibitem [{\citenamefont {Lin}, \citenamefont {Liang},\ and\ \citenamefont
  {Zhao}(2016)}]{Lin2016a}%
  \BibitemOpen
  \bibfield  {author} {\bibinfo {author} {\bibnamefont {Lin}, \bibfnamefont
  {Z.}}, \bibinfo {author} {\bibnamefont {Liang}, \bibfnamefont {D.}}, and\
  \bibinfo {author} {\bibnamefont {Zhao}, \bibfnamefont {M.}},\ }\bibfield
  {title} {\enquote {\bibinfo {title} {{Numerical Study of the Interaction
  Between Two Immersed Cylinders}},}\ }in\ \href@noop {} {\emph {\bibinfo
  {booktitle} {The 12th International Conference on Hydrodynamics}}}\ (\bibinfo
  {year} {2016})\ p.~\bibinfo {pages} {55}\BibitemShut {NoStop}%
\bibitem [{\citenamefont {Lin}, \citenamefont {Liang},\ and\ \citenamefont
  {Zhao}(2017)}]{Lin2017a}%
  \BibitemOpen
  \bibfield  {author} {\bibinfo {author} {\bibnamefont {Lin}, \bibfnamefont
  {Z.}}, \bibinfo {author} {\bibnamefont {Liang}, \bibfnamefont {D.}}, and\
  \bibinfo {author} {\bibnamefont {Zhao}, \bibfnamefont {M.}},\ }\bibfield
  {title} {\enquote {\bibinfo {title} {{Interaction Between Two Vibrating
  Cylinders Immersed in Fluid}},}\ }in\ \href@noop {} {\emph {\bibinfo
  {booktitle} {The 27th International Ocean and Polar Engineering
  Conference}}}\ (\bibinfo  {publisher} {International Society of Offshore and
  Polar Engineers},\ \bibinfo {address} {San Francisco, California, USA},\
  \bibinfo {year} {2017})\ p.~\bibinfo {pages} {8}\BibitemShut {NoStop}%
\bibitem [{\citenamefont {Lin}, \citenamefont {Liang},\ and\ \citenamefont
  {Zhao}(2018{\natexlab{a}})}]{Lin2018b}%
  \BibitemOpen
  \bibfield  {author} {\bibinfo {author} {\bibnamefont {Lin}, \bibfnamefont
  {Z.}}, \bibinfo {author} {\bibnamefont {Liang}, \bibfnamefont {D.}}, and\
  \bibinfo {author} {\bibnamefont {Zhao}, \bibfnamefont {M.}},\ }\bibfield
  {title} {\enquote {\bibinfo {title} {{Effects of Damping on Flow-Mediated
  Interaction Between Two Cylinders}},}\ }\href
  {https://doi.org/10.1115/1.4039712} {\bibfield  {journal} {\bibinfo
  {journal} {Journal of Fluids Engineering - ASME}\ } (\bibinfo {year}
  {2018}{\natexlab{a}}),\ 10.1115/1.4039712}\BibitemShut {NoStop}%
\bibitem [{\citenamefont {Lin}, \citenamefont {Liang},\ and\ \citenamefont
  {Zhao}(2018{\natexlab{b}})}]{Lin2018c}%
  \BibitemOpen
  \bibfield  {author} {\bibinfo {author} {\bibnamefont {Lin}, \bibfnamefont
  {Z.}}, \bibinfo {author} {\bibnamefont {Liang}, \bibfnamefont {D.}}, and\
  \bibinfo {author} {\bibnamefont {Zhao}, \bibfnamefont {M.}},\ }\bibfield
  {title} {\enquote {\bibinfo {title} {{Flow-Mediated Interaction Between a
  Vibrating Cylinder and an Elastically-Mounted Cylinder}},}\ }\href
  {https://doi.org/10.1016/j.oceaneng.2018.04.019} {\bibfield  {journal}
  {\bibinfo  {journal} {Ocean Engineering}\ } (\bibinfo {year}
  {2018}{\natexlab{b}}),\ 10.1016/j.oceaneng.2018.04.019}\BibitemShut {NoStop}%
\bibitem [{\citenamefont {Lin}, \citenamefont {Liang},\ and\ \citenamefont
  {Zhao}(2019)}]{Lin2019}%
  \BibitemOpen
  \bibfield  {author} {\bibinfo {author} {\bibnamefont {Lin}, \bibfnamefont
  {Z.}}, \bibinfo {author} {\bibnamefont {Liang}, \bibfnamefont {D.}}, and\
  \bibinfo {author} {\bibnamefont {Zhao}, \bibfnamefont {M.}},\ }\bibfield
  {title} {\enquote {\bibinfo {title} {{Effects of Reynolds Number on
  Flow-Mediated Interaction between Two Cylinders}},}\ }\href
  {https://doi.org/10.1061/(ASCE)EM.1943-7889.0001670} {\bibfield  {journal}
  {\bibinfo  {journal} {Journal of Engineering Mechanics - ASCE}\ }\textbf
  {\bibinfo {volume} {145}} (\bibinfo {year} {2019}),\
  10.1061/(ASCE)EM.1943-7889.0001670}\BibitemShut {NoStop}%
\bibitem [{\citenamefont {Lin}, \citenamefont {Liang},\ and\ \citenamefont
  {Zhao}(2022)}]{lin2022flow}%
  \BibitemOpen
  \bibfield  {author} {\bibinfo {author} {\bibnamefont {Lin}, \bibfnamefont
  {Z.}}, \bibinfo {author} {\bibnamefont {Liang}, \bibfnamefont {D.}}, and\
  \bibinfo {author} {\bibnamefont {Zhao}, \bibfnamefont {M.}},\ }\bibfield
  {title} {\enquote {\bibinfo {title} {{Flow-Mediated Interaction Between a
  Forced-Oscillating Cylinder and an Elastically-Mounted Cylinder in Less
  Regular Regimes}},}\ }\href@noop {} {\bibfield  {journal} {\bibinfo
  {journal} {Physics of Fluids}\ } (\bibinfo {year} {2022})}\BibitemShut
  {NoStop}%
\bibitem [{\citenamefont {Lindsey}(1978)}]{Lindsey1978}%
  \BibitemOpen
  \bibfield  {author} {\bibinfo {author} {\bibnamefont {Lindsey}, \bibfnamefont
  {C.~C.}},\ }\bibfield  {title} {\enquote {\bibinfo {title} {{Form, function,
  and locomotory habits in fish}},}\ }in\ \href
  {https://doi.org/10.1016/S1546-5098(08)60163-6} {\emph {\bibinfo {booktitle}
  {Fish Physiology}}},\ Vol.~\bibinfo {volume} {7}\ (\bibinfo {year}
  {1978})\BibitemShut {NoStop}%
\bibitem [{\citenamefont {Ma}, \citenamefont {Huang},\ and\ \citenamefont
  {Xu}(2019)}]{Ma2019}%
  \BibitemOpen
  \bibfield  {author} {\bibinfo {author} {\bibnamefont {Ma}, \bibfnamefont
  {M.}}, \bibinfo {author} {\bibnamefont {Huang}, \bibfnamefont {W.~X.}}, and\
  \bibinfo {author} {\bibnamefont {Xu}, \bibfnamefont {C.~X.}},\ }\bibfield
  {title} {\enquote {\bibinfo {title} {{A dynamic wall model for large eddy
  simulation of turbulent flow over complex/moving boundaries based on the
  immersed boundary method}},}\ }\href {https://doi.org/10.1063/1.5126853}
  {\bibfield  {journal} {\bibinfo  {journal} {Physics of Fluids}\ }\textbf
  {\bibinfo {volume} {31}} (\bibinfo {year} {2019}),\
  10.1063/1.5126853}\BibitemShut {NoStop}%
\bibitem [{\citenamefont {Maertens}, \citenamefont {Gao},\ and\ \citenamefont
  {Triantafyllou}(2017)}]{Maertens2017}%
  \BibitemOpen
  \bibfield  {author} {\bibinfo {author} {\bibnamefont {Maertens},
  \bibfnamefont {A.~P.}}, \bibinfo {author} {\bibnamefont {Gao}, \bibfnamefont
  {A.}}, and\ \bibinfo {author} {\bibnamefont {Triantafyllou}, \bibfnamefont
  {M.~S.}},\ }\bibfield  {title} {\enquote {\bibinfo {title} {{Optimal
  undulatory swimming for a single fish-like body and for a pair of
  interacting{\^{A}} swimmers}},}\ }\href
  {https://doi.org/10.1017/jfm.2016.845} {\bibfield  {journal} {\bibinfo
  {journal} {Journal of Fluid Mechanics}\ }\textbf {\bibinfo {volume} {813}}
  (\bibinfo {year} {2017}),\ 10.1017/jfm.2016.845}\BibitemShut {NoStop}%
\bibitem [{\citenamefont {Maertens}, \citenamefont {Triantafyllou},\ and\
  \citenamefont {Yue}(2015)}]{Maertens2015}%
  \BibitemOpen
  \bibfield  {author} {\bibinfo {author} {\bibnamefont {Maertens},
  \bibfnamefont {A.~P.}}, \bibinfo {author} {\bibnamefont {Triantafyllou},
  \bibfnamefont {M.~S.}}, and\ \bibinfo {author} {\bibnamefont {Yue},
  \bibfnamefont {D.~K.}},\ }\bibfield  {title} {\enquote {\bibinfo {title}
  {{Efficiency of fish propulsion}},}\ }\href
  {https://doi.org/10.1088/1748-3190/10/4/046013} {\bibfield  {journal}
  {\bibinfo  {journal} {Bioinspiration and Biomimetics}\ }\textbf {\bibinfo
  {volume} {10}} (\bibinfo {year} {2015}),\
  10.1088/1748-3190/10/4/046013}\BibitemShut {NoStop}%
\bibitem [{\citenamefont {Matchev}, \citenamefont {Matcheva},\ and\
  \citenamefont {Roman}(2022)}]{matchev2022analytical}%
  \BibitemOpen
  \bibfield  {author} {\bibinfo {author} {\bibnamefont {Matchev}, \bibfnamefont
  {K.~T.}}, \bibinfo {author} {\bibnamefont {Matcheva}, \bibfnamefont {K.}},
  and\ \bibinfo {author} {\bibnamefont {Roman}, \bibfnamefont {A.}},\
  }\bibfield  {title} {\enquote {\bibinfo {title} {{Analytical Modeling of
  Exoplanet Transit Spectroscopy with Dimensional Analysis and Symbolic
  Regression}},}\ }\href@noop {} {\bibfield  {journal} {\bibinfo  {journal}
  {The Astrophysical Journal}\ }\textbf {\bibinfo {volume} {930}},\ \bibinfo
  {pages} {33} (\bibinfo {year} {2022})}\BibitemShut {NoStop}%
\bibitem [{\citenamefont {Moriche}, \citenamefont {Flores},\ and\ \citenamefont
  {Garc{\'{i}}a-Villalba}(2016)}]{Moriche2016}%
  \BibitemOpen
  \bibfield  {author} {\bibinfo {author} {\bibnamefont {Moriche}, \bibfnamefont
  {M.}}, \bibinfo {author} {\bibnamefont {Flores}, \bibfnamefont {O.}}, and\
  \bibinfo {author} {\bibnamefont {Garc{\'{i}}a-Villalba}, \bibfnamefont
  {M.}},\ }\bibfield  {title} {\enquote {\bibinfo {title} {{Three-dimensional
  instabilities in the wake of a flapping wing at low Reynolds number}},}\
  }\href {https://doi.org/10.1016/j.ijheatfluidflow.2016.06.015} {\bibfield
  {journal} {\bibinfo  {journal} {International Journal of Heat and Fluid
  Flow}\ }\textbf {\bibinfo {volume} {62}} (\bibinfo {year} {2016}),\
  10.1016/j.ijheatfluidflow.2016.06.015}\BibitemShut {NoStop}%
\bibitem [{\citenamefont {Nair}\ and\ \citenamefont {Kanso}(2007)}]{NAIR2007}%
  \BibitemOpen
  \bibfield  {author} {\bibinfo {author} {\bibnamefont {Nair}, \bibfnamefont
  {S.}}and\ \bibinfo {author} {\bibnamefont {Kanso}, \bibfnamefont {E.}},\
  }\bibfield  {title} {\enquote {\bibinfo {title} {{Hydrodynamically Coupled
  Rigid Bodies}},}\ }\href {https://doi.org/10.1017/S002211200700849X}
  {\bibfield  {journal} {\bibinfo  {journal} {Journal of Fluid Mechanics}\
  }\textbf {\bibinfo {volume} {592}},\ \bibinfo {pages} {393--411} (\bibinfo
  {year} {2007})}\BibitemShut {NoStop}%
\bibitem [{\citenamefont {Nangia}\ \emph
  {et~al.}(2017{\natexlab{a}})\citenamefont {Nangia}, \citenamefont {Bale},
  \citenamefont {Chen}, \citenamefont {Hanna},\ and\ \citenamefont
  {Patankar}}]{Nangia2017a}%
  \BibitemOpen
  \bibfield  {author} {\bibinfo {author} {\bibnamefont {Nangia}, \bibfnamefont
  {N.}}, \bibinfo {author} {\bibnamefont {Bale}, \bibfnamefont {R.}}, \bibinfo
  {author} {\bibnamefont {Chen}, \bibfnamefont {N.}}, \bibinfo {author}
  {\bibnamefont {Hanna}, \bibfnamefont {Y.}}, and\ \bibinfo {author}
  {\bibnamefont {Patankar}, \bibfnamefont {N.~A.}},\ }\bibfield  {title}
  {\enquote {\bibinfo {title} {{Optimal specific wavelength for maximum thrust
  production in undulatory propulsion}},}\ }\href
  {https://doi.org/10.1371/journal.pone.0179727} {\bibfield  {journal}
  {\bibinfo  {journal} {PLoS ONE}\ }\textbf {\bibinfo {volume} {12}} (\bibinfo
  {year} {2017}{\natexlab{a}}),\ 10.1371/journal.pone.0179727}\BibitemShut
  {NoStop}%
\bibitem [{\citenamefont {Nangia}\ \emph
  {et~al.}(2017{\natexlab{b}})\citenamefont {Nangia}, \citenamefont {Johansen},
  \citenamefont {Patankar},\ and\ \citenamefont {Bhalla}}]{Nangia2017}%
  \BibitemOpen
  \bibfield  {author} {\bibinfo {author} {\bibnamefont {Nangia}, \bibfnamefont
  {N.}}, \bibinfo {author} {\bibnamefont {Johansen}, \bibfnamefont {H.}},
  \bibinfo {author} {\bibnamefont {Patankar}, \bibfnamefont {N.~A.}}, and\
  \bibinfo {author} {\bibnamefont {Bhalla}, \bibfnamefont {A.~P.~S.}},\
  }\bibfield  {title} {\enquote {\bibinfo {title} {{A moving control volume
  approach to computing hydrodynamic forces and torques on immersed bodies}},}\
  }\href {https://doi.org/10.1016/j.jcp.2017.06.047} {\bibfield  {journal}
  {\bibinfo  {journal} {Journal of Computational Physics}\ }\textbf {\bibinfo
  {volume} {347}} (\bibinfo {year} {2017}{\natexlab{b}}),\
  10.1016/j.jcp.2017.06.047}\BibitemShut {NoStop}%
\bibitem [{\citenamefont {Nangia}, \citenamefont {Patankar},\ and\
  \citenamefont {Bhalla}(2019)}]{Nangia2019}%
  \BibitemOpen
  \bibfield  {author} {\bibinfo {author} {\bibnamefont {Nangia}, \bibfnamefont
  {N.}}, \bibinfo {author} {\bibnamefont {Patankar}, \bibfnamefont {N.~A.}},
  and\ \bibinfo {author} {\bibnamefont {Bhalla}, \bibfnamefont {A.~P.~S.}},\
  }\bibfield  {title} {\enquote {\bibinfo {title} {{A DLM immersed boundary
  method based wave-structure interaction solver for high density ratio
  multiphase flows}},}\ }\href {https://doi.org/10.1016/j.jcp.2019.07.004}
  {\bibfield  {journal} {\bibinfo  {journal} {Journal of Computational
  Physics}\ }\textbf {\bibinfo {volume} {398}} (\bibinfo {year} {2019}),\
  10.1016/j.jcp.2019.07.004}\BibitemShut {NoStop}%
\bibitem [{\citenamefont {Ni}, \citenamefont {Huang},\ and\ \citenamefont
  {Xu}(2023)}]{ni2023mode}%
  \BibitemOpen
  \bibfield  {author} {\bibinfo {author} {\bibnamefont {Ni}, \bibfnamefont
  {J.-Y.}}, \bibinfo {author} {\bibnamefont {Huang}, \bibfnamefont {W.-X.}},
  and\ \bibinfo {author} {\bibnamefont {Xu}, \bibfnamefont {C.-X.}},\
  }\bibfield  {title} {\enquote {\bibinfo {title} {{Mode transition of a
  coupled rigid--flexible system in a uniform flow}},}\ }\href@noop {}
  {\bibfield  {journal} {\bibinfo  {journal} {Physics of Fluids}\ }\textbf
  {\bibinfo {volume} {35}} (\bibinfo {year} {2023})}\BibitemShut {NoStop}%
\bibitem [{\citenamefont {Pan}\ and\ \citenamefont
  {Dong}(2020)}]{pan2020computational}%
  \BibitemOpen
  \bibfield  {author} {\bibinfo {author} {\bibnamefont {Pan}, \bibfnamefont
  {Y.}}and\ \bibinfo {author} {\bibnamefont {Dong}, \bibfnamefont {H.}},\
  }\bibfield  {title} {\enquote {\bibinfo {title} {{Computational analysis of
  hydrodynamic interactions in a high-density fish school}},}\ }\href@noop {}
  {\bibfield  {journal} {\bibinfo  {journal} {Physics of Fluids}\ }\textbf
  {\bibinfo {volume} {32}},\ \bibinfo {pages} {121901} (\bibinfo {year}
  {2020})}\BibitemShut {NoStop}%
\bibitem [{\citenamefont {Pan}\ and\ \citenamefont
  {Dong}(2022)}]{pan2022effects}%
  \BibitemOpen
  \bibfield  {author} {\bibinfo {author} {\bibnamefont {Pan}, \bibfnamefont
  {Y.}}and\ \bibinfo {author} {\bibnamefont {Dong}, \bibfnamefont {H.}},\
  }\bibfield  {title} {\enquote {\bibinfo {title} {{Effects of phase difference
  on hydrodynamic interactions and wake patterns in high-density fish
  schools}},}\ }\href@noop {} {\bibfield  {journal} {\bibinfo  {journal}
  {Physics of Fluids}\ }\textbf {\bibinfo {volume} {34}},\ \bibinfo {pages}
  {111902} (\bibinfo {year} {2022})}\BibitemShut {NoStop}%
\bibitem [{\citenamefont {Partridge}(1981)}]{Partridge1981}%
  \BibitemOpen
  \bibfield  {author} {\bibinfo {author} {\bibnamefont {Partridge},
  \bibfnamefont {B.~L.}},\ }\bibfield  {title} {\enquote {\bibinfo {title}
  {{Internal dynamics and the interrelations of fish in schools}},}\ }\href
  {https://doi.org/10.1007/BF00612563} {\bibfield  {journal} {\bibinfo
  {journal} {Journal of Comparative Physiology A}\ }\textbf {\bibinfo {volume}
  {144}} (\bibinfo {year} {1981}),\ 10.1007/BF00612563}\BibitemShut {NoStop}%
\bibitem [{\citenamefont {Rosenthal}\ \emph {et~al.}(2015)\citenamefont
  {Rosenthal}, \citenamefont {Twomey}, \citenamefont {Hartnett}, \citenamefont
  {Wu},\ and\ \citenamefont {Couzin}}]{Rosenthal2015}%
  \BibitemOpen
  \bibfield  {author} {\bibinfo {author} {\bibnamefont {Rosenthal},
  \bibfnamefont {S.~B.}}, \bibinfo {author} {\bibnamefont {Twomey},
  \bibfnamefont {C.~R.}}, \bibinfo {author} {\bibnamefont {Hartnett},
  \bibfnamefont {A.~T.}}, \bibinfo {author} {\bibnamefont {Wu}, \bibfnamefont
  {H.~S.}}, and\ \bibinfo {author} {\bibnamefont {Couzin}, \bibfnamefont
  {I.~D.}},\ }\bibfield  {title} {\enquote {\bibinfo {title} {{Revealing the
  hidden networks of interaction in mobile animal groups allows prediction of
  complex behavioral contagion}},}\ }\href
  {https://doi.org/10.1073/pnas.1420068112} {\bibfield  {journal} {\bibinfo
  {journal} {Proceedings of the National Academy of Sciences of the United
  States of America}\ }\textbf {\bibinfo {volume} {112}} (\bibinfo {year}
  {2015}),\ 10.1073/pnas.1420068112}\BibitemShut {NoStop}%
\bibitem [{\citenamefont {Santo}\ \emph {et~al.}(2021)\citenamefont {Santo},
  \citenamefont {Goerig}, \citenamefont {Wainwright}, \citenamefont {Akanyeti},
  \citenamefont {Liao}, \citenamefont {Castro-Santos},\ and\ \citenamefont
  {Lauder}}]{Santo2021}%
  \BibitemOpen
  \bibfield  {author} {\bibinfo {author} {\bibnamefont {Santo}, \bibfnamefont
  {V.~D.}}, \bibinfo {author} {\bibnamefont {Goerig}, \bibfnamefont {E.}},
  \bibinfo {author} {\bibnamefont {Wainwright}, \bibfnamefont {D.~K.}},
  \bibinfo {author} {\bibnamefont {Akanyeti}, \bibfnamefont {O.}}, \bibinfo
  {author} {\bibnamefont {Liao}, \bibfnamefont {J.~C.}}, \bibinfo {author}
  {\bibnamefont {Castro-Santos}, \bibfnamefont {T.}}, and\ \bibinfo {author}
  {\bibnamefont {Lauder}, \bibfnamefont {G.~V.}},\ }\bibfield  {title}
  {\enquote {\bibinfo {title} {{Convergence of undulatory swimming kinematics
  across a diversity of fishes}},}\ }\href
  {https://doi.org/10.1073/pnas.2113206118} {\bibfield  {journal} {\bibinfo
  {journal} {Proceedings of the National Academy of Sciences of the United
  States of America}\ }\textbf {\bibinfo {volume} {118}} (\bibinfo {year}
  {2021}),\ 10.1073/pnas.2113206118}\BibitemShut {NoStop}%
\bibitem [{\citenamefont {Schwalbe}\ \emph {et~al.}(2019)\citenamefont
  {Schwalbe}, \citenamefont {Boden}, \citenamefont {Wise},\ and\ \citenamefont
  {Tytell}}]{Schwalbe2019}%
  \BibitemOpen
  \bibfield  {author} {\bibinfo {author} {\bibnamefont {Schwalbe},
  \bibfnamefont {M.~A.}}, \bibinfo {author} {\bibnamefont {Boden},
  \bibfnamefont {A.~L.}}, \bibinfo {author} {\bibnamefont {Wise}, \bibfnamefont
  {T.~N.}}, and\ \bibinfo {author} {\bibnamefont {Tytell}, \bibfnamefont
  {E.~D.}},\ }\bibfield  {title} {\enquote {\bibinfo {title} {{Red muscle
  activity in bluegill sunfish Lepomis macrochirus during forward
  accelerations}},}\ }\href {https://doi.org/10.1038/s41598-019-44409-7}
  {\bibfield  {journal} {\bibinfo  {journal} {Scientific Reports}\ }\textbf
  {\bibinfo {volume} {9}} (\bibinfo {year} {2019}),\
  10.1038/s41598-019-44409-7}\BibitemShut {NoStop}%
\bibitem [{\citenamefont {Sfakiotakis}, \citenamefont {Lane},\ and\
  \citenamefont {Davies}(1999)}]{Sfakiotakis1999}%
  \BibitemOpen
  \bibfield  {author} {\bibinfo {author} {\bibnamefont {Sfakiotakis},
  \bibfnamefont {M.}}, \bibinfo {author} {\bibnamefont {Lane}, \bibfnamefont
  {D.~M.}}, and\ \bibinfo {author} {\bibnamefont {Davies}, \bibfnamefont
  {J.~B.~C.}},\ }\bibfield  {title} {\enquote {\bibinfo {title} {{Review of
  fish swimming modes for aquatic locomotion}},}\ }\href
  {https://doi.org/10.1109/48.757275} {\bibfield  {journal} {\bibinfo
  {journal} {IEEE Journal of Oceanic Engineering}\ }\textbf {\bibinfo {volume}
  {24}} (\bibinfo {year} {1999}),\ 10.1109/48.757275}\BibitemShut {NoStop}%
\bibitem [{\citenamefont {Shao}\ \emph {et~al.}(2010)\citenamefont {Shao},
  \citenamefont {Pan}, \citenamefont {Deng},\ and\ \citenamefont
  {Yu}}]{Shao2010}%
  \BibitemOpen
  \bibfield  {author} {\bibinfo {author} {\bibnamefont {Shao}, \bibfnamefont
  {X.}}, \bibinfo {author} {\bibnamefont {Pan}, \bibfnamefont {D.}}, \bibinfo
  {author} {\bibnamefont {Deng}, \bibfnamefont {J.}}, and\ \bibinfo {author}
  {\bibnamefont {Yu}, \bibfnamefont {Z.}},\ }\bibfield  {title} {\enquote
  {\bibinfo {title} {{Hydrodynamic performance of a fishlike undulating foil in
  the wake of a cylinder}},}\ }\href {https://doi.org/10.1063/1.3504651}
  {\bibfield  {journal} {\bibinfo  {journal} {Physics of Fluids}\ }\textbf
  {\bibinfo {volume} {22}} (\bibinfo {year} {2010}),\
  10.1063/1.3504651}\BibitemShut {NoStop}%
\bibitem [{\citenamefont {Shrivastava}\ \emph {et~al.}(2017)\citenamefont
  {Shrivastava}, \citenamefont {Malushte}, \citenamefont {Agrawal},\ and\
  \citenamefont {Sharma}}]{Shrivastava2017}%
  \BibitemOpen
  \bibfield  {author} {\bibinfo {author} {\bibnamefont {Shrivastava},
  \bibfnamefont {M.}}, \bibinfo {author} {\bibnamefont {Malushte},
  \bibfnamefont {M.}}, \bibinfo {author} {\bibnamefont {Agrawal}, \bibfnamefont
  {A.}}, and\ \bibinfo {author} {\bibnamefont {Sharma}, \bibfnamefont {A.}},\
  }\bibfield  {title} {\enquote {\bibinfo {title} {{CFD study on hydrodynamics
  of three fish-like undulating hydrofoils in side-by-side arrangement}},}\
  }\href@noop {} {\bibfield  {journal} {\bibinfo  {journal} {Lecture Notes in
  Mechanical Engineering}\ } (\bibinfo {year} {2017})}\BibitemShut {NoStop}%
\bibitem [{\citenamefont {Taylor}, \citenamefont {Nudds},\ and\ \citenamefont
  {Thomas}(2003)}]{Taylor2003}%
  \BibitemOpen
  \bibfield  {author} {\bibinfo {author} {\bibnamefont {Taylor}, \bibfnamefont
  {G.~K.}}, \bibinfo {author} {\bibnamefont {Nudds}, \bibfnamefont {R.~L.}},
  and\ \bibinfo {author} {\bibnamefont {Thomas}, \bibfnamefont {A.~L.}},\
  }\bibfield  {title} {\enquote {\bibinfo {title} {{Flying and swimming animals
  cruise at a Strouhal number tuned for high power efficiency}},}\ }\href
  {https://doi.org/10.1038/nature02000} {\bibfield  {journal} {\bibinfo
  {journal} {Nature}\ }\textbf {\bibinfo {volume} {425}} (\bibinfo {year}
  {2003}),\ 10.1038/nature02000}\BibitemShut {NoStop}%
\bibitem [{\citenamefont {Thekkethil}, \citenamefont {Sharma},\ and\
  \citenamefont {Agrawal}(2018)}]{Thekkethil2018}%
  \BibitemOpen
  \bibfield  {author} {\bibinfo {author} {\bibnamefont {Thekkethil},
  \bibfnamefont {N.}}, \bibinfo {author} {\bibnamefont {Sharma}, \bibfnamefont
  {A.}}, and\ \bibinfo {author} {\bibnamefont {Agrawal}, \bibfnamefont {A.}},\
  }\bibfield  {title} {\enquote {\bibinfo {title} {{Unified hydrodynamics study
  for various types of fishes-like undulating rigid hydrofoil in a free stream
  flow}},}\ }\href {https://doi.org/10.1063/1.5041358} {\bibfield  {journal}
  {\bibinfo  {journal} {Physics of Fluids}\ }\textbf {\bibinfo {volume} {30}}
  (\bibinfo {year} {2018}),\ 10.1063/1.5041358}\BibitemShut {NoStop}%
\bibitem [{\citenamefont {Thekkethil}, \citenamefont {Sharma},\ and\
  \citenamefont {Agrawal}(2020)}]{Thekkethil2020}%
  \BibitemOpen
  \bibfield  {author} {\bibinfo {author} {\bibnamefont {Thekkethil},
  \bibfnamefont {N.}}, \bibinfo {author} {\bibnamefont {Sharma}, \bibfnamefont
  {A.}}, and\ \bibinfo {author} {\bibnamefont {Agrawal}, \bibfnamefont {A.}},\
  }\bibfield  {title} {\enquote {\bibinfo {title} {{Self-propulsion of
  fishes-like undulating hydrofoil: A unified kinematics based unsteady
  hydrodynamics study}},}\ }\href
  {https://doi.org/10.1016/j.jfluidstructs.2020.102875} {\bibfield  {journal}
  {\bibinfo  {journal} {Journal of Fluids and Structures}\ }\textbf {\bibinfo
  {volume} {93}} (\bibinfo {year} {2020}),\
  10.1016/j.jfluidstructs.2020.102875}\BibitemShut {NoStop}%
\bibitem [{\citenamefont {Thekkethil}\ \emph {et~al.}(2017)\citenamefont
  {Thekkethil}, \citenamefont {Shrivastava}, \citenamefont {Agrawal},\ and\
  \citenamefont {Sharma}}]{Thekkethil2017}%
  \BibitemOpen
  \bibfield  {author} {\bibinfo {author} {\bibnamefont {Thekkethil},
  \bibfnamefont {N.}}, \bibinfo {author} {\bibnamefont {Shrivastava},
  \bibfnamefont {M.}}, \bibinfo {author} {\bibnamefont {Agrawal}, \bibfnamefont
  {A.}}, and\ \bibinfo {author} {\bibnamefont {Sharma}, \bibfnamefont {A.}},\
  }\bibfield  {title} {\enquote {\bibinfo {title} {{Effect of wavelength of
  fish-like undulation of a hydrofoil in a free-stream flow}},}\ }\href
  {https://doi.org/10.1007/s12046-017-0619-7} {\bibfield  {journal} {\bibinfo
  {journal} {Sadhana - Academy Proceedings in Engineering Sciences}\ }\textbf
  {\bibinfo {volume} {42}} (\bibinfo {year} {2017}),\
  10.1007/s12046-017-0619-7}\BibitemShut {NoStop}%
\bibitem [{\citenamefont {Triantafyllou}, \citenamefont {Triantafyllou},\ and\
  \citenamefont {Gopalkrishnan}(1991)}]{Triantafyllou1991}%
  \BibitemOpen
  \bibfield  {author} {\bibinfo {author} {\bibnamefont {Triantafyllou},
  \bibfnamefont {M.~S.}}, \bibinfo {author} {\bibnamefont {Triantafyllou},
  \bibfnamefont {G.~S.}}, and\ \bibinfo {author} {\bibnamefont {Gopalkrishnan},
  \bibfnamefont {R.}},\ }\bibfield  {title} {\enquote {\bibinfo {title} {{Wake
  mechanics for thrust generation in oscillating foils}},}\ }\href
  {https://doi.org/10.1063/1.858173} {\bibfield  {journal} {\bibinfo  {journal}
  {Physics of Fluids A}\ }\textbf {\bibinfo {volume} {3}} (\bibinfo {year}
  {1991}),\ 10.1063/1.858173}\BibitemShut {NoStop}%
\bibitem [{\citenamefont {Triantafyllou}, \citenamefont {Weymouth},\ and\
  \citenamefont {Miao}(2016)}]{Triantafyllou2016}%
  \BibitemOpen
  \bibfield  {author} {\bibinfo {author} {\bibnamefont {Triantafyllou},
  \bibfnamefont {M.~S.}}, \bibinfo {author} {\bibnamefont {Weymouth},
  \bibfnamefont {G.~D.}}, and\ \bibinfo {author} {\bibnamefont {Miao},
  \bibfnamefont {J.}},\ }\bibfield  {title} {\enquote {\bibinfo {title}
  {{Biomimetic Survival Hydrodynamics and Flow Sensing}},}\ }\href
  {https://doi.org/10.1146/annurev-fluid-122414-034329} {\bibfield  {journal}
  {\bibinfo  {journal} {Annual Review of Fluid Mechanics}\ }\textbf {\bibinfo
  {volume} {48}} (\bibinfo {year} {2016}),\
  10.1146/annurev-fluid-122414-034329}\BibitemShut {NoStop}%
\bibitem [{\citenamefont {Tytell}(2004)}]{Tytell2004b}%
  \BibitemOpen
  \bibfield  {author} {\bibinfo {author} {\bibnamefont {Tytell}, \bibfnamefont
  {E.~D.}},\ }\bibfield  {title} {\enquote {\bibinfo {title} {{Kinematics and
  hydrodynamics of linear acceleration in eels, Anguilla rostrata}},}\ }\href
  {https://doi.org/10.1098/rspb.2004.2901} {\bibfield  {journal} {\bibinfo
  {journal} {Proceedings of the Royal Society B: Biological Sciences}\ }\textbf
  {\bibinfo {volume} {271}} (\bibinfo {year} {2004}),\
  10.1098/rspb.2004.2901}\BibitemShut {NoStop}%
\bibitem [{\citenamefont {Tytell}\ and\ \citenamefont
  {Lauder}(2008)}]{Tytell2008}%
  \BibitemOpen
  \bibfield  {author} {\bibinfo {author} {\bibnamefont {Tytell}, \bibfnamefont
  {E.~D.}}and\ \bibinfo {author} {\bibnamefont {Lauder}, \bibfnamefont
  {G.~V.}},\ }\bibfield  {title} {\enquote {\bibinfo {title} {{Hydrodynamics of
  the escape response in bluegill sunfish, Lepomis macrochirus}},}\ }\href
  {https://doi.org/10.1242/jeb.020917} {\bibfield  {journal} {\bibinfo
  {journal} {Journal of Experimental Biology}\ }\textbf {\bibinfo {volume}
  {211}} (\bibinfo {year} {2008}),\ 10.1242/jeb.020917}\BibitemShut {NoStop}%
\bibitem [{\citenamefont {{Van Buren}}\ \emph {et~al.}(2017)\citenamefont {{Van
  Buren}}, \citenamefont {Floryan}, \citenamefont {Quinn},\ and\ \citenamefont
  {Smits}}]{VanBuren2017}%
  \BibitemOpen
  \bibfield  {author} {\bibinfo {author} {\bibnamefont {{Van Buren}},
  \bibfnamefont {T.}}, \bibinfo {author} {\bibnamefont {Floryan}, \bibfnamefont
  {D.}}, \bibinfo {author} {\bibnamefont {Quinn}, \bibfnamefont {D.}}, and\
  \bibinfo {author} {\bibnamefont {Smits}, \bibfnamefont {A.~J.}},\ }\bibfield
  {title} {\enquote {\bibinfo {title} {{Nonsinusoidal gaits for unsteady
  propulsion}},}\ }\href {https://doi.org/10.1103/PhysRevFluids.2.053101}
  {\bibfield  {journal} {\bibinfo  {journal} {Physical Review Fluids}\ }\textbf
  {\bibinfo {volume} {2}} (\bibinfo {year} {2017}),\
  10.1103/PhysRevFluids.2.053101}\BibitemShut {NoStop}%
\bibitem [{\citenamefont {Wang}\ \emph {et~al.}(2021)\citenamefont {Wang},
  \citenamefont {Xu}, \citenamefont {Sung},\ and\ \citenamefont
  {Huang}}]{Wang2021}%
  \BibitemOpen
  \bibfield  {author} {\bibinfo {author} {\bibnamefont {Wang}, \bibfnamefont
  {L.~H.}}, \bibinfo {author} {\bibnamefont {Xu}, \bibfnamefont {C.~X.}},
  \bibinfo {author} {\bibnamefont {Sung}, \bibfnamefont {H.~J.}}, and\ \bibinfo
  {author} {\bibnamefont {Huang}, \bibfnamefont {W.~X.}},\ }\bibfield  {title}
  {\enquote {\bibinfo {title} {{Wall-attached structures over a traveling wavy
  boundary: Turbulent velocity fluctuations}},}\ }\href
  {https://doi.org/10.1103/PhysRevFluids.6.034611} {\bibfield  {journal}
  {\bibinfo  {journal} {Physical Review Fluids}\ }\textbf {\bibinfo {volume}
  {6}} (\bibinfo {year} {2021}),\ 10.1103/PhysRevFluids.6.034611}\BibitemShut
  {NoStop}%
\bibitem [{\citenamefont {Webb}(1984)}]{Webb1984}%
  \BibitemOpen
  \bibfield  {author} {\bibinfo {author} {\bibnamefont {Webb}, \bibfnamefont
  {P.~W.}},\ }\bibfield  {title} {\enquote {\bibinfo {title} {{Form and
  Function in Fish Swimming}},}\ }\href
  {https://doi.org/10.1038/scientificamerican0784-72} {\bibfield  {journal}
  {\bibinfo  {journal} {Scientific American}\ }\textbf {\bibinfo {volume}
  {251}} (\bibinfo {year} {1984}),\
  10.1038/scientificamerican0784-72}\BibitemShut {NoStop}%
\bibitem [{\citenamefont {Wei}\ \emph {et~al.}(2022)\citenamefont {Wei},
  \citenamefont {Hu}, \citenamefont {Zhang},\ and\ \citenamefont
  {Zeng}}]{Wei2022}%
  \BibitemOpen
  \bibfield  {author} {\bibinfo {author} {\bibnamefont {Wei}, \bibfnamefont
  {C.}}, \bibinfo {author} {\bibnamefont {Hu}, \bibfnamefont {Q.}}, \bibinfo
  {author} {\bibnamefont {Zhang}, \bibfnamefont {T.}}, and\ \bibinfo {author}
  {\bibnamefont {Zeng}, \bibfnamefont {Y.}},\ }\bibfield  {title} {\enquote
  {\bibinfo {title} {{Passive hydrodynamic interactions in minimal fish
  schools}},}\ }\href {https://doi.org/10.1016/j.oceaneng.2022.110574}
  {\bibfield  {journal} {\bibinfo  {journal} {Ocean Engineering}\ }\textbf
  {\bibinfo {volume} {247}} (\bibinfo {year} {2022}),\
  10.1016/j.oceaneng.2022.110574}\BibitemShut {NoStop}%
\bibitem [{\citenamefont {Weihs}(1973)}]{weihs1973hydromechanics}%
  \BibitemOpen
  \bibfield  {author} {\bibinfo {author} {\bibnamefont {Weihs}, \bibfnamefont
  {D.}},\ }\bibfield  {title} {\enquote {\bibinfo {title} {{Hydromechanics of
  Fish Schooling}},}\ }\href@noop {} {\bibfield  {journal} {\bibinfo  {journal}
  {Nature}\ }\textbf {\bibinfo {volume} {241}},\ \bibinfo {pages} {290--291}
  (\bibinfo {year} {1973})}\BibitemShut {NoStop}%
\bibitem [{\citenamefont {Yu}\ and\ \citenamefont
  {Huang}(2021)}]{yu2021scaling}%
  \BibitemOpen
  \bibfield  {author} {\bibinfo {author} {\bibnamefont {Yu}, \bibfnamefont
  {Y.-L.}}and\ \bibinfo {author} {\bibnamefont {Huang}, \bibfnamefont
  {K.-J.}},\ }\bibfield  {title} {\enquote {\bibinfo {title} {{Scaling law of
  fish undulatory propulsion}},}\ }\href@noop {} {\bibfield  {journal}
  {\bibinfo  {journal} {Physics of Fluids}\ }\textbf {\bibinfo {volume} {33}},\
  \bibinfo {pages} {61905} (\bibinfo {year} {2021})}\BibitemShut {NoStop}%
\bibitem [{\citenamefont {Yucel}, \citenamefont {Sahin},\ and\ \citenamefont
  {Unal}(2022)}]{Yucel2022}%
  \BibitemOpen
  \bibfield  {author} {\bibinfo {author} {\bibnamefont {Yucel}, \bibfnamefont
  {S.~B.}}, \bibinfo {author} {\bibnamefont {Sahin}, \bibfnamefont {M.}}, and\
  \bibinfo {author} {\bibnamefont {Unal}, \bibfnamefont {M.~F.}},\ }\bibfield
  {title} {\enquote {\bibinfo {title} {{Propulsive performance of plunging
  airfoils in biplane configuration}},}\ }\href
  {https://doi.org/10.1063/5.0083040} {\bibfield  {journal} {\bibinfo
  {journal} {Physics of Fluids}\ }\textbf {\bibinfo {volume} {34}} (\bibinfo
  {year} {2022}),\ 10.1063/5.0083040}\BibitemShut {NoStop}%
\bibitem [{\citenamefont {Zheng}\ \emph {et~al.}(2005)\citenamefont {Zheng},
  \citenamefont {Kashimori}, \citenamefont {Hoshino}, \citenamefont {Fujita},\
  and\ \citenamefont {Kambara}}]{Zheng2005}%
  \BibitemOpen
  \bibfield  {author} {\bibinfo {author} {\bibnamefont {Zheng}, \bibfnamefont
  {M.}}, \bibinfo {author} {\bibnamefont {Kashimori}, \bibfnamefont {Y.}},
  \bibinfo {author} {\bibnamefont {Hoshino}, \bibfnamefont {O.}}, \bibinfo
  {author} {\bibnamefont {Fujita}, \bibfnamefont {K.}}, and\ \bibinfo {author}
  {\bibnamefont {Kambara}, \bibfnamefont {T.}},\ }\bibfield  {title} {\enquote
  {\bibinfo {title} {{Behavior pattern (innate action) of individuals in fish
  schools generating efficient collective evasion from predation}},}\ }\href
  {https://doi.org/10.1016/j.jtbi.2004.12.025} {\bibfield  {journal} {\bibinfo
  {journal} {Journal of Theoretical Biology}\ }\textbf {\bibinfo {volume}
  {235}} (\bibinfo {year} {2005}),\ 10.1016/j.jtbi.2004.12.025}\BibitemShut
  {NoStop}%
\end{thebibliography}%


%

\pagebreak

\appendix

\section{Flow structure maps in detail}
\label{appendix_flow_map_detail}

\begin{sidewaysfigure}
	\centering
	\includegraphics[width=1\linewidth]{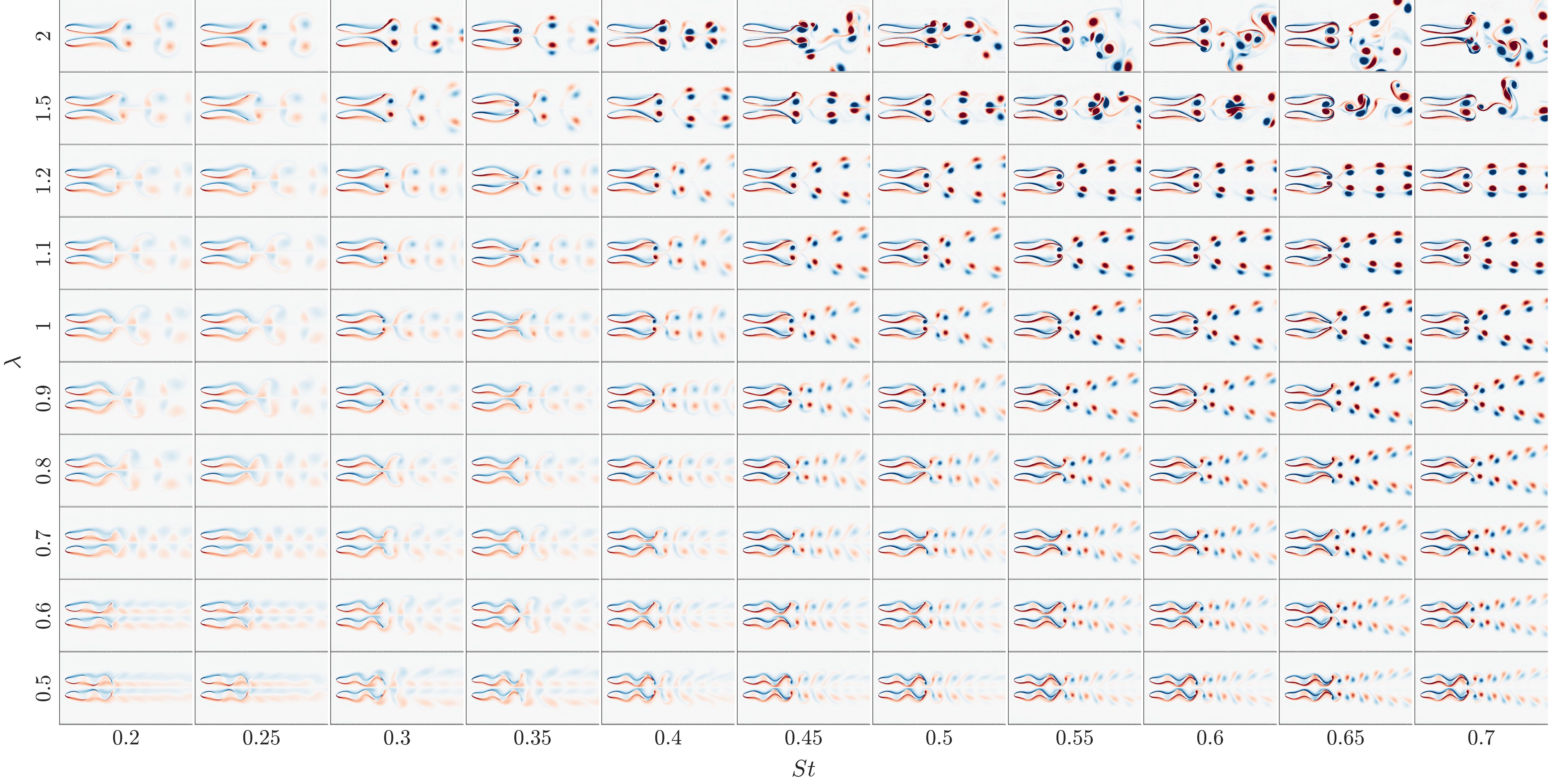}
	\caption{Flow structure visualised by vorticity contours at $ Re = 1000 $ with $ St = 0.2 - 0.7 $ and $ \lam = 0.5 - 2 $.}
	\label{fig:vor_map_re_1000}
\end{sidewaysfigure}

\begin{sidewaysfigure}
	\centering
	\includegraphics[width=1\linewidth]{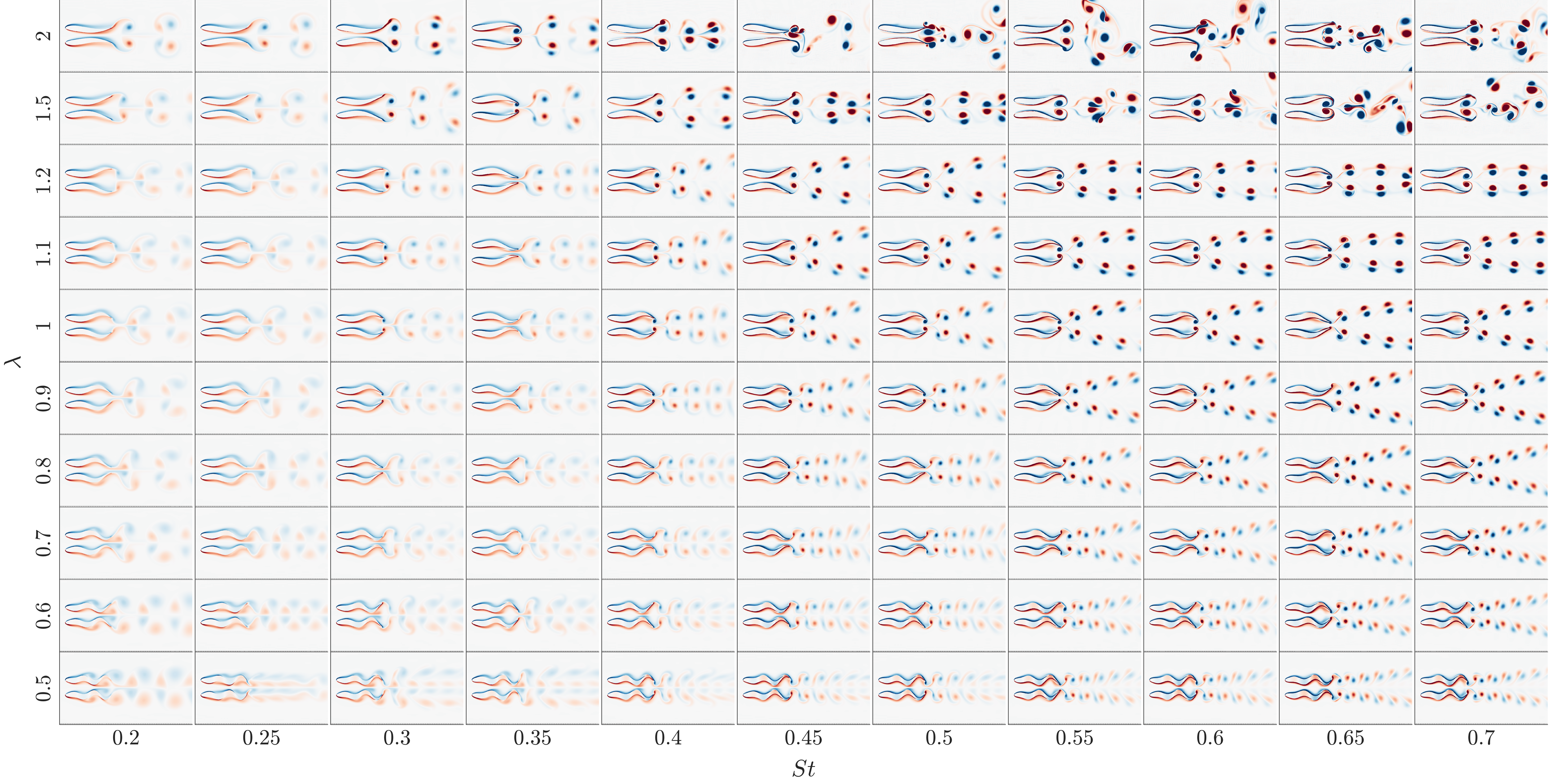}
	\caption{Flow structure visualised by vorticity contours at $ Re = 1250 $ with $ St = 0.2 - 0.7 $ and $ \lam = 0.5 - 2 $.}
	\label{fig:vor_map_re_1250}
\end{sidewaysfigure}

\begin{sidewaysfigure}
	\centering
	\includegraphics[width=1\linewidth]{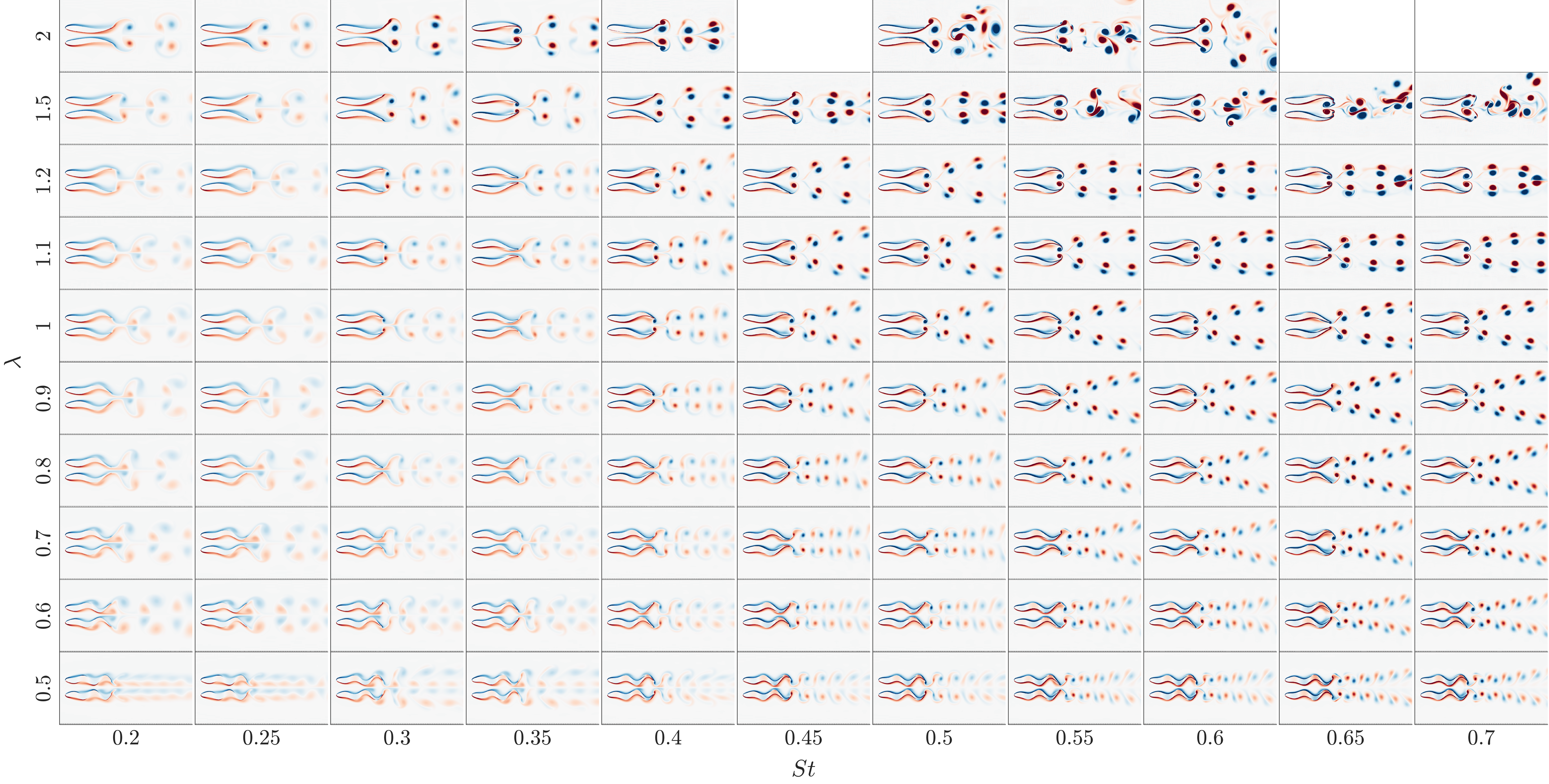}
	\caption{Flow structure visualised by vorticity contours at $ Re = 1500 $ with $ St = 0.2 - 0.7 $ and $ \lam = 0.5 - 2 $.}
	\label{fig:vor_map_re_1500}
\end{sidewaysfigure}

\begin{sidewaysfigure}
	\centering
	\includegraphics[width=1\linewidth]{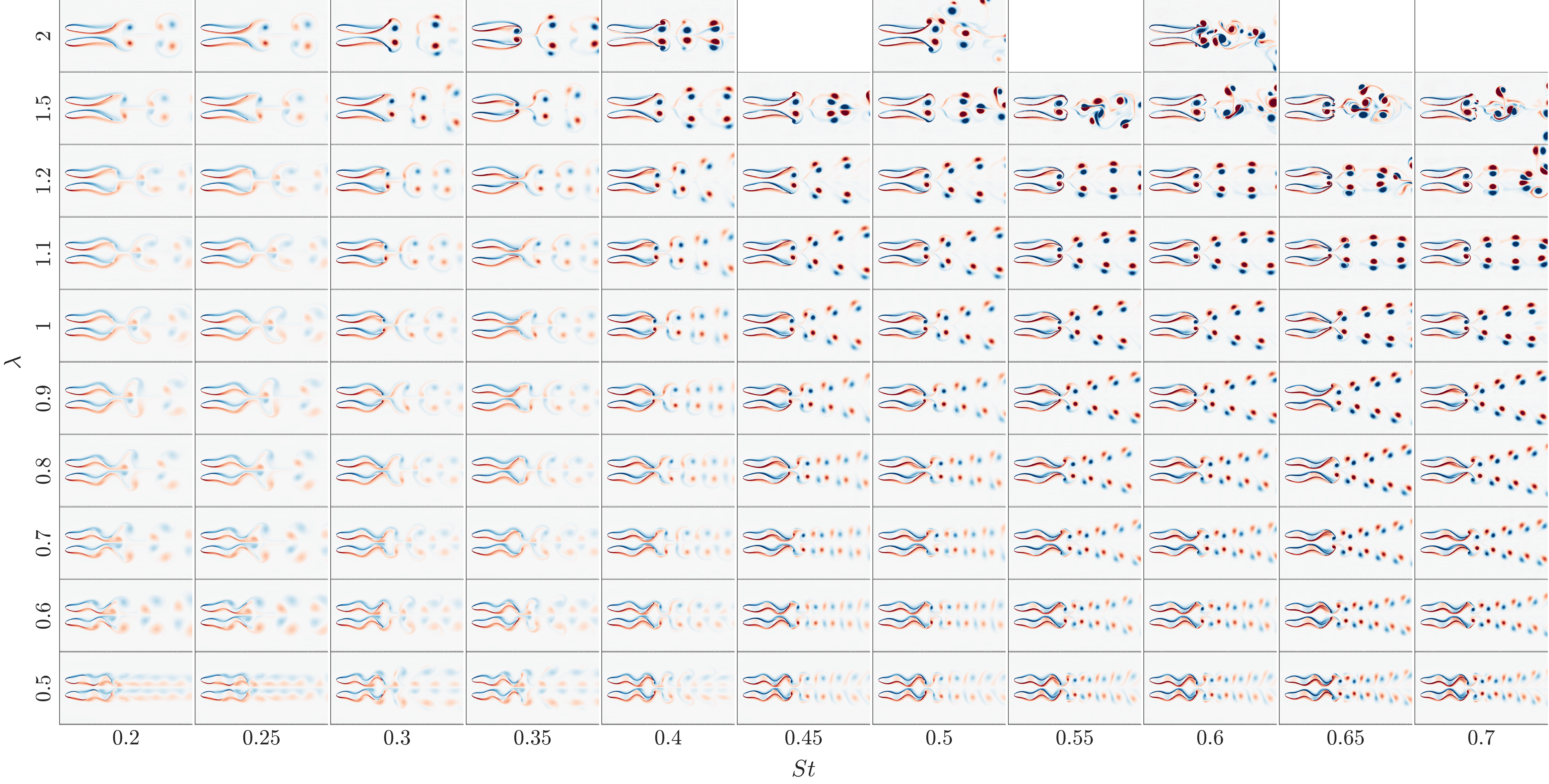}
	\caption{Flow structure visualised by vorticity contours at $ Re = 1750 $ with $ St = 0.2 - 0.7 $ and $ \lam = 0.5 - 2 $.}
	\label{fig:vor_map_re_1750}
\end{sidewaysfigure}

\begin{sidewaysfigure}
	\centering
	\includegraphics[width=1\linewidth]{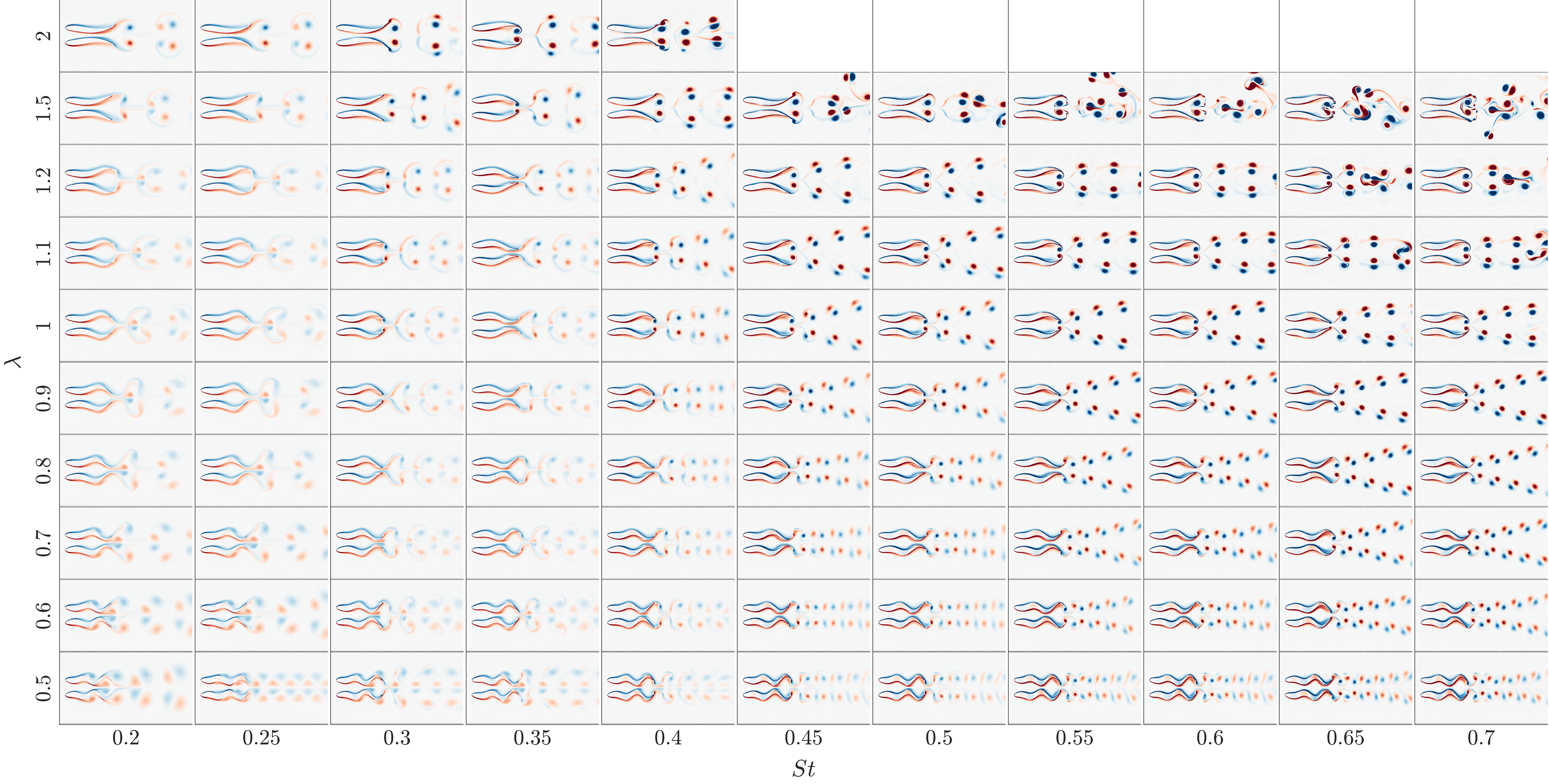}
	\caption{Flow structure visualised by vorticity contours at $ Re = 2000 $ with $ St = 0.2 - 0.7 $ and $ \lam = 0.5 - 2 $.}
	\label{fig:vor_map_re_2000}
\end{sidewaysfigure}

This section supplements the detailed flow structure maps in the tested parametric space of our present study.
Vorticity contours are drawn to illustrate the flow structures at the Reynolds numbers of $ \re = 1000-2000 $, as seen in \Cref{fig:vor_map_re_1000,fig:vor_map_re_1250,fig:vor_map_re_1500,fig:vor_map_re_1750,fig:vor_map_re_2000}. These flow structures can be classified into six types, as exemplified by \Cref{fig:example_flow_struct} and the type distribution is illustrated by \Cref{fig:map_flow_struct}, as discussed in \Cref{sec_flow_map}.

\clearpage
\FloatBarrier
\section{Derivation of Thrust Amplification Factor and Metrics for Thrust \& Propulsive Efficiency}
This section expounds on the derivation of the thrust amplification factor and elucidates our choice of metric for propulsive efficiency. The metric for net Froude or propulsive efficiency employed herein is congruent with that used by \cite{Akanyeti2017}, denoting the efficacy with which the thrust force is generated.

The justification for our computation of the thrust amplification factor lies in the following: given a situation where the net average thrust coefficient for a single foil, $\ctms$, converges to zero — that is, in a steady swimming scenario — the denominator of the thrust amplification factor, $ \ctm/\ctms $, would correspondingly converge to zero. This occurrence results in an exorbitantly large value for the amplification factor. However, our data-processing procedures duly account for this. In the course of our post-processing computation of the thrust amplification factor $ \ctm/\ctms $ and efficiency amplification factor $ \eta_{\rm pair} / \eta_{\rm single} $, we calculate the amplification factor only if $ \ctm > 0.01 $ and $ \ctms > 0.01 $ — both of which are larger than a pre-established threshold. If these conditions are not met, the amplification factor is set as zero in the heat maps. The absence of this restriction would have the amplification factor reaching an excessive scale of $ 10^3 $ owing to its proximity to zero and numerical discrepancies. The choice of 0.01 ensures a satisfactory distance from zero while considering the potential range of numerical variances.

Delving further into specifics, we discerned that $ \ctm = 0.2289 $ while $ \ctms = 0.0165 $. We thus regard the single swimmer \textit{net} thrust $ \ctms = 0.0165 $ as sufficiently larger than the steady swimming condition $ \ctms \approx 0 $, especially in comparison to other results, where the net thrust can approach zero as $ \ctms = 2.24 \sci{-5} \approx 0 $.

This section also examines the definitions of thrust and efficiency. Our study principally addresses the imbalanced situation of linear acceleration, deviating from the steady swimming conditions examined in prior works \citep{pan2020computational}. Consequently, our chosen metrics are the \textit{net} thrust coefficient and the \textit{net} Froude or propulsive efficiency — that is, the efficiency calculated from net thrust. In the context of this work, the \textit{net} thrust coefficient, $ \ctm $, signifies the net force propelling the swimmer, while the \textit{net} Froude efficiency indicates the proficiency with which the force or acceleration is generated. For an in-depth exploration, we kindly direct the reader to our previous publication which focuses on acceleration \cite{lin2022swimming}, where we have extensively justified our chosen metrics.

The concept of efficiency is identical between the present paper and that by \cite{Dong2007}. The metric is fundamentally the Froude efficiency derived from net thrust as:

\begin{equation}\label{equ:net_froude}
\eta = \frac{\rm Thrust_{ave} \times U}{\rm Power_{ave}} = \frac{\bar{C_T}}{\bar{C_P}}
\end{equation}

In their paper "Efficiency of Fish Propulsion", \cite{Maertens2015} reviewed numerous extant metrics of "fish swimming efficiency". The "propulsive efficiency" adopted in this paper equates to the "net propulsive efficiency" referenced in \cite{Maertens2015}, defined in their work as $ \eta = {\bar{C_T}}/{\bar{C_P}} $. The optimal "net propulsive efficiency" equates to the minimum energy consumption required to attain a given acceleration. This is distinct from the optimal steady swimming efficiency, which sustains a specific velocity with zero acceleration. \cite{Maertens2015} proposed a novel metric, quasi-propulsive efficiency, as $ \eta = ({\bar{C_T} + C_D})/{\bar{C_P}} $, which incorporates a separately measured drag term. However, we posit that this new metric is better suited for gauging overall swimming performance rather than acceleration performance \cite{Maertens2015}. For more detailed discussions, please refer to the appendix in our previous publication \citep{lin2022swimming}.

\cite{Maertens2017} employed quasi-propulsive efficiency, where the "thrust" used for calculation is determined by "towing a rigid body in a static flow with a prescribed velocity". This methodology circumvents the issue of zero net thrust force and consequent zero efficiency during steady swimming— the central focus of the \cite{Maertens2017} paper. Contrarily, the current manuscript prioritises acceleration over steady swimming conditions, wherein the net force is non-zero. Hence, we compute the propulsive efficiency using net thrust instead of the method used by \cite{Maertens2017}.

\cite{pan2020computational} likewise focused on steady swimming. They used \textit{pure} thrust— that is, without considering the drag— rather than \textit{net} thrust, to calculate a modified form of Froude efficiency as $ \eta = {FU}/{(FU + P_{wake})} $, where $ P_{wake} $ denotes the power in the wake.

In summary, the calculation of thrust amplification factor and the chosen efficiency metric of net Froude or propulsive efficiency is apt for investigating the current issue pertaining to linear acceleration.

\end{document}